%% file: main.tex
\RequirePackage{lineno}
\documentclass[twocolumn,epjc3,superscriptaddress,unsortedaddress,floatfix]{svjour3}
\RequirePackage{amsmath}
\RequirePackage{amssymb}
\pdfoutput=1
\journalname{Eur. Phys. J. C}

\input{common/preamble}
\begin{document}

\title{Prospects for Beyond the Standard Model Physics Searches at the Deep Underground Neutrino Experiment}

\date{\today}
\authorrunning{DUNE Collaboration}
\input{Latex-svjour}

\onecolumn
\maketitle
\twocolumn
\sloppy

\begin{abstract}
The Deep Underground Neutrino Experiment (DUNE) will be a powerful tool for a variety of physics topics.  The high-intensity proton beams provide a large neutrino flux, sampled by a near detector system consisting of a combination of capable precision detectors, and by the massive far detector system located deep underground. This configuration sets up DUNE as a machine for discovery, as it enables opportunities not only to perform precision neutrino measurements that may uncover deviations from the present three-flavor mixing paradigm, but also to  discover new particles and unveil new interactions and symmetries beyond those predicted in the \dword{sm}. Of the many potential \dword{bsm} topics DUNE will probe, this paper presents a selection of studies quantifying DUNE's sensitivities to sterile neutrino mixing, heavy neutral leptons, non-standard interactions, CPT symmetry violation, Lorentz invariance violation, neutrino trident production, dark matter from both beam induced and cosmogenic sources, baryon number violation, and other new physics topics that complement those at high-energy colliders and significantly extend the present reach.
\end{abstract}

\input{paper.tex}

\bibliographystyle{utphys}
\bibliography{common/tdr-citedb}

\end{document}

%% file: Latex-svjour.tex
\author {
The DUNE collaboration\\\\
    B.~Abi\thanksref{Oxford}
	 \and R.~Acciarri\thanksref{Fermi}
	 \and M.~A.~Acero\thanksref{Atlantico}
	 \and G.~Adamov\thanksref{Georgian}
	 \and D.~Adams\thanksref{Brookhaven}
	 \and M.~Adinolfi\thanksref{Bristol}
	 \and Z.~Ahmad\thanksref{VariableEnergy}
	 \and J.~Ahmed\thanksref{Warwick}
	 \and T.~Alion\thanksref{Sussex}
	 \and S.~Alonso Monsalve\thanksref{CERN}
% ...
}

\thankstext{e1}{E-mail: lkoerner@central.uh.edu}
\thankstext{e2}{E-mail: alex.sousa@uc.edu}
\thankstext{e3}{E-mail: jaehoon@uta.edu}

\institute {
% ...
\pagebreak[0] Universidad del Atl{\'a}ntico, Barranquilla, Atl{\'a}ntico, Colombia\label{Atlantico}
% ...
\and\pagebreak[0] University of Bristol, Bristol BS8 1TL, United Kingdom\label{Bristol}
	 \and\pagebreak[0] Brookhaven National Laboratory, Upton, NY 11973, USA\label{Brookhaven}
% ...
\and\pagebreak[0] CERN, The European Organization for Nuclear Research, 1211 Meyrin, Switzerland\label{CERN}
% ...
\and\pagebreak[0] Fermi National Accelerator Laboratory, Batavia, IL 60510, USA\label{Fermi}
% ...
\and\pagebreak[0] Georgian Technical University, Tbilisi, Georgia\label{Georgian}
% ...
\and\pagebreak[0] University of Oxford, Oxford, OX1 3RH, United Kingdom\label{Oxford}
% ...
\and\pagebreak[0] University of Sussex, Brighton, BN1 9RH, United Kingdom\label{Sussex}
% ...
\and\pagebreak[0] Variable Energy Cyclotron Centre, 700 064 West Bengal, India\label{VariableEnergy}
% ...
\and\pagebreak[0] University of Warwick, Coventry CV4 7AL, United Kingdom\label{Warwick}
% ...
}

%% file: paper.tex
 \section{Introduction}
\label{sec:intro}

The \dword{dune} is a next-generation, \dword{lbl} neutrino oscillation experiment, designed to be sensitive to $\numu$ to $\nue$ oscillation. The experiment consists of a high-power, broadband neutrino beam, a powerful precision \dword{nd} complex located at Fermi National Accelerator Laboratory, in Batavia, Illinois, USA, and a massive \dword{lartpc} \dword{fd} located at the 4850~ft level of Sanford Underground Research Facility (SURF), in Lead, South Dakota, USA. The baseline of 1285~km provides sensitivity, in a single experiment, to all parameters governing \dword{lbl} neutrino oscillation. The deep underground location of the \dword{fd} facilitates sensitivity to nucleon decay and other rare processes including low-energy neutrino detection enabling, for instance, observation of neutrinos from a core-collapse supernova. 

Owing to the high-power proton beam facility, the \dword{nd} consisting of precision detectors capable of off-axis data taking and the massive FD, DUNE provides enormous opportunities to probe phenomena beyond the \dword{sm} traditionally difficult to reach in neutrino experiments. Of such vast, rich physics topics that profoundly expand those probed in the past neutrino experiments, this paper reports a selection of studies of DUNE's sensitivity to a variety of \dword{bsm} particles and effects, initially presented in the physics volume of the DUNE \dword{tdr}~\cite{Abi:2020evt} recently made available.
Some of these phenomena impact the \dword{lbl} oscillation measurement, while others may be detected by \dword{dune} using specific analyses. 

Section~\ref{sec:analysis} describes some of the common assumptions and tools used in these analyses. Section~\ref{sec:sterile} discusses sensitivity to sterile neutrinos, Section~\ref{sec:nonUnitarity} looks into the effect of non-unitary of the neutrino mixing matrix, Section~\ref{sec:nsi} describes sensitivity to non-standard neutrino interactions, Section~\ref{sect:cpt} discusses sensitivity to CPT and Lorentz violation, Section~\ref{sec:tridents} describes the sensitivity to new physics by measuring neutrino trident production, Section~\ref{sec:DM} discusses various dark matter searches that could be performed by DUNE, Section~\ref{sect:bnv} describes sensitivity to baryon number violation by one and two units, and Section~\ref{sec:otheropps} lists some other possible avenues for \dword{bsm} physics searches.

These studies reveal that \dword{dune} can probe a rich and diverse BSM phenomenology at the discovery level, as in the case of searches for dark matter created in the high-power proton beam interactions and from cosmogenic sources, or by significantly improving existing constraints, as in the cases of sterile neutrino mixing, non-standard neutrino interactions, CPT violation, new physics enhancing neutrino trident production, and nucleon decay. 

\section{Analysis Details}
\label{sec:analysis}
The \dword{bsm} searches presented in this paper span a wide variety of physics topics and techniques.  The analyses rely on neutrino beam data taken at the \dword{nd} and/or \dword{fd}, atmospheric or other astrophysical sources of neutrinos, or signal from the detector material itself, as in nucleon decay searches.  This section summarizes some of the common assumptions and tools used in the analyses, with more details provided in the following sections.

\subsection{Detector Assumptions}

The \dword{dune} \dword{fd} will consist of four \nominalmodsize fiducial mass \lartpc modules with integrated \dwords{pds}~\cite{Acciarri:2016crz,Acciarri:2015uup,Acciarri:2016ooe}. In these analyses, we assume all four modules have identical responses.  All of the analyses described will use data from the \dword{fd}, except for the analyses presented in Sections~\ref{sec:tridents}, \ref{sec:darkmatter_ND}, and \ref{sec:hnl}, which use data exclusively from the \dword{nd}.

The \dword{nd} will be located at a distance of \ndfromtarget from the target. 
The \dword{nd} concept consists of a modular \lartpc, a magnetized high-pressure gas argon TPC and a beam monitor.
The combination of the first two detectors is planned to be movable to sample the off-axis neutrino spectrum to reduce flux uncertainties, a concept called DUNE-PRISM~\cite{Abi:2020evt}.
Since the \dword{nd} configuration, however, was not yet finalized at the time these studies were performed, we adopted only the \dword{lartpc} component of the detector and its fiducial volume. 
In the analyses presented here, 
the \dword{lartpc} is assumed to be \SI{7}{m} wide, \SI{3}{m} high, and \SI{5}{m} long. 
The fiducial volume is assumed to include the detector volume up to 50 cm of each face of the detector.
The \dword{nd} properties are given in Table~\ref{tabND}. The signal and background efficiencies vary with the physics model being studied. Detailed signal and background efficiencies for each physics topic are discussed along with each analysis.  

\begin{table}[htp]
\centering
\caption{LArTPC ND properties used in some of the BSM physics analyses.}
    \label{tabND}
\begin{tabular}{ l | c }\hline
   Properties & Values\\ \hline\hline
    Active volume &  \SI{7}{m} wide, \SI{3}{m} high, \SI{5}{m} long \\ 
    Fiducial volume & \SI{6}{m} wide, \SI{2}{m} high, \SI{4}{m} long\\ 
    Total mass  & 147 ton \\ 
    Fiducial mass & 67.2 ton \\ 
    Distance from target & \ndfromtarget \\ \hline
    \end{tabular}
 \end{table} 

\subsection{Neutrino Beam Assumptions}
The analyses described in Sections~\ref{sec:sterile}, \ref{sec:nonUnitarity}, \ref{sec:nsi}, and \ref{sect:cpt} are based on analysis of neutrino beam data at both the \dword{nd} and \dword{fd}.
The DUNE neutrino beam is produced using protons from Fermilab's Main Injector and a traditional horn-focusing system~\cite{Abi:2020wmh}. The polarity of the focusing magnets may be reversed to produce a neutrino- or antineutrino-dominated beam.  This optimized beam configuration includes a three-horn focusing system with a 1~m long target embedded within the first horn and a decay pipe with \SI{194}{m} length and \SI{4}{m} diameter. The neutrino flux produced by this beamline is simulated at a distance of \SI{574}{m} downstream of the neutrino target for the \dword{nd} and \SI{1285}{km} for the \dword{fd}. Fluxes have been generated for both neutrino mode and antineutrino mode using G4LBNF~\cite{Aliaga:2016oaz,Abi:2020evt}, a \textsc{Geant4}-based simulation~\cite{Agostinelli:2002hh,Allison:2006ve,Allison:2016lfl}. 

Results based on beam neutrino data are given for a 300~\ktMWyr exposure. With the current deployment plan~\cite{Abi:2020evt}, this exposure will be achieved in approximately 7~years once the beam is operational. For results not based on beam data, the exposure is given in units of \ktyr in each relevant section.

\subsection{Tools}

In the analyses presented in Sections~\ref{sec:sterile}, \ref{sec:nonUnitarity}, \ref{sec:nsi}, and \ref{sect:cpt}, the simulation of the DUNE experimental setup was performed with the \dword{globes} software~\cite{Huber:2004ka,Huber:2007ji}. Unless otherwise noted, the neutrino fluxes used in the \dword{bsm} physics analysis are the same as those used in the DUNE \dword{lbl} three-flavor analysis~\cite{Abi:2020evt}. The configuration of the beam used in \dword{nd} analyses is assumed to be a 120~GeV proton beam with 1.2~MW beam power at 56\% uptime, providing $1.1\times10^{21}$ POT/year. Cross-section files describing \dword{nc} and \dword{cc} interactions with argon are generated using \dword{genie}~\cite{Andreopoulos:2009rq,Andreopoulos:2015wxa} version 2.8.4. The true-to-reconstructed smearing matrices and the selection efficiency as a function of energy for various signal and background modes are generated using nominal DUNE MC simulation. A \fdfiducialmass fiducial mass is assumed for the \dword{fd}, exposed to a \SI{120}{GeV}, \SI{1.2}{MW} beam. 
The $\nu_{e}$ and $\bar\nu_{e}$ appearance signal modes have independent normalization uncertainties of $2\%$ each, while $\nu_{\mu}$ and $\bar{\nu}_{\mu}$ disappearance signal modes have independent normalization uncertainties of $5\%$. The background normalization uncertainties range from $5\%$ to $20\%$ and include correlations among various sources of background. More details can be found in Ref.~\cite{Abi:2020evt}. 

The neutrino trident search presented in Section~\ref{sec:tridents} and the baryon number violation analyses presented in Section~\ref{sect:bnv} use samples of simulated and reconstructed signal and background events, produced using standard \dword{dune} detection simulation and reconstruction software. Further details are given in those sections.

For analyses that use neither \dword{globes} nor the standard DUNE simulation and reconstruction software, such as the dark matter analyses described in Section~\ref{sec:DM} and several of the analyses described in Section~\ref{sec:otheropps}, details are given in the relevant sections.

\section{Sterile Neutrino Mixing}
\label{sec:sterile}

Experimental results in tension with the three-neutrino-flavor paradigm, which may be interpreted as mixing between the known active neutrinos and one or more sterile states, have led to a rich and diverse program of searches for oscillations into sterile neutrinos~\cite{ref:tension,Gariazzo:2017fdh}. DUNE is sensitive over a broad range of potential sterile neutrino mass splittings by looking for disappearance of \dword{cc} and \dword{nc}  interactions over the long distance separating the \dword{nd} and \dword{fd}, as well as over the short baseline of the \dword{nd}. With a longer baseline, a more intense beam, and a high-resolution large-mass \dword{fd}, compared to previous experiments, DUNE provides a unique opportunity to improve significantly on the sensitivities of the existing probes, and greatly enhance the ability to map the extended parameter space if a sterile neutrino is discovered.
In the sterile neutrino mixing studies presented here, we assume a minimal 3+1 oscillation scenario with three active neutrinos and one sterile neutrino, which includes a new independent neutrino mass-squared difference, $\Delta m^2_{41}$, and for which the mixing matrix is extended with three new mixing angles, $\theta_{14}$, $\theta_{24}$, $\theta_{34}$, and two additional phases $\delta_{14}$ and $\delta_{24}$.

Disappearance of the beam neutrino flux between the \dword{nd} and \dword{fd} results from the quadratic suppression of the sterile mixing angle measured in appearance experiments, $\theta_{\mu e}$, with respect to its disappearance counterparts, $\theta_{\mu\mu}\approx\theta_{24}$ for \dword{lbl} experiments, and $\theta_{ee}\approx\theta_{14}$ for reactor experiments. These disappearance effects have not yet been observed and are in tension with appearance results~\cite{ref:tension,Gariazzo:2017fdh} when global fits of all available data are carried out. The exposure of DUNE's high-resolution \dword{fd} to the high-intensity LBNF beam will also allow direct probes of non-standard electron (anti)neutrino appearance.

DUNE will look for active-to-sterile neutrino mixing using the reconstructed energy spectra of both \dword{nc} and \dword{cc}  neutrino interactions  in the \dword{fd}, and their comparison to the extrapolated predictions from the \dword{nd} measurement. Since \dword{nc} cross sections and interaction topologies are the same for all three active neutrino flavors, the \dword{nc} spectrum is insensitive to standard neutrino mixing. However, should there be oscillations into a fourth light neutrino, an energy-dependent depletion of the neutrino flux would be observed at the \dword{fd}, as the sterile neutrino would not interact in the detector volume. Furthermore, if sterile neutrino mixing is driven by a large mass-square difference $\Delta m^2_{\rm{41}}$ $\sim$1\,eV$^{2}$, the \dword{cc} spectrum will be distorted at energies higher than the energy corresponding to the standard oscillation maximum. Therefore, \dword{cc} disappearance is also a powerful probe of sterile neutrino mixing at long baselines.

\begin{figure}[htb]
\setlength{\lineskip}{5pt}
    \centering
	  	\includegraphics[width=0.98\columnwidth]{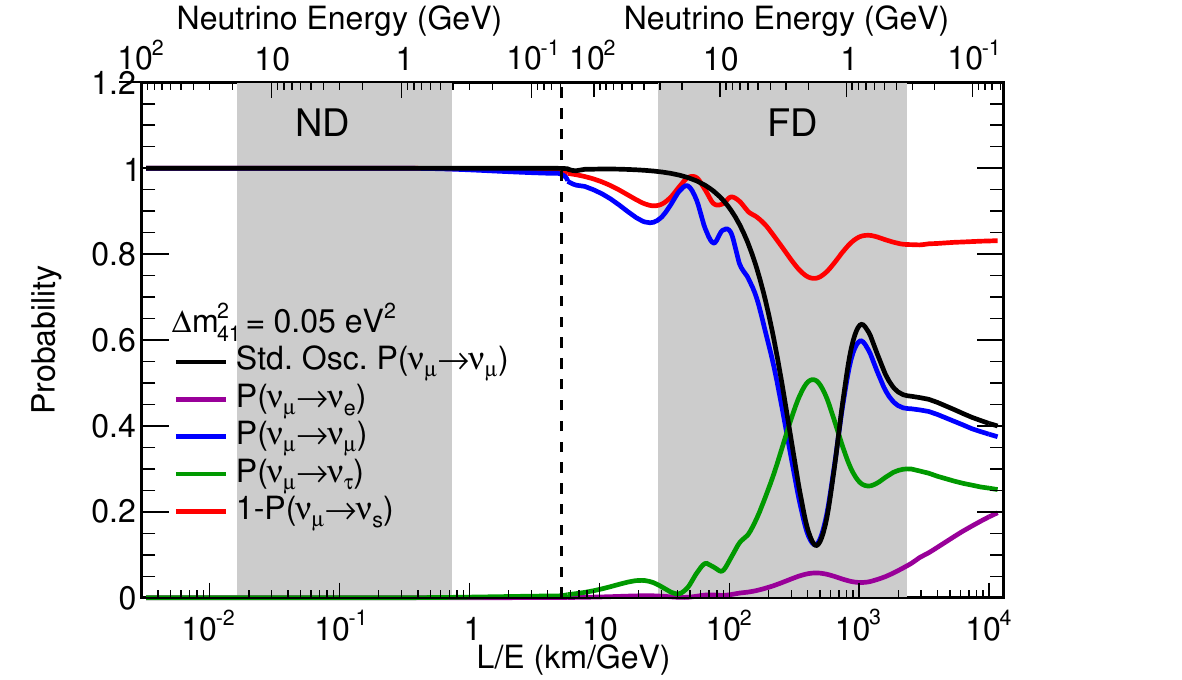}
        \includegraphics[width=0.98\columnwidth]{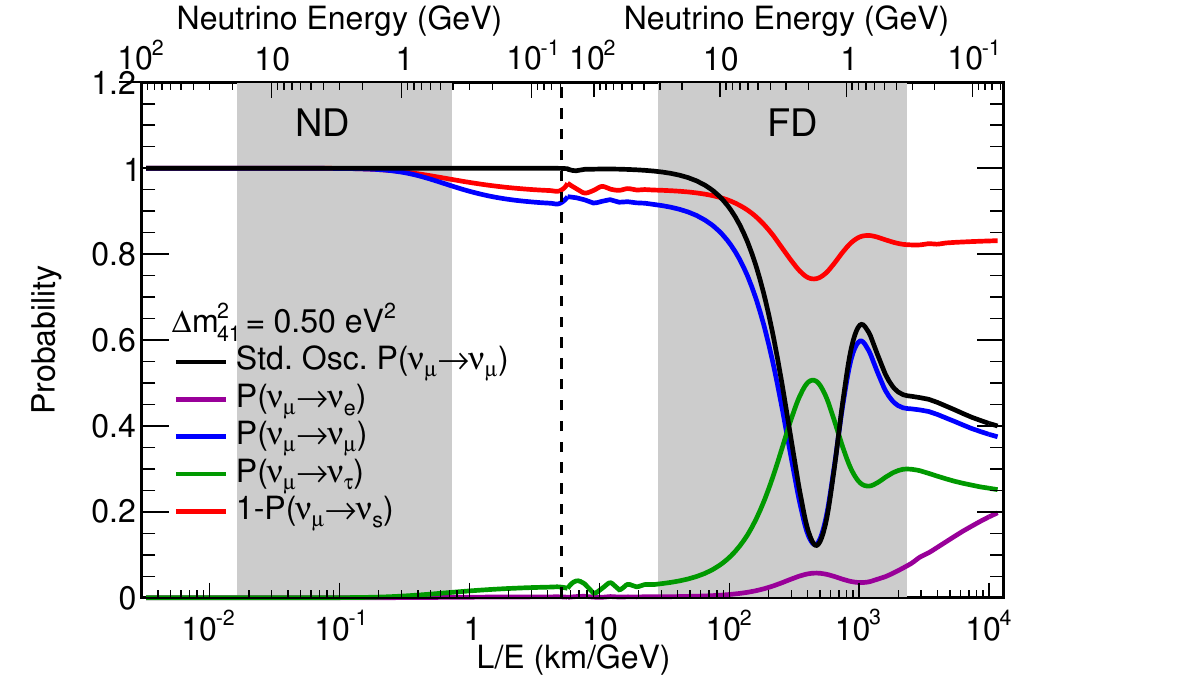}
        \includegraphics[width=0.98\columnwidth]{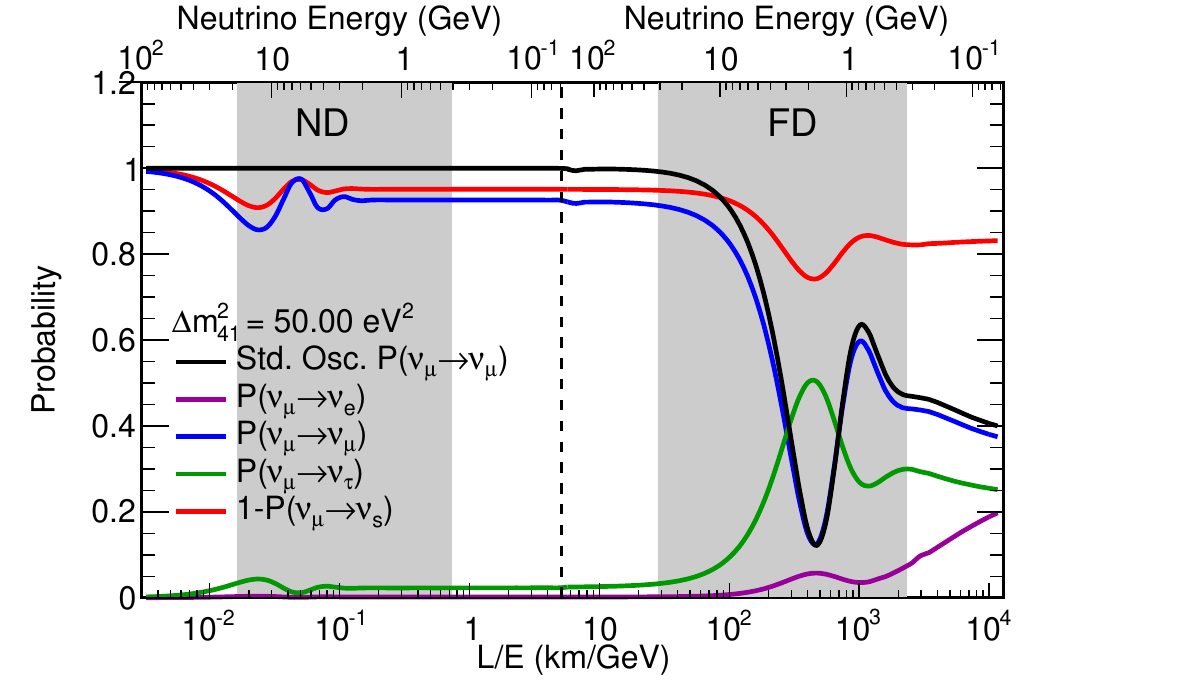}
\caption[Regions of $L/E$ probed by DUNE for  
3-flavor and 3+1-flavor $\nu$ oscillations]
{Regions of $L/E$ probed by the DUNE detector compared to 3-flavor and 3+1-flavor neutrino disappearance and appearance probabilities. The gray-shaded areas show the range of true neutrino energies probed by the \dword{nd} and \dword{fd}. The top axis shows true neutrino energy, increasing from right to left. The top plot shows the probabilities assuming mixing with one sterile neutrino with $\Delta m^2_{\rm{41}}=0.05$~eV$^2$, corresponding to the slow oscillations regime. The middle plot assumes mixing with one sterile neutrino with $\Delta m^2_{\rm{41}}=0.5$~eV$^2$, corresponding to the intermediate oscillations regime. The bottom plot includes mixing with one sterile neutrino with $\Delta m^2_{\rm{41}}=50$~eV$^2$, corresponding to the rapid oscillations regime. As an example, the slow sterile oscillations cause visible distortions in the three-flavor \numu~survival probability (blue curve) for neutrino energies $\sim10$\,GeV, well above the three-flavor oscillation minimum.}
\label{fig:regimes}
\end{figure}

We assume the mixing matrix augmented with one sterile state is parameterized by $U\nobreak=\nobreak R_{34}S_{24}S_{14}R_{23}S_{13}R_{12}$~\cite{Harari:1986xf}, where $R_{ij}$ is the rotational matrix for the mixing angle $\theta_{ij}$, and $S_{ij}$ represents a complex rotation by the mixing angle $\theta_{ij}$ and the $CP$-violating phase $\delta_{ij}$. At long baselines the \dword{nc} disappearance probability to first order for small mixing angles is then approximated by:

\begin{equation}\label{eq:numu_nus}
\begin{aligned}
1 - P(\nu_{\mu} \rightarrow \nu_s) & \approx 1 - \cos^4\theta_{14}\cos^2\theta_{34}\sin^{2}2\theta_{24}\sin^2\Delta_{41} \\
& - \sin^2\theta_{34}\sin^22\theta_{23}\sin^2\Delta_{31} \\
& + \frac{1}{2}\sin\delta_{24}\sin\theta_{24}\sin2\theta_{23}\sin\Delta_{31},
\end{aligned}
\end{equation}
where $\Delta_{ji} = \frac{\Delta m^2_{ji}L}{4E}$. 
The relevant oscillation probability for \numu~\dword{cc} disappearance is the \numu~survival probability, similarly approximated by:

\begin{equation}
\begin{aligned}
P(\nu_{\mu} \rightarrow \nu_{\mu}) &\approx 1 - \sin^22\theta_{23}\sin^2\Delta_{31} \\
& + 2\sin^22\theta_{23}\sin^2\theta_{24}\sin^2\Delta_{31} \\ 
& - \sin^22\theta_{24}\sin^2\Delta_{41}.
\label{eq:NuMuDisFull}
\end{aligned}
\end{equation}
Finally, the disappearance of $\overset{(-)}\nu\!\!_e$~\dword{cc} is described by: 
\begin{equation}
\begin{aligned}
P(\overset{(-)}\nu\!\!_e \rightarrow \overset{(-)}\nu\!\!_e) &\approx 1 - \sin^22\theta_{13}\sin^2\Delta_{31} \\
& - \sin^22\theta_{14}\sin^2\Delta_{41}.
\label{eq:NueDisFull}
\end{aligned}
\end{equation}
Figure~\ref{fig:regimes} shows how the standard three-flavor oscillation probability is distorted at neutrino energies above the standard oscillation peak when oscillations into sterile neutrinos are included.

The sterile neutrino effects have been implemented in  \dword{globes} via the existing plug-in for sterile neutrinos and non-standard interactions~\cite{Joachim}. As described above, the \dword{nd} will play a very important role in the sensitivity to sterile neutrinos both directly, for rapid oscillations with $\Delta m_{41}^2 > 1$~eV$^2$ where the sterile oscillation matches the \dword{nd} baseline, and indirectly, at smaller values of $\Delta m_{41}^2$ where the \dword{nd} is crucial to reduce the systematic uncertainties affecting the \dword{fd} to increase its sensitivity. To include these \dword{nd} effects in these studies, the most recent \dword{globes} DUNE configuration files describing the \dword{fd} were modified by adding a \dword{nd} with correlated systematic errors with the \dword{fd}. As a first approximation, the \dword{nd} is assumed to be an identical scaled-down version of the TDR \dword{fd}, with identical efficiencies, backgrounds and energy reconstruction. The systematic uncertainties originally defined in the \dword{globes} DUNE \dword{cdr} configuration already took into account the effect of the \dword{nd} constraint. Thus, since we are now explicitly simulating the \dword{nd}, larger uncertainties have been adopted but partially correlated between the different channels in the \dword{nd} and \dword{fd}, so that their impact is reduced by the combination of both data sets. The full set of systematic uncertainties employed in the sterile neutrino studies is listed in Table~\ref{tab:sterile_sys}.
%%%%%%%%%%%%%%%%%%%%%
\begin{table*}[htp]
\caption{List of systematic errors assumed in the sterile neutrino studies. }
\label{tab:sterile_sys}
\begin{center}
\begin{tabular}{  l@{\quad\quad}  c@{\quad} c@{\quad\quad} c@{\quad}      }
\hline
   Type of error  &  Value & affects & ND/FD correlated?\\ \hline\hline
ND fiducial volume  & 0.01 & all ND events & no \\
FD fiducial volume  & 0.01 & all FD events & no \\
flux signal component  & 0.08 & all events from signal comp. & yes \\
flux background component  & 0.15 & all events from bckg comp. & yes \\
flux signal component n/f & 0.004 & all events from signal comp. in ND & no \\
flux background component n/f & 0.02 & all events from bckg comp. in ND & no \\
CC cross section (each flav.)  & 0.15 & all events of that flavor & yes \\
NC cross section  & 0.25 & all NC events & yes \\
CC cross section (each flav.) n/f & 0.02 & all events of that flavor in ND & no \\
NC cross section n/f & 0.02 & all NC events in ND & no \\
 \hline\hline
\end{tabular}
\end{center}
\end{table*}
%%%%%%%%%%%%%%%%%%%%%

Finally, for oscillations observed at the \dword{nd}, the uncertainty on the production point of the neutrinos can play an important role. We have included an additional $20\%$ energy smearing, which produces a similar effect given the $L/E$ dependence of oscillations. We implemented this smearing in the \dword{nd} through multiplication of the migration matrices provided with the \dword{globes} files by an additional matrix with the $20\%$ energy smearing obtained by integrating the Gaussian

\begin{equation}
R^c(E,E')\equiv\frac{1}{\sigma(E)\sqrt{2\pi}}e^{-\frac{(E-E')^2}{2(\sigma(E))^2}},
\label{R_mat}
\end{equation}
with $\sigma(E)=0.2 E$ in reconstructed energy $E'$, where $E$ is the true neutrino energy from simulation.

By default, \dword{globes} treats all systematic uncertainties included in the fit as normalization shifts. However, depending on the value of $\Delta m^2_{41}$, sterile mixing will induce shape distortions in the measured energy spectrum beyond simple normalization shifts. As a consequence, shape uncertainties are very relevant for sterile neutrino searches, particularly in regions of parameter space where the \dword{nd}, with virtually infinite statistics, has a dominant contribution. The correct inclusion of systematic uncertainties affecting the shape of the energy spectrum in the two-detector fit \dword{globes} framework used for this analysis posed technical and computational challenges beyond the scope of the study.
Therefore, for each limit plot, we present two limits bracketing the expected DUNE sensitivity limit, namely: the black limit line, a best-case scenario, where only normalization shifts are considered in a \dword{nd}+\dword{fd} fit, where the ND statistics and shape have the strongest impact; and the grey limit line, corresponding to a worst-case scenario where only the \dword{fd} is considered in the fit, together with a rate constraint from the \dword{nd}.

\begin{figure}
\centering
\includegraphics[width=0.9\columnwidth]{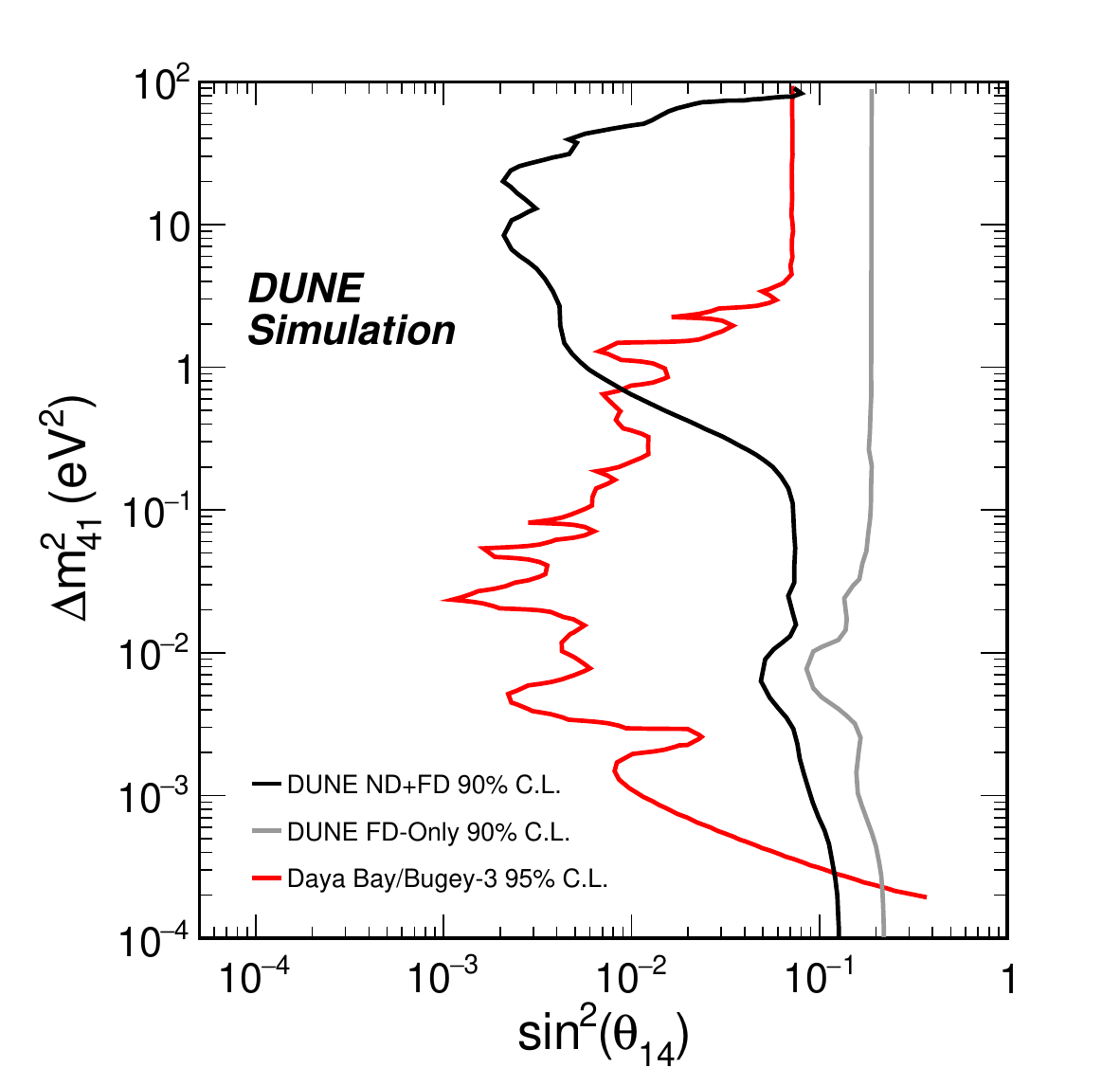}
\includegraphics[width=0.9\columnwidth]{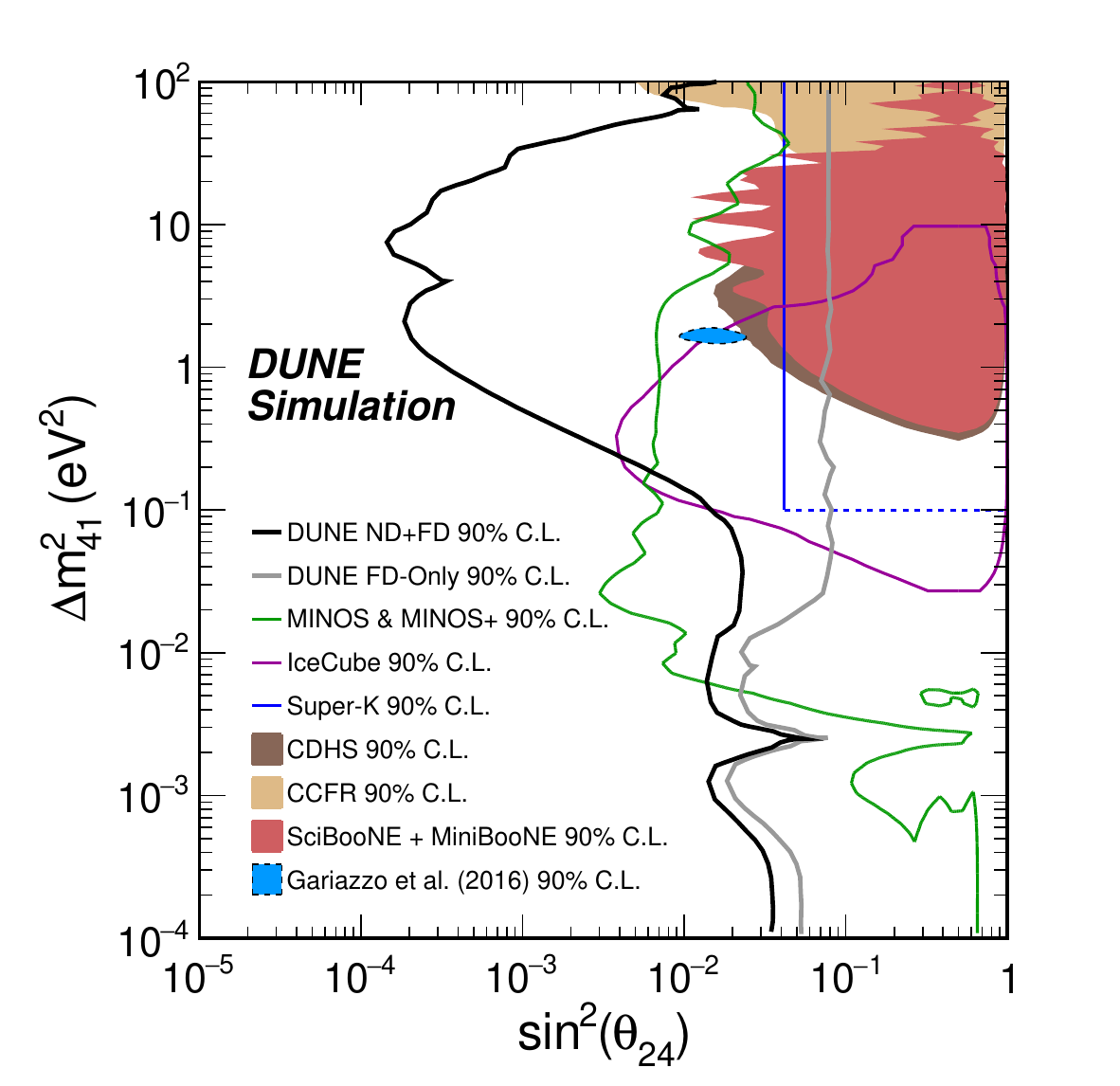}
\caption{The top plot shows the DUNE sensitivities to $\theta_{14}$ from the $\nu_e$ \dword{cc} samples at the \dword{nd} and \dword{fd}, along with a comparison with the combined reactor result from Daya Bay and Bugey-3. The bottom plot is adapted from Ref.~\cite{Todd:2018hin} and displays sensitivities to $\theta_{24}$ using the $\nu_\mu$ \dword{cc} and \dword{nc} samples at both detectors, along with a comparison with previous and existing experiments. In both cases, regions to the right of the contours are excluded.}
\label{fig:th_14+th_24}
\end{figure}

Studying the sensitivity to $\theta_{14}$, the dominant channels are those regarding $\nu_e$ disappearance. Therefore, only the $\nu_e$ \dword{cc} sample is analyzed and the channels for \dword{nc} and $\nu_{\mu}$ \dword{cc} disappearance are not taken into account, as they do not influence greatly the sensitivity and they slow down the simulations. The sensitivity at the 90\% \dword{cl}, taking into account the systematic uncertainties mentioned above, is shown in Fig.~\ref{fig:th_14+th_24}, along with a comparison to current constraints.

For the $\theta_{24}$ mixing angle, we analyze the $\nu_{\mu}$ \dword{cc} disappearance and the \dword{nc} samples, which are the main contributors to the sensitivity. 
The results are shown in Fig.~\ref{fig:th_14+th_24}, along with comparisons with present constraints.

\begin{figure} 
\centering
\includegraphics[width=0.9\columnwidth]{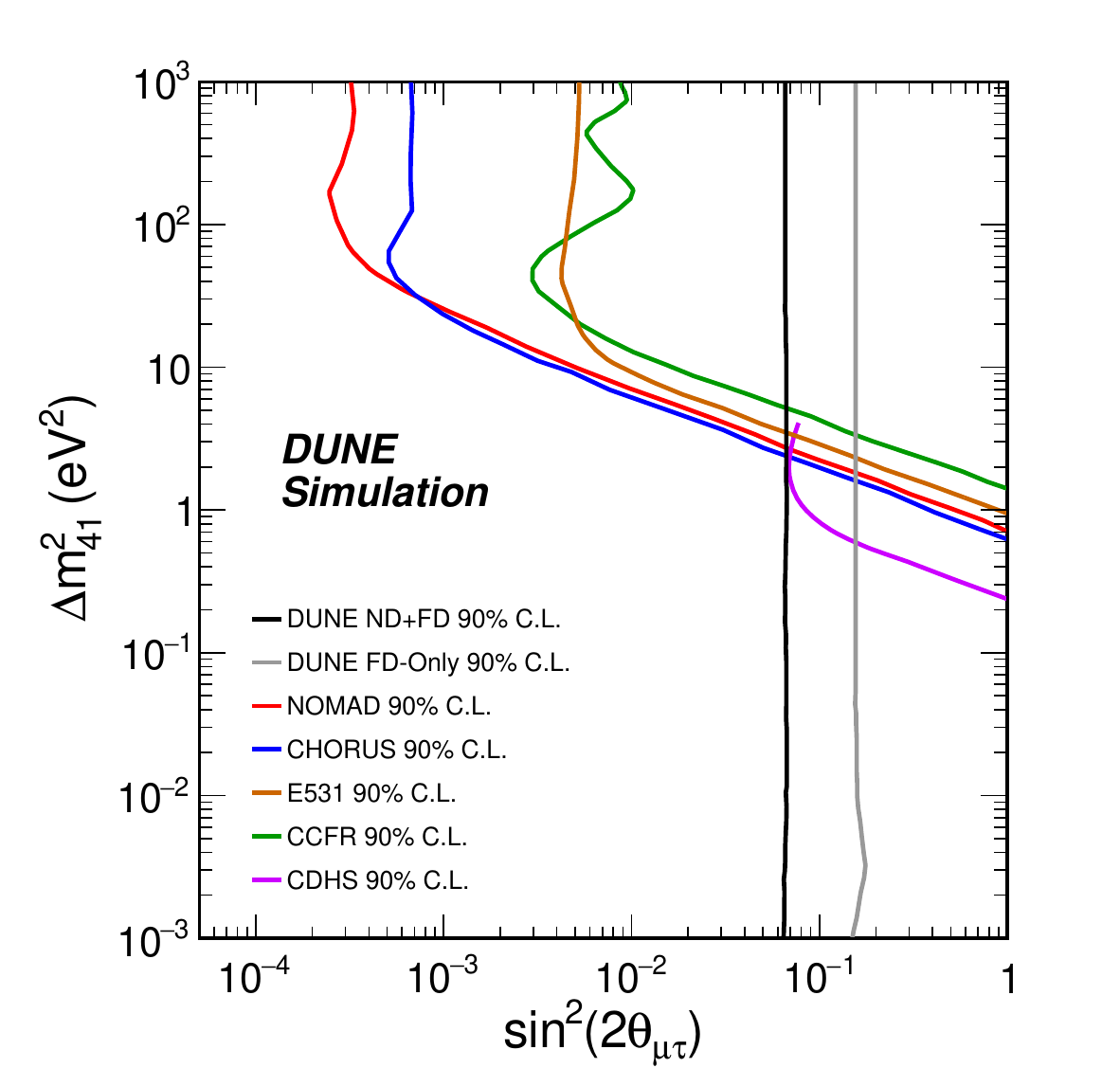}
\caption{Comparison of the DUNE sensitivity to $\theta_{34}$ using the \dword{nc} samples at the \dword{nd} and \dword{fd} with previous and existing experiments. Regions to the right of the contour are excluded.}
\label{fig:th_34} 
\end{figure}

In the case of the $\theta_{34}$ mixing angle, we look for disappearance in the \dword{nc} sample, the only contributor to this sensitivity. The results are shown in Fig.~\ref{fig:th_34}. Further, a comparison with previous experiments sensitive to \numu, \nutau~mixing with large mass-squared splitting is possible by considering an effective mixing angle $\theta_{\mu\tau}$, such that $\sin^2{2\theta_{\mu\tau}}\equiv 4|U_{\tau4}|^2|U_{\mu 4}|^2=\cos^4\theta_{14}\sin^22\theta_{24}\sin^2\theta_{34}$, and assuming conservatively that $\cos^4\theta_{14}=1$, and $\sin^22\theta_{24}=1$. This comparison with previous experiments is also shown in Fig.~\ref{fig:th_34}.
The sensitivity to $\theta_{34}$ is largely independent of 
$\Delta m^2_{41}$, since the term with $\sin^2\theta_{34}$ in Eq.~\eqref{eq:numu_nus}, the expression describing ${P(\nu_{\mu}\rightarrow\nu_s)}$, depends solely on the $\Delta m^2_{31}$ mass splitting.

Another quantitative comparison of our results for $\theta_{24}$ and $\theta_{34}$ with existing constraints can be made for projected upper limits on the sterile mixing angles assuming no evidence for sterile oscillations is found, and picking the value of  $\Delta m^2_{41} = 0.5$~eV$^2$ corresponding to the simpler counting experiment regime. For the $3+1$ model, upper limits of $\theta_{24}$\,$<$\,$1.8^{\circ}$ $(15.1^{\circ})$ and $\theta_{34}$\,$<$\,$15.0^{\circ}$ $(25.5^{\circ})$ are obtained at the 90\% \dword{cl} from the presented best(worst)-case scenario DUNE sensitivities. If expressed in terms of the relevant matrix elements

\begin{align}
\begin{split}
|U_{\mu4}|^2 =&\,\,\cos^2\theta_{14}\sin^2\theta_{24} \\
|U_{\tau4}|^2= & \,\,\cos^2\theta_{14}\cos^2\theta_{24}\sin^2\theta_{34},
\end{split}
\label{eq:DisapToApp}
\end{align}
these limits become $|U_{\mu4}|^{2}$\,$<$\,0.001 (0.068) and $|U_{\tau4}|^{2}$\,$<$\,0.067 (0.186) at the 90\% \dword{cl}, where we conservatively assume $\cos^2\theta_{14}$\,=\,1 in both cases, and additionally $\cos^2\theta_{24}$\,=\,1 in the second case.

\begin{figure}
$\vcenter{\hbox{\includegraphics[width=0.9\columnwidth]{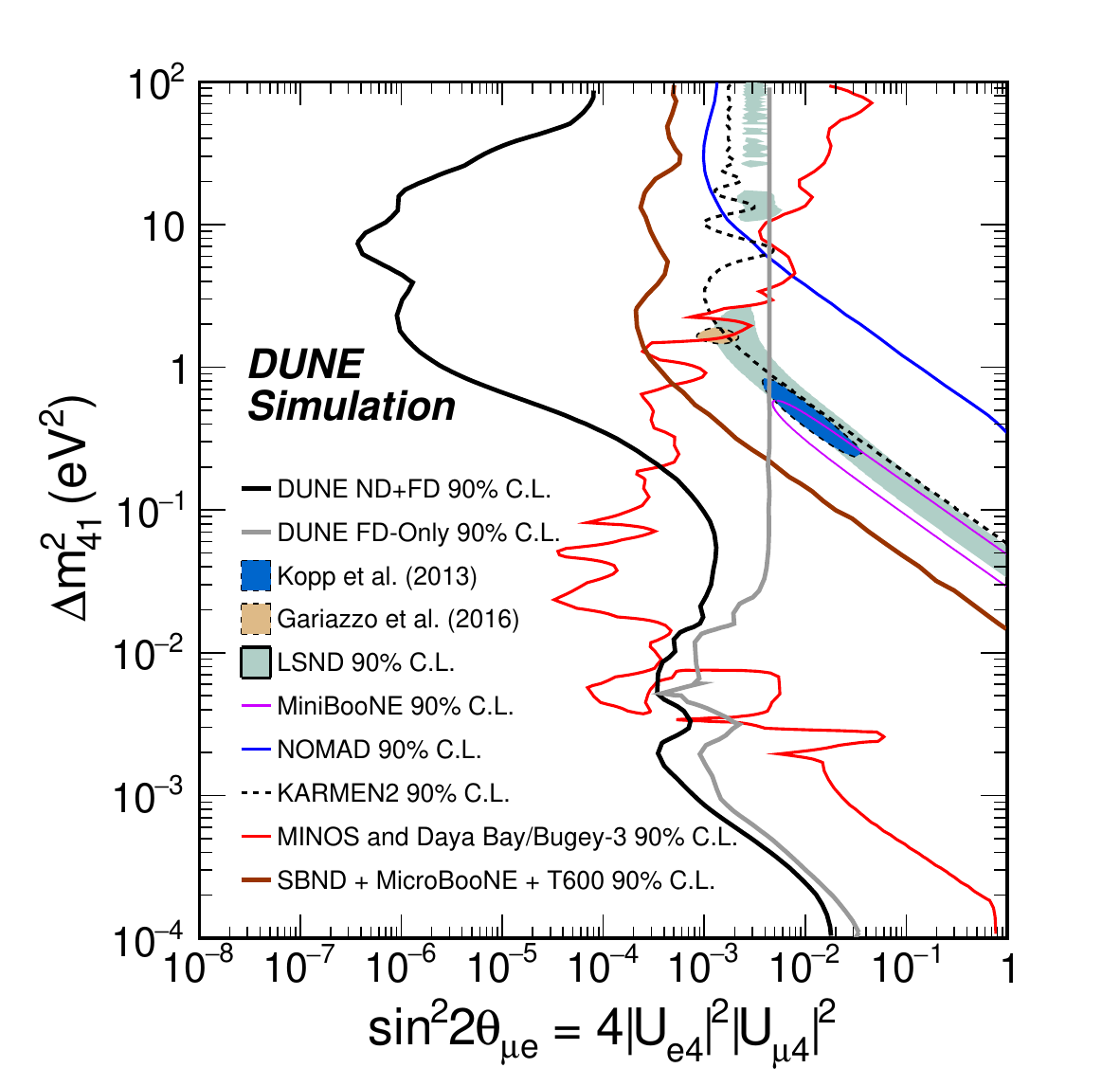}}}$
$\vcenter{\hbox{\includegraphics[width=0.9\columnwidth]{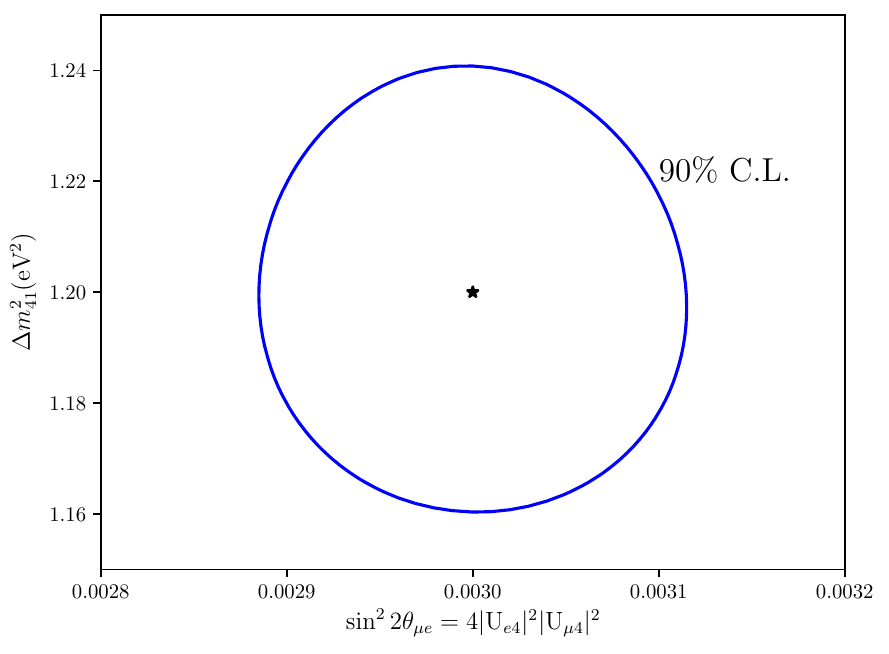}}}$
\caption{DUNE sensitivities to $\theta_{\mu e}$ from the appearance and disappearance samples at the \dword{nd} and \dword{fd} are shown on the top plot, along with a comparison with previous existing experiments and the sensitivity from the future \dword{sbn} program. Regions to the right of the DUNE contours are excluded. The plot is adapted from Ref.~\cite{Todd:2018hin}. In the bottom plot, the ellipse displays the DUNE discovery potential assuming $\theta_{\mu e}$ and $\Delta m_{41}^2$ set at the best-fit point determined by \dword{lsnd}~\cite{LSNDSterile} (represented by the star) for the best-case scenario referenced in the text.}
\label{fig:th_me}
\end{figure}

Finally, sensitivity to the $\theta_{\mu e}$ effective mixing angle, defined as $\sin^2{2\theta_{\mu e}}\equiv 4|U_{e4}|^2|U_{\mu 4}|^2=\sin^22\theta_{14}\sin^2\theta_{24}$, is shown in Fig.~\ref{fig:th_me}, which also displays a comparison with the allowed regions from the \dword{lsnd} and MiniBooNE, as well as with present constraints and projected constraints from the \fnal \dword{sbn} program.

As an illustration, Fig.~\ref{fig:th_me} also shows DUNE's discovery potential  for a scenario with one sterile neutrino governed by the \dword{lsnd} best-fit parameters: \\ $\left(\Delta m_{41}^2= 1.2\;\text{eV}^2;\,\,\sin^2{2\theta_{\mu e}}=0.003\right)$~\cite{LSNDSterile}. 
A small 90\% \dword{cl} allowed region is obtained, which can be compared with the \dword{lsnd} allowed region in the same figure. 

\section{Non-Unitarity of the Neutrino Mixing Matrix}
\label{sec:nonUnitarity}

A generic characteristic of most models explaining the neutrino mass
pattern is the presence of heavy neutrino states, additional to the
three light states of the \dword{sm}  of particle
physics~\cite{Mohapatra:1998rq,Valle:2015pba,Fukugita:2003en}. These
types of models imply that the $3 \times 3$ \dword{pmns} matrix is not unitary due to mixing with additional states.  Besides the type-I seesaw
mechanism~\cite{GellMann:1980vs,Yanagida:1979as,Mohapatra:1979ia,Schechter:1980gr},
different low-scale seesaw models include right-handed neutrinos that are relatively not-so-heavy, with mass of 1-10~TeV~\cite{Mohapatra:1986bd}, and perhaps detectable at collider experiments.

These additional heavy leptons would mix with the light neutrino states and, as a result, the complete unitary mixing matrix would be a squared $n \times n$ matrix, with $n$ the total number of neutrino
states. Therefore, the usual $3 \times 3$ \dword{pmns} matrix, which we dub $N$ to stress its non-standard nature, will be
non-unitary. One possible general way to parameterize these unitarity deviations in $N$ is through a triangular matrix~\cite{Escrihuela:2015wra}\footnote{For a similar parameterization corresponding to a $(3+1)$ and a $(3+3)$-dimensional mixing matrix,  see Refs.~\cite{Xing:2007zj,Xing:2011ur}}

 \begin{equation}
  N = 
 \left\lgroup
 \begin{array}{ccc} 
 1-\alpha_{ee} & 0 & 0 \\
 \alpha_{\mu e} & 1-\alpha_{\mu \mu} & 0 \\
  \alpha_{\tau e} & \alpha_{\tau \mu} & 1-\alpha_{\tau \tau}
 \end{array}
 \right \rgroup U \,,
 \label{eq:triangular}
 \end{equation}
with $U$ representing the unitary \dword{pmns} matrix, and the $\alpha_{ij}$ representing the non-unitary parameters.\footnote{The original parameterization in Ref.~\cite{Escrihuela:2015wra} uses $\alpha_{ii}$ instead of $\alpha_{\beta\gamma}$. The equivalence between the two notations is as follows: $\alpha_{ii} = 1-\alpha_{\beta\beta}$ and $\alpha_{ij} = \alpha_{\beta\gamma}$.} In the limit where $\alpha_{ij}=0$, $N$ becomes the usual \dword{pmns} mixing matrix.

The triangular matrix in this equation accounts for the non-unitarity of the $3 \times 3$ matrix for any number of extra neutrino species. This parameterization has been shown to be particularly well-suited for oscillation searches~\cite{Escrihuela:2015wra,Blennow:2016jkn} since, compared to other alternatives, it minimizes the departures of its unitary component $U$ from the mixing angles that are directly measured in neutrino oscillation experiments when unitarity is assumed.

The phenomenological implications of a non-unitary leptonic mixing matrix have been extensively studied in flavor and electroweak precision observables as well as in the neutrino oscillation phenomenon~\cite{Shrock:1980vy,Schechter:1980gr,Shrock:1980ct,Shrock:1981wq,Langacker:1988ur,Bilenky:1992wv,Nardi:1994iv,Tommasini:1995ii,Antusch:2006vwa,FernandezMartinez:2007ms,Antusch:2008tz,Biggio:2008in,Antusch:2009pm,Forero:2011pc,Alonso:2012ji,Antusch:2014woa,Abada:2015trh,Fernandez-Martinez:2015hxa,Escrihuela:2015wra,Parke:2015goa,Miranda:2016wdr,Fong:2016yyh,Escrihuela:2016ube}. For recent global fits to all flavor and electroweak precision data summarizing present bounds on non-unitarity see Refs.~\cite{Antusch:2014woa,Fernandez-Martinez:2016lgt}.

Recent studies have shown that DUNE can constrain the non-unitarity parameters~\cite{Blennow:2016jkn,Escrihuela:2016ube}. The summary of the $90 \%$~\dword{cl}  bounds on the different $\alpha_{ij}$ elements profiled over all other parameters is given in Table~\ref{tab:bounds}.

\begin{table}[htp]
\centering
\caption{Expected $90\%$~\dword{cl} constraints on the non-unitarity parameters $\alpha$ from DUNE.}
\label{tab:bounds}
\begin{tabular}{c | c}\hline
Parameter & Constraint \\ \hline\hline
$\alpha_{ee}$ & $0.3$   \\ 
$\alpha_{\mu\mu}$ & $0.2$ \\ 
$\alpha_{\tau\tau}$ & $0.8$ \\ 
$\alpha_{\mu e}$ & $0.04$ \\ 
$\alpha_{\tau e}$ & $0.7$ \\ 
$\alpha_{\tau\mu}$ & $0.2$ \\\hline
\end{tabular}
\end{table}
These bounds are comparable with other constraints from present oscillation experiments, although they are not competitive with those obtained from flavor and electroweak precision data.
For this analysis, and  
those presented below, we have used the \dword{globes} software~\cite{Huber:2004ka,Huber:2007ji} with the DUNE \dword{tdr} configuration presented in Ref.~\cite{Abi:2020evt} and assumed a data exposure of 300~\ktMWyr. The standard (unitary) oscillation parameters have also been treated as in~\cite{Abi:2020evt}. The unitarity deviations have been included both by an independent code (used to obtain the results shown in Ref.~\cite{Escrihuela:2016ube}) and via the Monte Carlo Utility Based Experiment Simulator (MonteCUBES)~\cite{Blennow:2009pk} plug-in to cross validate our results.

\begin{figure*}[htp]
\centering
\includegraphics[width=0.9\textwidth]{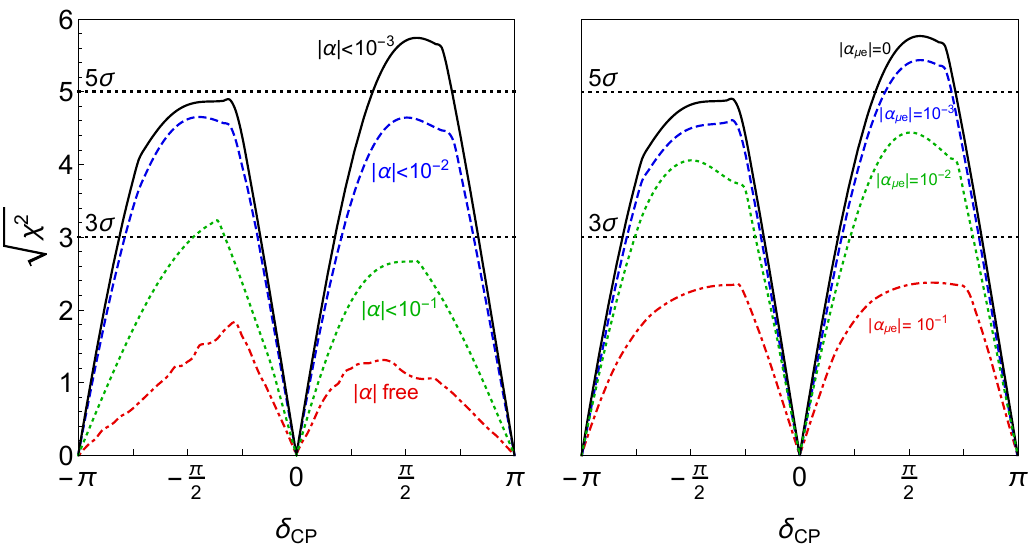}
\caption{The impact of non-unitarity on the DUNE \dword{cpv} discovery potential. See the text for details.}
\label{fig:CPsens}
\end{figure*}

Conversely, the presence of non-unitarity may affect the determination of the
Dirac \dword{cp}-violating phase $\delta_{CP}$ in \dword{lbl} experiments~\cite{Miranda:2016wdr,Fernandez-Martinez:2016lgt,Escrihuela:2016ube}.
Indeed, when allowing for unitarity deviations, the expected \dword{cp} discovery potential for DUNE could be significantly reduced.
However, the situation is alleviated when a combined analysis with the constraints on non-unitarity from other experiments is considered. This is illustrated in Fig.~\ref{fig:CPsens}. In the left panel, the discovery potential for \dword{cpv} is computed when the non-unitarity parameters introduced in Eq.~(\ref{eq:triangular}) are allowed in the fit. While for the Asimov data all $\alpha_{ij}=0$, the non-unitary parameters are allowed to vary in the fit with $1 \sigma$ priors of $10^{-1}$, $10^{-2}$ and $10^{-3}$ for the dotted green, dashed blue and solid black lines respectively. For the dot-dashed red line no prior information on the non-unitarity parameters has been assumed. As can be observed, without additional priors on the non-unitarity parameters, the capabilities of DUNE to discover \dword{cpv} from $\delta_{CP}$ would be seriously compromised~\cite{Escrihuela:2016ube}. However, with priors of order $10^{-2}$ matching the present constraints from other neutrino oscillation experiments~\cite{Escrihuela:2016ube,Blennow:2016jkn}, the sensitivity expected in the three-flavor model is almost recovered. If the more stringent priors of order $10^{-3}$ stemming from flavor and electroweak precision observables are added~\cite{Antusch:2014woa,Fernandez-Martinez:2016lgt}, the standard sensitivity is obtained.   

The right panel of Fig.~\ref{fig:CPsens} concentrates on the impact of the phase of the element $\alpha_{\mu e}$ in the discovery potential of \dword{cpv} from $\delta_{CP}$, since this element has a very important impact in the $\nu_e$ appearance channel. In this plot the modulus of $\alpha_{ee}$, $\alpha_{\mu \mu}$ and $\alpha_{\mu e}$ have been fixed to $10^{-1}$, $10^{-2}$, $10^{-3}$ and 0 for the dot-dashed red, dotted green, dashed blue and solid black lines respectively. All other non-unitarity parameters have been set to zero and the phase of $\alpha_{\mu e}$ has been allowed to vary both in the fit and in the Asimov data, showing the most conservative curve obtained. As for the right panel, it can be seen that a strong deterioration of the \dword{cp} discovery potential could be induced by the phase of $\alpha_{\mu e}$ (see Ref.~\cite{Escrihuela:2016ube}). However, for unitarity deviations of order $10^{-2}$, as required by present neutrino oscillation data constraints, the effect is not too significant in the range of $\delta_{CP}$ for which a $3 \sigma$ exclusion of \dword{cp} conservation would be possible and it becomes negligible if the stronger $10^{-3}$ constraints from flavor and electroweak precision data are taken into account.  

\begin{figure*}[htp]
\centering
\includegraphics[width=0.9\textwidth]{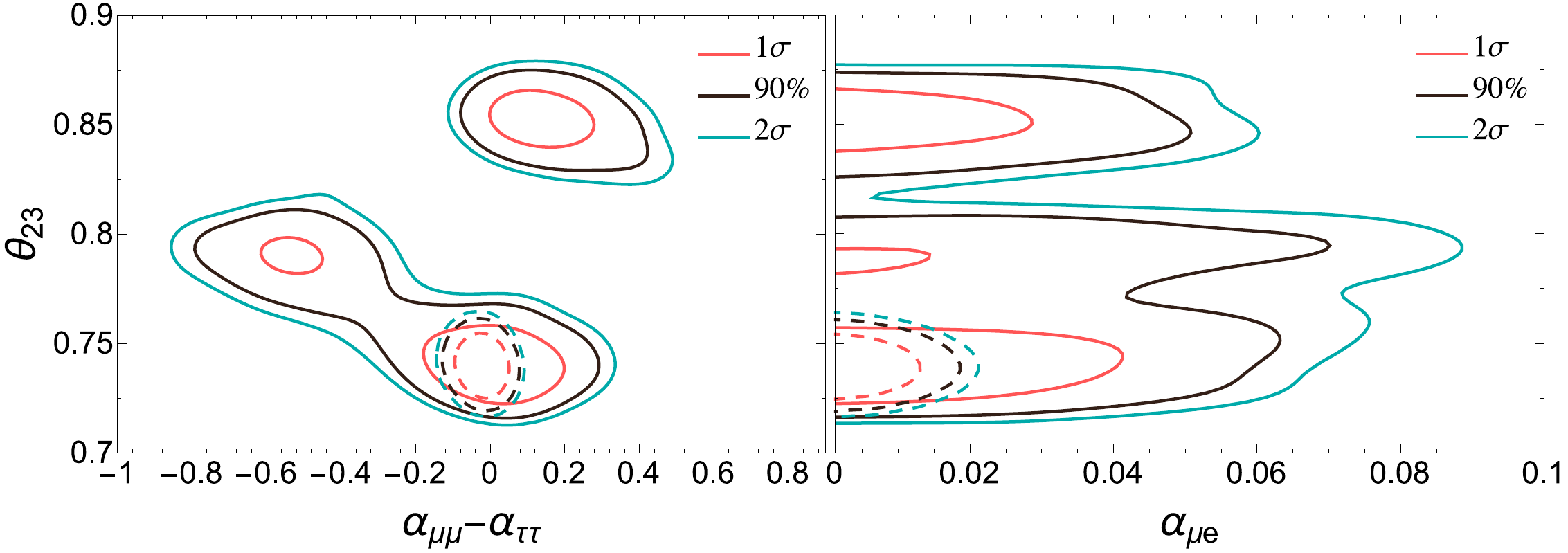}
\caption{Expected frequentist allowed regions at the $1 \sigma$, $90\%$ and $2\sigma$ \dword{cl}\ for DUNE. All new physics parameters are assumed to be zero so as to obtain the expected non-unitarity sensitivities. A value $\theta_{23} = 0.235 \pi \approx 0.738$ rad is assumed.  The solid lines correspond to the analysis of DUNE data alone, while the dashed lines include the present constraints on non-unitarity. The values of $\theta_{23}$ are shown in radians.}
 \label{fig:octant}
\end{figure*}

Similarly, the presence of non-unitarity worsens degeneracies involving $\theta_{23}$, making the determination of the octant or even its maximality challenging.
This situation is shown in Fig.~\ref{fig:octant} where an input value of $\theta_{23} = 42.3^\circ$ was assumed. As can be seen, the fit in presence of non-unitarity (solid lines) introduces degeneracies for the wrong octant and even for maximal mixing~\cite{Blennow:2016jkn}. However, these degeneracies are resolved upon the inclusion of present priors on the non-unitarity parameters from other oscillation data (dashed lines) and a clean determination of the standard oscillation parameters following DUNE expectations is again recovered.   

The sensitivity that DUNE would provide to the non-unitarity parameters is comparable to that from present oscillation experiments, while not competitive to that from flavor and electroweak precision observables, which are roughly an order of magnitude more stringent. 
On the other hand, the capability of DUNE to determine the standard oscillation parameters such as \dword{cpv} from $\delta_{CP}$ or the octant or maximality of $\theta_{23}$ would be seriously compromised by unitarity deviations in the \dword{pmns} matrix. This negative impact is however significantly reduced when priors on the size of these deviations from other oscillation experiments are considered, and disappears altogether if the more stringent constraints from flavor and electroweak precision data are added instead.

\section{Non-Standard Neutrino Interactions}
\label{sec:nsi}

\Dword{nsi}, affecting neutrino propagation through the Earth, can significantly modify the data to be collected by DUNE as long as the new physics parameters are large enough~\cite{Farzan:2017xzy}. Leveraging its very long baseline and wide-band beam, DUNE is uniquely sensitive to these probes. \dword{nsi} may impact the determination of current unknowns such as \dword{cpv}~\cite{Masud:2015xva,Masud:2016bvp}, mass hierarchy~\cite{Masud:2016gcl,Capozzi:2019iqn} and octant of $\theta_{23}$~\cite{Agarwalla:2016fkh}. If the DUNE data are consistent with the standard oscillation for three massive neutrinos, off-diagonal \dword{nc} \dword{nsi} effects of order 0.1 $G_F$ can be ruled out at the 68 to 95\% \dword{cl}~\cite{deGouvea:2015ndi,Coloma:2015kiu}. We note that DUNE might improve current constraints on $|\epsilon^m_{e \tau}|$ and $|\epsilon^m_{e \mu}|$, the electron flavor-changing NSI intensity parameters (see Eq.~\ref{eq:epsmatrix}), by a factor 2-5~\cite{Ohlsson:2012kf,Miranda:2015dra,Farzan:2017xzy}. New  \dword{cc}  interactions can also lead to modifications in the production, at the beam source, and the detection of neutrinos. The findings on source and detector \dword{nsi} studies at DUNE are presented in~\cite{Blennow:2016etl,Bakhti:2016gic}, in which DUNE does not have sensitivity to discover or to improve bounds on source/detector \dword{nsi}. In particular, the simultaneous impact on the measurement of $\delta_{\rm CP}$ and $\theta_{23}$ is investigated in detail. Depending on the assumptions, such as the use of the \dword{nd}  and whether \dword{nsi} at production and detection are the same, the impact of source/detector \dword{nsi} at DUNE may be relevant. We focus our attention on the propagation, based on the results from~\cite{Blennow:2016etl}.  

\dword{nc} \dword{nsi} can be understood as non-standard
matter effects that are visible only in an \dword{fd} at a sufficiently long baseline. They can be parameterized as new contributions
to the matter potential in the \dword{msw} \cite{Mikheev:1986gs,Wolfenstein:1977ue,Guzzo:1991hi, Guzzo:1991cp, Roulet:1991sm,Valle:1987gv} matrix in the neutrino-propagation Hamiltonian:

\begin{equation}
  H = U \left( \begin{array}{ccc}
           0 &                    & \\
             & \Delta m_{21}^2/2E & \\
             &                    & \Delta m_{31}^2/2E
         \end{array} \right) U^\dag + \tilde{V}_{\rm MSW} \,,
\end{equation}

with

\begin{equation}
  \tilde{V}_{\rm MSW} = \sqrt{2} G_F N_e
\left(
  \begin{array}{ccc}
    1 + \epsilon^m_{ee}       & \epsilon^m_{e\mu}       & \epsilon^m_{e\tau}  \\
        \epsilon^{m*}_{e\mu}  & \epsilon^m_{\mu\mu}     & \epsilon^m_{\mu\tau} \\
        \epsilon^{m*}_{e\tau} & \epsilon^{m*}_{\mu\tau} & \epsilon^m_{\tau\tau}
  \end{array} 
\right)
\label{eq:epsmatrix}
\end{equation}
Here, $U$ is the standard \dword{pmns} leptonic mixing matrix, for which we use the standard parameterization found, e.g., in~\cite{Agashe:2014kda}, 
and the $\epsilon$-parameters give the
magnitude of the \dword{nsi} relative to standard weak interactions.  For new physics scales of a few hundred GeV,  a value of $|\epsilon|$ of the order
0.01 or less is
expected~\cite{Davidson:2003ha,GonzalezGarcia:2007ib,Biggio:2009nt}. The DUNE
baseline provides an advantage in the detection of \dword{nsi} relative
to existing beam-based experiments with shorter baselines.
Only atmospheric-neutrino experiments have longer baselines, but the sensitivity of these experiments to \dword{nsi} is limited by systematic effects~\cite{Adams:2013qkq}.

In this analysis, we use \dword{globes} with the MonteCUBES C library, a plugin that replaces the deterministic  \dword{globes} minimizer by a Markov Chain Monte Carlo (MCMC) method that is able to handle higher dimensional parameter spaces. In the simulations we use the configuration for the DUNE \dword{tdr}~\cite{Abi:2020evt}. Each point scanned by the MCMC is stored and a frequentist $\chi^2$ analysis is performed with the results. The analysis assumes an exposure of 300~\ktMWyr.

\begin{figure*}[htp]
\centering
\includegraphics[width=0.98\textwidth]{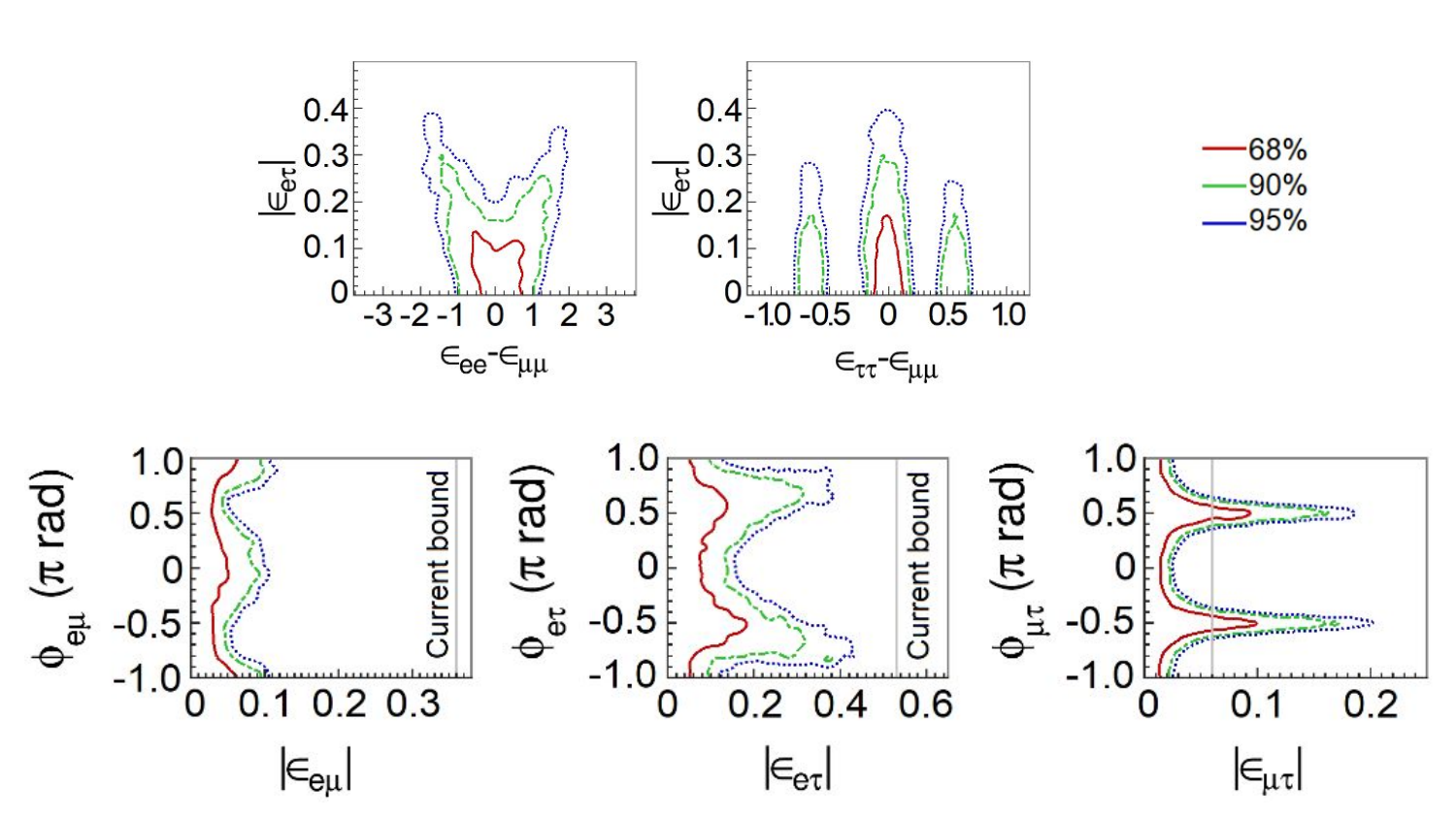}
\caption{Allowed regions of the non-standard oscillation parameters in which we see important degeneracies (top) and the complex non-diagonal ones (bottom). We conduct the analysis considering all the \dword{nsi} parameters as non-negligible. The sensitivity regions are for 68\% CL [red line (left)], 90\% CL [green dashed line (middle)], and 95\% CL [blue dotted line (right)]. Current bounds are taken from~\cite{Gonzalez-Garcia:2013usa}.}
\label{fig:nsi}
\end{figure*}

In an analysis with all the \dword{nsi} parameters free to vary, we obtain the sensitivity regions in Fig.~\ref{fig:nsi}. We omit the superscript $m$ that appears in Eq.~\eqref{eq:epsmatrix}. 
The credible regions are shown for different confidence levels.
We note, however, that constraints on $\epsilon_{\tau\tau}-\epsilon_{\mu\mu}$ coming from global fit analysis~\cite{Gonzalez-Garcia:2013usa,Miranda:2015dra,Farzan:2017xzy,Esteban:2018ppq} can remove the left and right solutions of $\epsilon_{\tau\tau}-\epsilon_{\mu\mu}$ in Fig.~\ref{fig:nsi}.

In order to constrain the standard oscillation parameters when  \dword{nsi} are %is 
present, we use the fit for three-neutrino mixing from~\cite{Gonzalez-Garcia:2013usa} and implement prior constraints to restrict the region sampled by the MCMC. The sampling of the parameter space is explained in~\cite{Coloma:2015kiu} and the priors that we use can be found in Table~\ref{tab:priors1}.
\begin{table}
\centering
\caption{Oscillation parameters and priors implemented in MCMC for calculation of Fig.~\ref{fig:nsi}.} 
\label{tab:priors1}
\begin{tabular}{c | c | c }\hline
Parameter & Nominal & 1$\sigma$ Range ($\pm$) \\ \hline\hline 
$\theta_{12}$ &0.19$\pi$&2.29\%\\ 
$\sin^2(2\theta_{13})$ &0.08470&0.00292\\ 
$\sin^2(2\theta_{23})$ &0.9860&0.0123\\ 
$\Delta m^2_{21} $ &7.5 $\times10^{-5}\textrm{eV}^2$&2.53\%\\ 
$\Delta m^2_{31} $ &2.524 $\times10^{-3}\textrm{eV}^2$&free\\ 
$\delta_{\rm CP} $ &1.45$\pi$&free\\ \hline
\end{tabular}
\end{table}

\begin{figure*}[htp]
\centering
\includegraphics[width=0.65\textwidth]{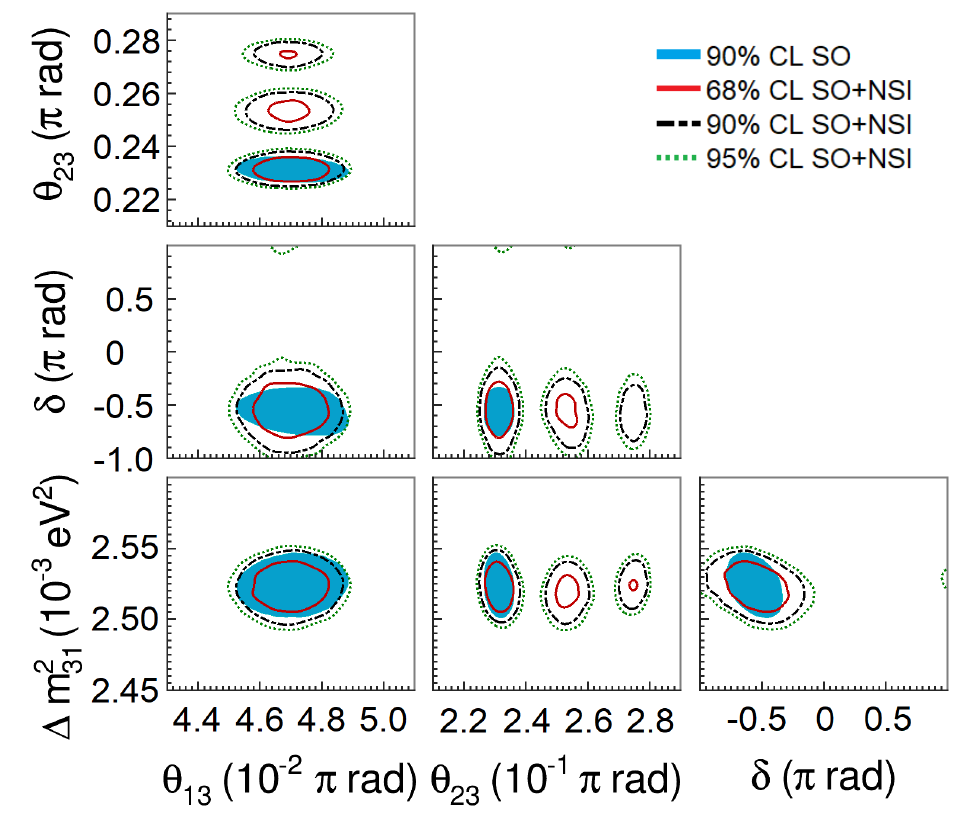}
\caption{Projections of the standard oscillation parameters with nonzero \dword{nsi}. The sensitivity regions are for 68\%, 90\%, and 95\% \dword{cl}. The allowed regions considering negligible \dword{nsi} (standard oscillation (SO) at 90\% \dword{cl}) are superposed to the SO+NSI.}
\label{fig:standar-nsi}
\end{figure*}

The effects of \dword{nsi} on the measurements of the standard oscillation parameters at DUNE are explicit in Fig.~\ref{fig:standar-nsi}, where we superpose the allowed regions with non-negligible  \dword{nsi} and the standard-only credible regions at 90\% \dword{cl}. 
In the blue filled areas we assume only standard oscillation. In the regions delimited by the red, black dashed, and green dotted lines we constrain standard oscillation parameters allowing NSI to vary freely.

An important degeneracy appears in the measurement of the mixing angle $\theta_{23}$.
Notice that this degeneracy appears because of the constraints obtained for $\epsilon_{\tau\tau}-\epsilon_{\mu\mu}$ shown in Fig.~\ref{fig:nsi}.
We also see that the sensitivity of the \dword{cp} phase is strongly affected.

The effects of matter density variation and its average along the beam path from \fnal to \surf  were studied considering the standard neutrino oscillation framework with three flavors~\cite{Roe:2017zdw,Kelly:2018kmb}. In order to obtain the results of Figs.~\ref{fig:nsi} and~\ref{fig:standar-nsi}, we use a high-precision calculation for the baseline of \SI{1285}{km} and the average density of \SI{2.848}{g/cm^3}~\cite{Roe:2017zdw}.

The DUNE collaboration has been using the so-called PREM~\cite{Dziewonski:1981xy,PREM2} density profile to consider matter density variation. With this assumption, the neutrino beam crosses a few constant density layers.
However, a more detailed density map is available for the USA with more than 50 layers and $0.25 \times 0.25$ degree cells of latitude and longitude: The Shen-Ritzwoller or S.R. profile~\cite{SR:2016,Roe:2017zdw}. Comparing the S.R. with the PREM profiles, Ref.~\cite{Kelly:2018kmb} shows that in the standard oscillation paradigm, DUNE is not highly sensitive to the density profile and that the only oscillation parameter with its measurement slightly impacted by the average density true value is \deltacp{}.
\dword{nsi}, however, may be sensitive to the profile, particularly considering the phase $\phi_{e\tau}$~\cite{Chatterjee:2018dyd}, where $\epsilon_{e\tau}=|\epsilon_{e\tau}|e^{i\phi_{e\tau}}$, to which DUNE will have a high sensitivity~\cite{Ohlsson:2012kf,Miranda:2015dra,deGouvea:2015ndi,Coloma:2015kiu,Farzan:2017xzy}, as we also see in Fig.~\ref{fig:nsi}.

\begin{figure*}[htp]
\centering
\includegraphics[width=0.98\textwidth]{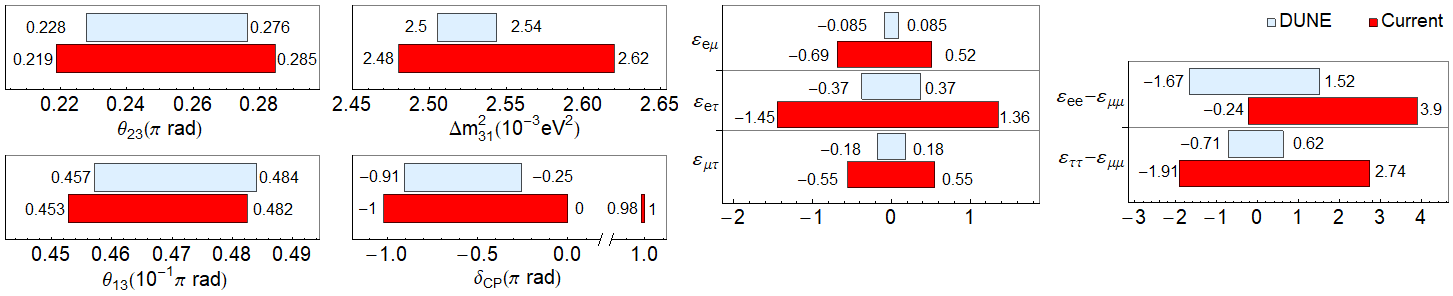}
\caption{One-dimensional DUNE constraints compared with current constraints calculated in Ref.~\cite{Farzan:2017xzy}. The left half of the figure shows constraints on the standard oscillation parameters, written in the bottom of each comparison. The five comparisons in the right half show constraints on non-standard interaction parameters.}
\label{fig:bars}
\end{figure*}

In order to compare the results of our analysis predictions for DUNE with the constraints from other experiments, we use the results from~\cite{Farzan:2017xzy}. There are differences in the nominal parameter values used for calculating the $\chi^2$ function and other assumptions. This is the reason why the regions in Fig.~\ref{fig:bars} do not have the same central values, but this comparison gives a good view of how DUNE can substantially improve the bounds on, for example, $\varepsilon_{\tau\tau}-\varepsilon_{\mu\mu}$, $\Delta m^2_{31}$, and the non-diagonal \dword{nsi} parameters.

\dword{nsi} can significantly impact the determination of current unknowns such as \dword{cpv} and the octant of $\theta_{23}$. Clean determination of the intrinsic \dword{cp} phase at \dword{lbl} experiments, such as DUNE, in the presence of \dword{nsi}, is a formidable task~\cite{Rout:2017udo}. A feasible strategy to disambiguate physics scenarios at DUNE using high-energy beams was suggested in~\cite{Masud:2017bcf}.
The conclusion here is that, using a tunable beam, it is possible to disentangle scenarios with NSI. Constraints from other experiments can also solve the \dword{nsi} induced degeneracy on $\theta_{23}$.

\section{CPT and Lorentz Violation}
\label{sect:cpt}

\Dword{cpt} 
%Charge conjugation, parity, and time reversal symmetry (CPT)
is a cornerstone of our model-building strategy. 
DUNE can improve the present limits on Lorentz and \dword{cpt} violation by several orders of magnitude~\cite{Streater:1989vi,Barenboim:2002tz,Kostelecky:2003cr,Diaz:2009qk,Kostelecky:2011gq,Barenboim:2017ewj,Barenboim:2018lpo,Barenboim:2018ctx}, contributing as a very important experiment to test these fundamental assumptions underlying quantum field theory.

\dword{cpt} invariance is one of the predictions of major importance of local, relativistic quantum field theory. One of the predictions of \dword{cpt} invariance is that particles and antiparticles have the same masses and, if unstable, the same lifetimes. To prove the \dword{cpt} theorem one needs only three ingredients~\cite{Streater:1989vi}: Lorentz invariance, hermiticity of the Hamiltonian, and locality.

Experimental bounds on \dword{cpt} invariance can be derived using the neutral kaon system~\cite{Schwingenheuer:1995uf}:

\begin{equation}
  \frac{|m(K^0) - m(\overline{K}^0)|}{m_K} < 0.6 \times 10^{-18}\,. 
  \label{eq:mK}
\end{equation}

This result, however, should be interpreted very carefully for two reasons. First, we do not have a complete theory of \dword{cpt} violation, and it is therefore arbitrary to take the kaon mass as a scale. Second, since kaons are bosons, the term entering the Lagrangian is the mass squared and not the mass itself. With this in mind, we can rewrite the previous bound as:
$ |m^2(K^0) - m^2(\overline{K}^0)| < 0.3~\mbox{eV}^2 \, $.
Modeling CPT violation as differences in the usual oscillation parameters between neutrinos and antineutrinos, we see here that neutrinos can test the predictions of the \dword{cpt} theorem to an unprecedented extent and could, therefore, provide stronger limits than the ones regarded as the most stringent ones
to date.\footnote{\dword{cpt} was tested also using charged leptons. However, these measurements involve a combination
of mass and charge and are not a direct \dword{cpt} test. Only neutrinos can provide \dword{cpt} tests on an elementary mass not contaminated by charge.}

In the absence of a solid model of flavor, not to mention one of \dword{cpt} violation, the spectrum  of neutrinos and antineutrinos can differ both  in the mass eigenstates themselves as well as in the flavor composition of each of these states. It is important to notice then that neutrino oscillation experiments can only test \dword{cpt} in the mass differences and mixing angles. An overall shift between the neutrino and antineutrino spectra will be missed by oscillation experiments.  Nevertheless, such a pattern can be bounded by cosmological data~\cite{Barenboim:2017vlc}. Unfortunately direct searches for neutrino mass (past, present, and future) involve only antineutrinos and hence cannot be used to draw any conclusion on \dword{cpt} invariance on the absolute mass scale, either.
Therefore, using neutrino oscillation data, we will compare the mass splittings and mixing angles of  neutrinos with those of antineutrinos. Differences in the neutrino and antineutrino spectrum would imply the violation of the \dword{cpt} theorem.

In Ref.~\cite{Barenboim:2017ewj} the authors derived the most up-to-date bounds on \dword{cpt} invariance from the neutrino sector
using the same data that was used in the global fit to neutrino oscillations in Ref.~\cite{deSalas:2017kay}. 
Of course, experiments that cannot distinguish between neutrinos and antineutrinos, such as atmospheric data from \superk~\cite{Abe:2017aap}, IceCube-DeepCore~\cite{Aartsen:2014yll,Aartsen:2017nmd} and ANTARES~\cite{AdrianMartinez:2012ph} were not included. The complete data set used, as well as the parameters to which they are sensitive, are
(1) from solar neutrino data~\cite{Cleveland:1998nv,Kaether:2010ag,Abdurashitov:2009tn,hosaka:2005um,Cravens:2008aa,Abe:2010hy,Nakano:PhD,Aharmim:2008kc,Aharmim:2009gd,Bellini:2013lnn}:  $\theta_{12}$, $\Delta m_{21}^2$, and $\theta_{13}$;
 (2) from neutrino mode in \dword{lbl} experiments K2K~\cite{Ahn:2006zza}, MINOS~\cite{Adamson:2013whj,Adamson:2014vgd}, T2K~\cite{Abe:2017uxa,Abe:2017bay}, and NO$\nu$A~\cite{Adamson:2017qqn,Adamson:2017gxd}:  $\theta_{23}$, $\Delta m_{31}^2$, and $\theta_{13}$;
 (3) from KamLAND reactor antineutrino data~\cite{Gando:2010aa}: $\overline{\theta}_{12}$, $\Delta \overline{m}_{21}^2$, and $\overline{\theta}_{13}$;
 (4) from short-baseline reactor antineutrino experiments Daya Bay~\cite{An:2016ses}, RENO~\cite{RENO:2015ksa}, and Double Chooz~\cite{Abe:2014bwa}:     $\overline{\theta}_{13}$ and $\Delta \overline{m}_{31}^2$; and 
 (5) from antineutrino mode in \dword{lbl} experiments MINOS~\cite{Adamson:2013whj,Adamson:2014vgd} and T2K~\cite{Abe:2017uxa,Abe:2017bay}: $\overline{\theta}_{23}$, $\Delta \overline{m}_{31}^2$, and
$\overline{\theta}_{13}$.\footnote{The K2K experiment took  data only in neutrino mode, while the \nova experiment had not published data in the antineutrino mode when these bounds were calculated.}

From the analysis of all previous data samples, one can derive the most up-to-date (3$\sigma$) bounds on \dword{cpt} violation:

\begin{align}
 |\Delta m_{21}^2-\Delta \overline{m}_{21}^2| &< 4.7\times 10^{-5} \,  \text{eV}^2,\,\,
  \nonumber \\
 |\Delta m_{31}^2-\Delta \overline{m}_{31}^2| &< 3.7\times 10^{-4} \, \text{eV}^2,\,\,
 \nonumber \\
 |\sin^2\theta_{12}-\sin^2\overline{\theta}_{12}| &< 0.14\,,\,\,
 \nonumber \\
 |\sin^2\theta_{13}-\sin^2\overline{\theta}_{13}| &< 0.03\,, \,\,
  \nonumber \\
 |\sin^2\theta_{23}-\sin^2\overline{\theta}_{23}| &< 0.32\,.
\end{align} 

At the moment it is not possible to set any bound on $|\delta-\overline{\delta}|$, since all possible values of
$\delta$ or $\overline{\delta}$ are allowed by data. The preferred intervals of $\delta$ obtained in Ref.~\cite{deSalas:2017kay} can only be obtained after combining the neutrino and antineutrino data samples. 
The limits  on $\Delta(\Delta m_{31}^2)$ and $\Delta(\Delta m_{21}^2)$  are already better than the one derived from the neutral kaon system and should be regarded as the best current bounds on \dword{cpt} violation on the mass squared.  
Note that these results were derived assuming the same mass ordering for neutrinos and antineutrinos. If the ordering was different for neutrinos and antineutrinos, this would be an indication for \dword{cpt} violation on its own. In the following we show how DUNE could improve this bound.

Sensitivity of the DUNE experiment to measure \dword{cpt} violation in the neutrino sector is studied by analyzing neutrino and antineutrino oscillation parameters separately. We assume the neutrino oscillations being parameterized by the usual \dword{pmns} matrix $U_{\text{PMNS}}$, with parameters $\theta_{12},\theta_{13},\theta_{23},\Delta m_{21}^2,\Delta m_{31}^2, {\rm and}~\delta$, while the antineutrino oscillations are parameterized by a matrix $\overline{U}_{\text{PMNS}}$ with parameters $\overline{\theta}_{12},\overline{\theta}_{13},\overline{\theta}_{23},\Delta \overline{m}_{21}^2,\Delta \overline{m}_{31}^2, {\rm and}~\overline{\delta}$. Hence, antineutrino oscillation is described  by the same probability functions as neutrinos with the neutrino parameters replaced by their antineutrino counterparts.\footnote{Note that the antineutrino oscillation probabilities also include the standard change of sign in the \dword{cp} phase.} 
To simulate the expected neutrino data signal in DUNE, we assume the true values for neutrinos and antineutrinos to be as listed in Table~\ref{tab:par2}.
Then, in the statistical analysis, we vary freely all the oscillation parameters, except the solar ones, which are fixed to their best fit values throughout the simulations. Given the great precision in the determination of the reactor mixing angle by the short-baseline reactor experiments~\cite{An:2016ses,RENO:2015ksa,Abe:2014bwa}, in our analysis we use a prior on $\overline{\theta}_{13}$, but not on $\theta_{13}$. We also consider three different values for the atmospheric angles, as indicated in Table~\ref{tab:par2}. The exposure considered in the analysis corresponds to 300~\ktMWyr.

\begin{table}
\centering
\caption{Oscillation parameters used to simulate neutrino and antineutrino data for the DUNE CPT sensitivity analysis.}
\label{tab:par2}
\begin{tabular}{c | c }\hline
Parameter & Value \\ \hline\hline
    $\Delta m^2_{21}$& $7.56\times 10^{-5}\text{eV}^2$\\ 
    $\Delta m^2_{31}$&  $2.55\times 10^{-3}\text{eV}^2$\\
    $\sin^2\theta_{12}$ & 0.321\\  
    $\sin^2\theta_{23}$ &  0.43, 0.50, 0.60\\  
    $\sin^2\theta_{13}$ & 0.02155\\  
    $\delta$ & 1.50$\pi$\\\hline
\end{tabular}
\end{table}

\begin{figure}[htb]
 \centering
        \includegraphics[width=0.9\columnwidth]{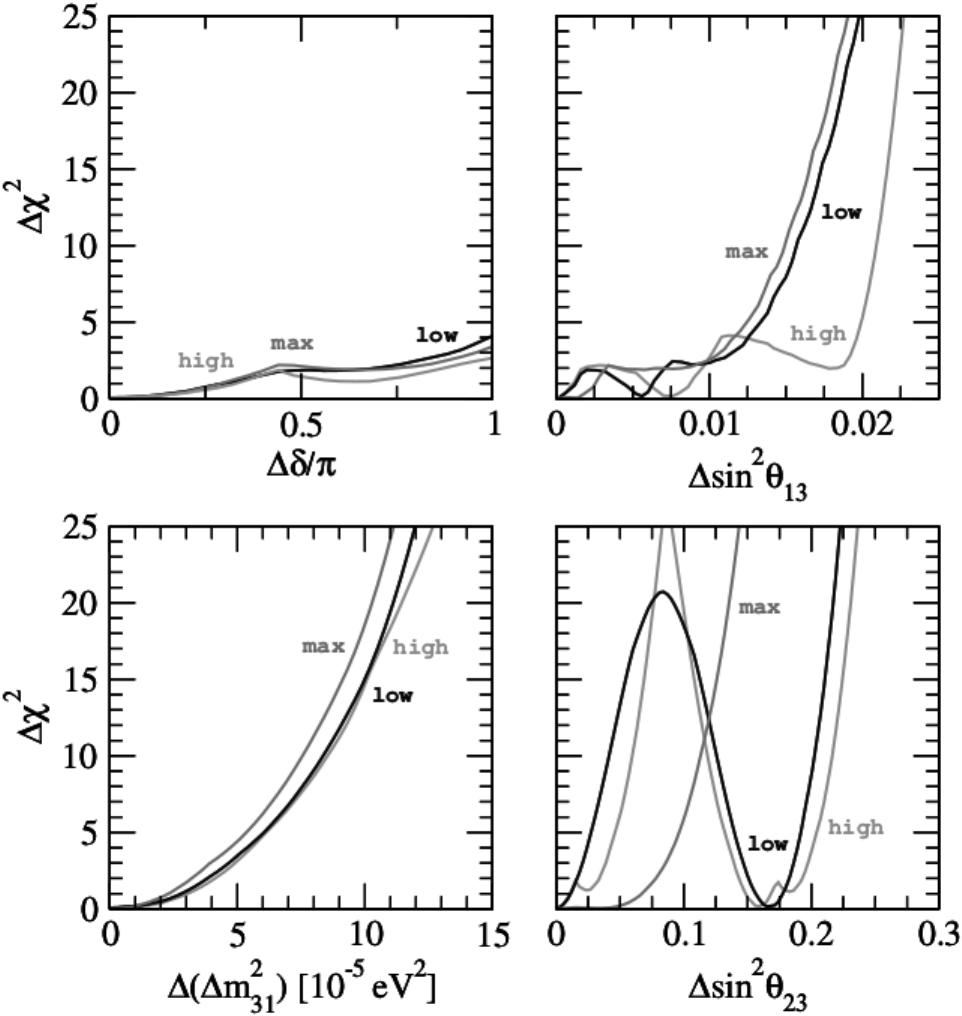}
        \caption[Sensitivities to the difference of neutrino and antineutrino parameters]{The sensitivities of DUNE to the difference of neutrino and antineutrino parameters: 
        $\Delta\delta$, $\Delta(\Delta m_{31}^2)$, $\Delta(\sin^2\theta_{13})$ and $\Delta(\sin^2\theta_{23})$  
        for the atmospheric angle in the lower octant (black line),  in the upper octant (light gray line) and for maximal mixing (dark gray line).}
	\label{fig:sensitivity-CPT}
\end{figure}

Therefore, to test the sensitivity at DUNE we perform the simulations assuming $\Delta x = |x-\overline{x}| = 0$, where $x$ is any of the oscillation parameters. Then we estimate the sensitivity to $\Delta x\neq 0$. To do so, we calculate two $\chi^2$-grids, one for neutrinos and one for antineutrinos, varying the four parameters of interest, in this case the atmospheric oscillation parameters. After minimizing over all parameters except $x$ and $\overline{x}$, we calculate 

\begin{equation}
 \chi^2(\Delta x) = \chi^2(|x-\overline{x}|) = \chi^2(x)+\chi^2(\overline{x}),
 \label{eq:chi2-nu-nubar}
\end{equation}
where we have considered all the possible combinations of $|x-\overline{x}|$. The results are presented in Fig.~\ref{fig:sensitivity-CPT}, where we plot three different lines, labelled as ``high'', ``max'' and ``low.'' These refer to the assumed value for the atmospheric angle: in the lower octant (low), maximal mixing (max) or in the upper octant (high). Here we can see that there is sensitivity neither to $\Delta(\sin^2\theta_{13})$, where the 3$\sigma$ bound would be of the same order 
as the current measured value for $\sin^2\overline{\theta}_{13}$, nor to $\Delta\delta$, where no single value of the parameter would be excluded at more than 2$\sigma$.

On the contrary, interesting results for $\Delta(\Delta m_{31}^2)$ and $\Delta(\sin^2\theta_{23})$ are obtained. First, we see that DUNE can put stronger bounds on the difference of the atmospheric mass splittings, namely $\Delta(\Delta m_{31}^2) < 8.1\times 10^{-5}$, improving the current neutrino bound by one order of magnitude. For the atmospheric angle, we obtain different results depending on the true value assumed in the simulation of DUNE data. In the lower right panel of Fig.~\ref{fig:sensitivity-CPT} we see the different behavior obtained for $\theta_{23}$ with the values of $\sin^2\theta_{23}$ from Table~\ref{tab:par2}, i.e., lying in the lower octant, being maximal, and lying in the upper octant.
As one might expect, the sensitivity increases with $\Delta\sin^2\theta_{23}$ in the case of maximal mixing. However, if the true value lies in the lower or upper octant, a degenerate solution appears in the complementary octant.

In  
some types of neutrino oscillation experiments, e.g., accelerator experiments, neutrino and antineutrino data are obtained in separate experimental runs. The usual procedure followed by the experimental collaborations, as well as the global oscillation fits as for example Ref.~\cite{deSalas:2017kay}, assumes \dword{cpt} invariance and analyzes the full data sample in a joint way.
However, if \dword{cpt} is violated in nature, the outcome of the joint data analysis might give rise to what we call an ``imposter'' solution, i.e., one that does not correspond to the true solution of any channel.

\begin{figure}[htb]
 \centering
        \includegraphics[width=0.9\columnwidth]{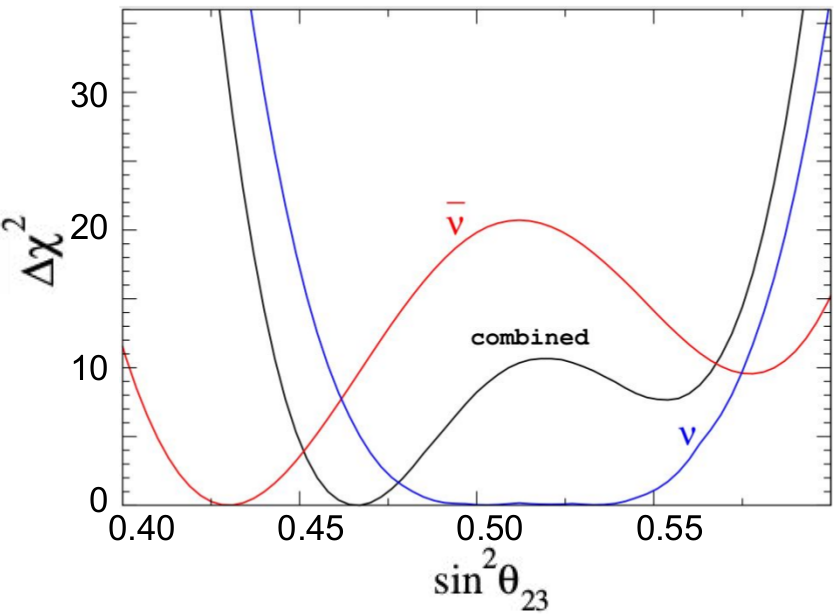}
        \caption[Sensitivity to $\theta_{23}$ for (anti)neutrinos, and combination under CPT conservation]{DUNE sensitivity to the atmospheric angle for neutrinos (blue), antineutrinos (red), and to the combination of both under the assumption of \dword{cpt} conservation (black).
         }
	\label{fig:imposter-sq23}
\end{figure}

Under the assumption of \dword{cpt} conservation, the $\chi^2$ functions are computed according to

\begin{equation}
 \chi^2_{\text{total}}=\chi^2(\nu)+\chi^2(\overline{\nu})\, ,
 \label{eq:CPT-cons}
\end{equation}
and assuming that the same parameters describe neutrino and antineutrino flavor oscillations. In contrast, in Eq.~(\ref{eq:chi2-nu-nubar}) we first profiled over the parameters in neutrino and antineutrino mode separately and then added the profiles. Here, we shall assume \dword{cpt} to be violated in nature, but perform our analysis as if it were conserved. As an example, we assume that the true value for the atmospheric neutrino mixing is $\sin^2\theta_{23}=0.5$, while the antineutrino mixing angle is given by $\sin^2\overline{\theta}_{23}=0.43$. The rest of the oscillation parameters are set to the values in Table~\ref{tab:par2}. Performing the statistical analysis in the \dword{cpt}-conserving way, as indicated in Eq.~(\ref{eq:CPT-cons}), we obtain the profile of the atmospheric mixing angle presented in Fig.~\ref{fig:imposter-sq23}. The profiles for the individual reconstructed results (neutrino and antineutrino) are also shown in the figure for comparison.
The result is a new best fit value at $\sin^2\theta^\text{comb}_{23}=0.467$, disfavoring the true values for neutrino and antineutrino parameters at approximately 3$\sigma$ and more than 5$\sigma$, respectively. 

Atmospheric neutrinos are a unique tool for studying neutrino oscillations: the oscillated flux contains all flavors of neutrinos and antineutrinos, is very sensitive to matter effects and to both \dm{} parameters, and covers a wide range of $L/E$. In principle, all oscillation parameters could be measured, with high
complementarity to measurements performed with a neutrino beam. 
Studying \dword{dune} atmospheric neutrinos is also a promising approach
to search for \dword{bsm} effects such as Lorentz and \dword{cpt} violation.
The \dword{dune} \dword{fd}, with its large mass and the overburden to protect it from atmospheric muon background, is an ideal tool for these studies.

The effective field theory describing \dword{cpt} violation is the  \dword{sme} \cite{ck97}, where \dword{cpt} violation is accompanied by Lorentz violation. This approach introduces a large set of neutrino coefficients governing corrections to standard neutrino-neutrino and antineutrino-antineutrino mixing probabilities, oscillations between neutrinos and antineutrinos, and modifications of oscillation-free propagation, all of which incorporate unconventional dependencies on the magnitudes and directions of momenta and spin.
For \dword{dune} atmospheric neutrinos,
the long available baselines,
the comparatively high energies accessible,
and the broad range of momentum directions
offer advantages that can make possible great
improvements 
in sensitivities to certain types of Lorentz and \dword{cpt} violation~\cite{Kostelecky:2003cr,Kostelecky:2011gq,Kostelecky:2003xn,Kostelecky:2004hg,Diaz:2009qk,Diaz:2013saa,Diaz:2013wia}.
To date,
experimental searches for Lorentz and \dword{cpt} violation
with atmospheric neutrinos have been published 
by the IceCube and \superk collaborations~\cite{Abbasi:2010kx,Abe:2014wla,Aartsen:2017ibm}.
Similar studies are possible with \dword{dune},
and many \dword{sme} coefficients can be measured that remain unconstrained to date.

\begin{figure}[htb]
\centering
\includegraphics[width=0.98\columnwidth]{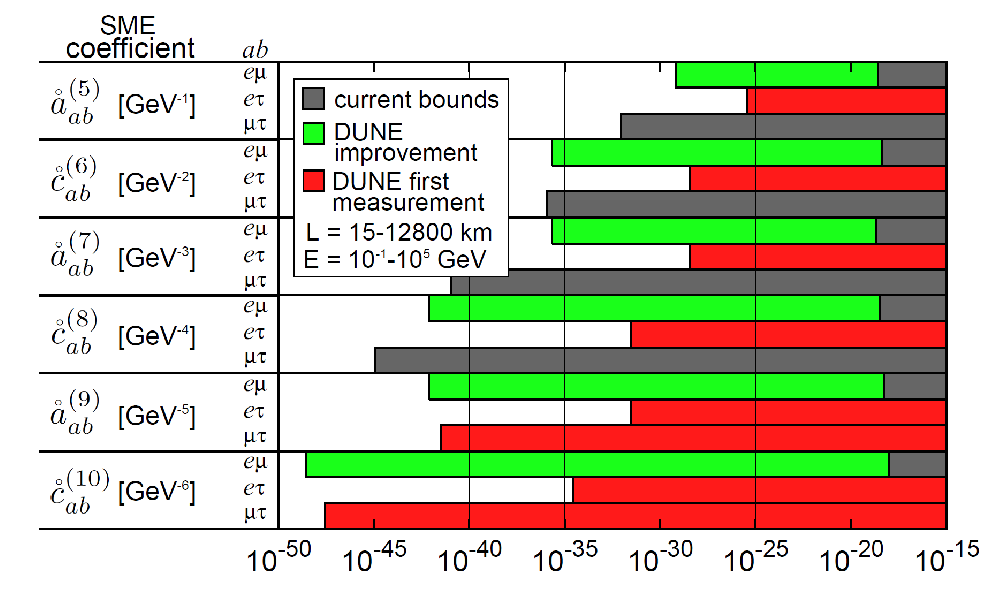}
\caption{Estimated sensitivity to Lorentz and CPT violation with atmospheric neutrinos in the non-minimal isotropic Standard Model Extension. The sensitivities are estimated by requiring that the Lorentz/CPT-violating effects are comparable in size to
those from conventional neutrino oscillations.}
\label{fig:atm}
\end{figure}

An example of the potential reach of studies with \dword{dune}
is shown in Fig.~\ref{fig:atm},
which displays estimated sensitivities
from atmospheric neutrinos in \dword{dune} to a subset of SME coefficients 
controlling isotropic (rotation-invariant) violations 
in the Sun-centered frame~\cite{Kostelecky:2002hh}.
The sensitivities are estimated by requiring that the
Lorentz/\dword{cpt}-violating effects are comparable in size to
those from conventional neutrino oscillations. 
The eventual \dword{dune} constraints will be determined by the
ultimate precision of the experiment (which is set in
part by the exposure).  The gray bars in Fig.~\ref{fig:atm} show existing limits.  These conservative sensitivity estimates show that \dword{dune} can achieve first measurements (red) on some coefficients that have never previously been measured and improved measurements (green) on others,
that have already been constrained in previous experiments but that can be measured with greater sensitivity with \dword{dune}.

\begin{figure}[htb]
\centering
\includegraphics[width=0.98\columnwidth]{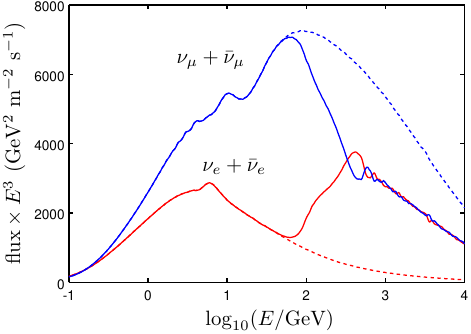}
\caption{Atmospheric fluxes of neutrinos and antineutrinos as a function of energy for conventional oscillations (dashed line) and in the non-minimal isotropic Standard Model Extension (solid line).}
\label{fig:atm2}
\end{figure}

To illustrate an \dword{sme} modification of oscillation probabilities,
consider a measurement of the atmospheric neutrino and antineutrino flux
as a function of energy.
For definiteness,
we adopt atmospheric neutrino fluxes~\cite{Honda:2015fha},
evaluated using the NRLMSISE-00 global atmospheric model~\cite{Picone},
that result from a production event at an altitude of \SI{20}{\km}.
Assuming conventional oscillations with standard three-flavor oscillation parameter values from the \dword{pdg}~\cite{Tanabashi:2018oca},
the fluxes at the \dword{fd} are shown in Fig.~\ref{fig:atm2}.
The sum of the \nue and \anue fluxes
is shown as a function of energy as a red dashed line, 
while the sum of the \numu and \anumu fluxes 
is shown as a blue dashed line. 
Adding an isotropic non-minimal coefficient for Lorentz violation
of magnitude $\mathaccent'27 c^{(6)}_{e \mu} = \SI{1e-28}{\per\GeV\square}$
changes the fluxes from the dashed lines to the solid ones.
This coefficient is many times smaller
than the current experimental limit.
Nonetheless,
the flux spectrum is predicted to change significantly 
at energies over approximately \SI{100}{\GeV}, changing the expected number of events.

\section{Neutrino Tridents at the Near Detector}
\label{sec:tridents}
\begin{figure}[!hb]
\centering
\includegraphics[width=0.45\columnwidth]{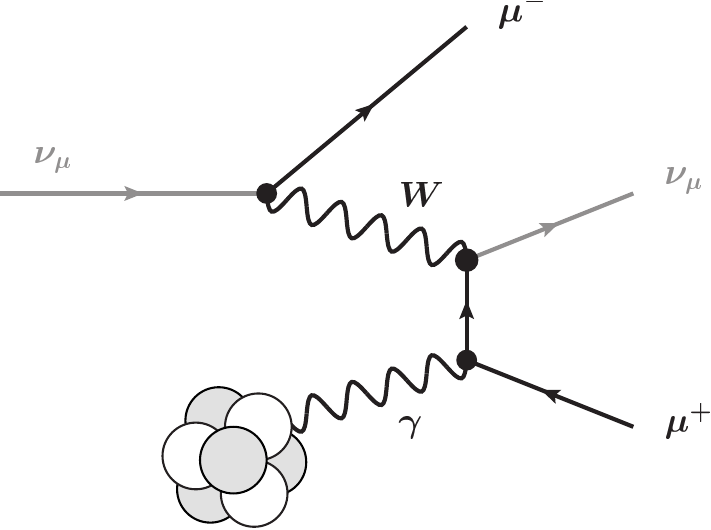} \qquad
\includegraphics[width=0.45\columnwidth]{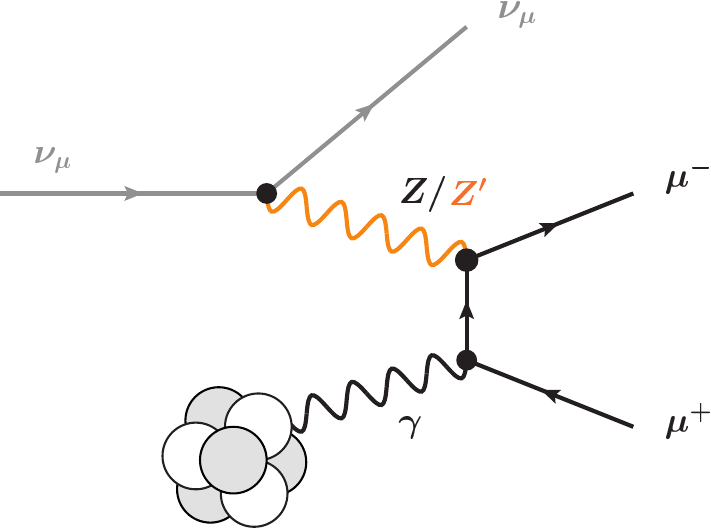} \\[\baselineskip]
\includegraphics[width=0.45\columnwidth]{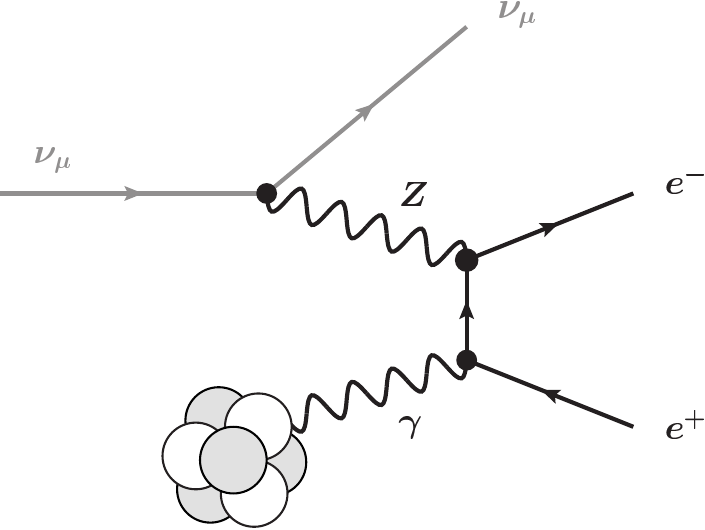} \qquad
\includegraphics[width=0.45\columnwidth]{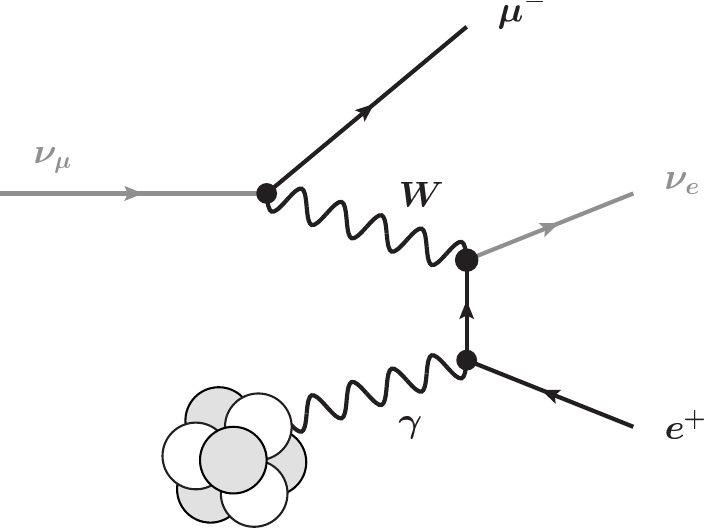} \\[\baselineskip]
\caption[Example diagrams for $\numu$-induced trident processes in the SM]{Example diagrams for muon-neutrino-induced trident processes in the Standard Model. A second set of diagrams where the photon couples to the negatively charged leptons is not shown. Analogous diagrams exist for processes induced by different neutrino flavors and by antineutrinos. A diagram illustrating trident interactions mediated by a new $Z'$ gauge boson, discussed in the text, is shown on the top right.}
\label{fig:diagrams}
\end{figure}

%%%
Neutrino trident production is a weak process in which a neutrino, scattering off the Coulomb field of a heavy nucleus, generates a pair of charged leptons~\cite{Czyz:1964zz,Lovseth:1971vv,Fujikawa:1971nx,Koike:1971tu,Koike:1971vg,Brown:1973ih,Belusevic:1987cw,Zhou:2019vxt,Beacom:2019pzs}, as shown in Fig.~\ref{fig:diagrams}.

Measurements of muonic neutrino tridents ($\nu_\mu \to \nu_\mu~\mu^+~\mu^-$) were carried out at the CHARM-II~\cite{Geiregat:1990gz}, CCFR~\cite{Mishra:1991bv} and NuTeV~\cite{Adams:1999mn} experiments:
%%%
\[
\frac{\sigma(\nu_\mu \to \nu_\mu \mu^+\mu^-)_\text{exp}}{\sigma(\nu_\mu \to \nu_\mu \mu^+\mu^-)_\text{SM}} = 
\begin{cases}
1.58 \pm 0.64         & \text{(CHARM-II)} \\ 
0.82 \pm 0.28         & \text{(CCFR)} \\
0.72 ^{+1.73}_{-0.72} & \text{(NuTeV)} 
\end{cases}
\]
%%%
The high-intensity muon-neutrino flux at the DUNE \dword{nd}  will lead to a sizable production rate of trident events (see Table~\ref{tab:trident_rates}), offering excellent prospects to improve the above measurements~\cite{Altmannshofer:2019zhy,Ballett_2019,Ballett_2019-zp}. A deviation from the event rate predicted by the \dword{sm} could be an indication of new interactions mediated by the corresponding new gauge bosons~\cite{Altmannshofer:2014pba}. 

\begin{table}[tb]
\centering
\caption{Expected number of \dword{sm} $\nu_\mu$ and $\bar\nu_\mu$-induced trident events at the LArTPC of the DUNE \dword{nd} per metric ton of argon and year of operation.}
\label{tab:trident_rates}
\begin{tabular}{lcc}
\hline
Process & Coherent & Incoherent \\ \hline
$\nu_\mu \to \nu_\mu \mu^+\mu^-$ & $1.17 \pm 0.07$ & $0.49 \pm 0.15$ \\
$\nu_\mu \to \nu_\mu e^+e^-$ & $2.84 \pm 0.17$ & $0.18 \pm 0.06$\\
$\nu_\mu \to \nu_e e^+\mu^-$ & $9.8 \pm 0.6$ & $1.2 \pm 0.4$ \\
$\nu_\mu \to \nu_e \mu^+e^-$ & $0$ & $0$ \\ \hline
$\bar\nu_\mu \to \bar\nu_\mu \mu^+\mu^-$ & $0.72 \pm 0.04$ & $0.32 \pm 0.10$ \\
$\bar\nu_\mu \to \bar\nu_\mu e^+e^-$ & $2.21 \pm 0.13$ & $0.13 \pm 0.04$ \\
$\bar\nu_\mu \to \bar\nu_e e^+\mu^-$ & $0$ & $0$ \\
$\bar\nu_\mu \to \bar\nu_e \mu^+e^-$ & $7.0 \pm 0.4$ & $0.9 \pm 0.3$ \\ \hline
\end{tabular}
\end{table}

The main challenge in obtaining a precise measurement of the muonic trident cross section will be the copious backgrounds, mainly consisting of \dword{cc} single-pion production events, $\nu_\mu N \to \mu \pi N^\prime$, as muon and pion tracks can be easily confused in LArTPC detectors. The discrimination power of the DUNE \dword{nd} LArTPC was evaluated using large simulated data sets of signal and background. Each simulated event represents a different neutrino-argon interaction in the active volume of the detector. Signal events were generated using a standalone code \cite{Altmannshofer:2019zhy} that simulates trident production of muons and electrons through the scattering of $\nu_{\mu}$ and $\nu_e$ on argon nuclei. The generator considers both the coherent scattering on the full nucleus (the dominant contribution) and the incoherent scattering on individual nucleons. Background events, consisting of several \dword{sm} neutrino interactions, were generated using \dword{genie}. Roughly $38\%$ of the generated events have a charged pion in the final state, leading to two charged tracks with muon-like energy deposition pattern ($\mathrm{d}E/\mathrm{d}x$), as in the trident signal. All final-state particles produced in the interactions were propagated through the detector geometry using the \textsc{Geant4}-based  simulation of the DUNE \dword{nd}. Charge collection and readout were not simulated, and possible inefficiencies due to mis-reconstruction effects or event pile-up were disregarded for simplicity.

\begin{figure*}[!tb]
\centering
\includegraphics[width=0.35\textwidth]{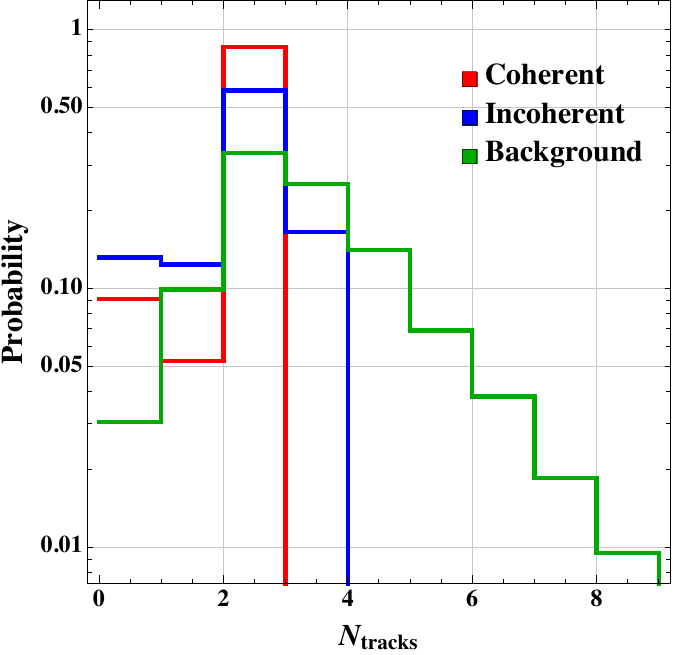}
\includegraphics[width=0.35\textwidth]{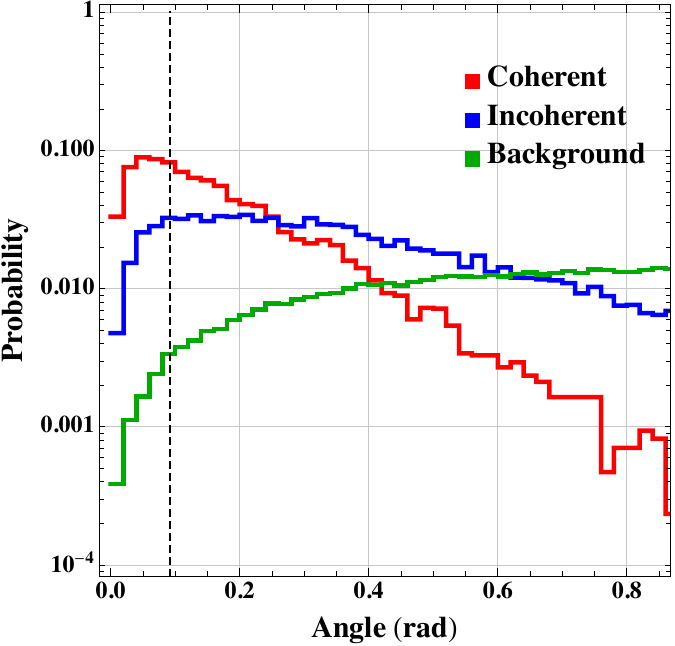} \\[0.75\baselineskip]
\includegraphics[width=0.35\textwidth]{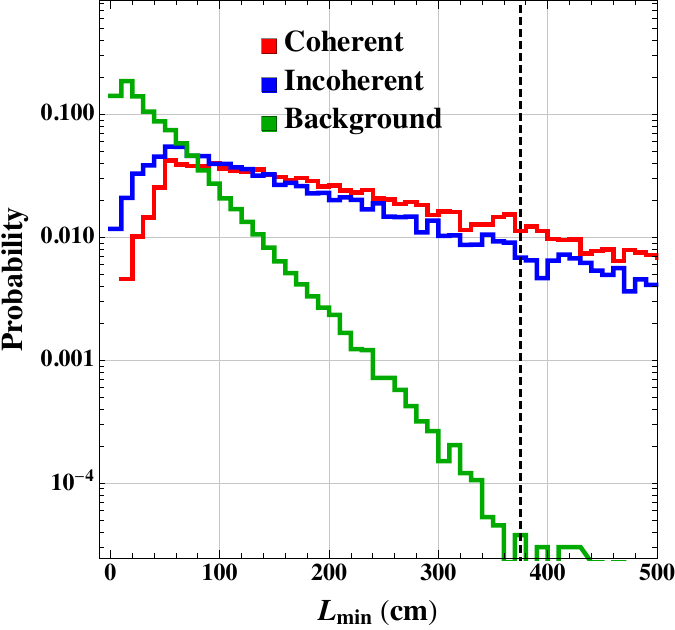}
\includegraphics[width=0.35\textwidth]{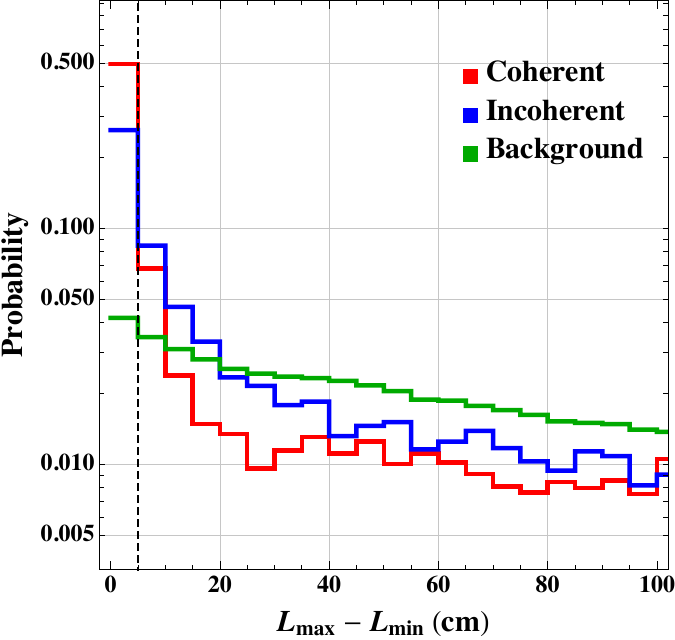}
\caption[
Signal and background  
for selecting  
muonic trident interactions in ND LArTPC]{Event kinematic distributions of signal and background considered for the selection of muonic trident interactions in the \dword{nd} \dword{lartpc}: number of tracks (top left), angle between the two main tracks (top right), length of the shortest track (bottom left), and the difference in length between the two main tracks (bottom right). The dashed, black vertical lines indicate the optimal cut values used in the analysis.} \label{fig:trident_kinematics}
\end{figure*}

Figure~\ref{fig:trident_kinematics} shows the distribution (area normalized) for signal and background of the different kinematic variables used in our analysis for the discrimination between signal and background. As expected, background events tend to contain a higher number of tracks than the signal. The other distributions also show a clear discriminating power: the angle between the two tracks is typically much smaller in the signal than in the background. Moreover, the signal tracks (two muons) tend to be longer than tracks in the background (mainly one muon plus one pion).

\begin{figure}[htb]
\centering
\includegraphics[width=0.9\columnwidth]{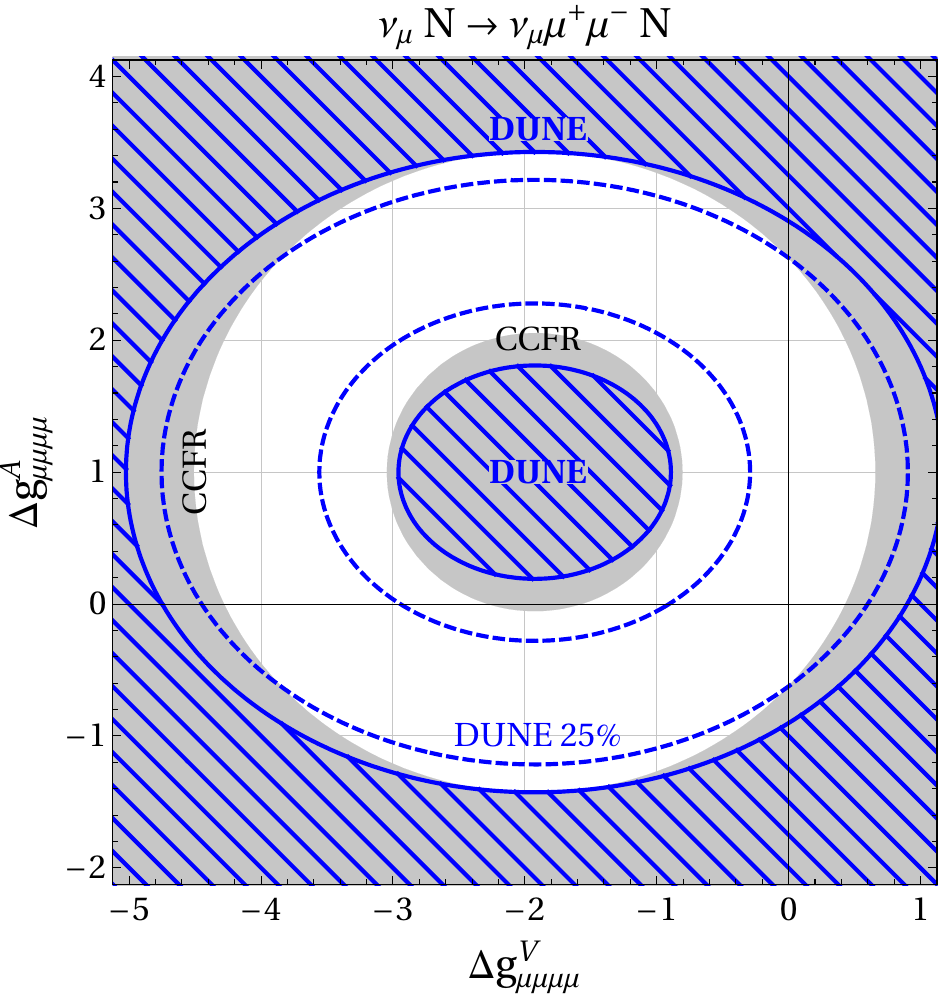}
\caption[$\nu_\mu N \to \nu_\mu \mu^+\mu^- N$ cross section at ND and (axial-)vector couplings of \numu{} to muons]
{95\% CL. sensitivity of a 40\% (blue hashed regions) and a 25\% (dashed contours) uncertainty measurement of the $\nu_\mu N \to \nu_\mu \mu^+\mu^- N$ cross section at the DUNE near detector to modifications of the vector and axial-vector couplings of muon-neutrinos to muons. The gray regions are excluded at 95\% CL by existing measurements of the cross section by the CCFR Collaboration. The intersection of the thin black lines indicates the \dword{sm} point. A 40\% precision measurement could be possible with 6
~years of data taking in neutrino mode.}
\label{fig:trident_gVgA}
\end{figure}

The sensitivity of neutrino tridents to heavy new physics (i.e., heavy compared to the momentum transfer in the process) can be parameterized in a model-independent way using a modification of the effective four-fermion interaction Hamiltonian.  Focusing on the case of muon neutrinos interacting with muons, the vector and axial-vector couplings can be written as
%%%%%%%%%%
\begin{eqnarray}
g_{\mu\mu\mu\mu}^V & = & 1 + 4 \sin^2\theta_W + \Delta g_{\mu\mu\mu\mu}^V \quad \mathrm{and} \\
\quad g_{\mu\mu\mu\mu}^A & = & -1 + \Delta g_{\mu\mu\mu\mu}^A ~, \nonumber
\end{eqnarray}
%%%%%%%%%%
where $\Delta g_{\mu\mu\mu\mu}^V$ and $\Delta g_{\mu\mu\mu\mu}^A$ represent possible new physics contributions. Couplings involving other combinations of lepton flavors can be modified analogously. Note, however, that for interactions that involve electrons, very strong constraints can be derived from LEP bounds on electron contact interactions~\cite{Schael:2013ita}. The modified interactions of the muon-neutrinos with muons alter the cross section of the $\nu_\mu N \to \nu_\mu \mu^+\mu^- N$ trident process. In Fig.~\ref{fig:trident_gVgA} we show the regions in the $\Delta g^V_{\mu\mu\mu\mu}$ vs.\ $\Delta g^A_{\mu\mu\mu\mu}$ plane that are excluded by the existing CCFR measurement $\sigma_\text{CCFR} / \sigma_\text{CCFR}^\text{SM} = 0.82 \pm 0.28$~\cite{Mishra:1991bv} at the 95\% \dword{cl} in gray. A measurement of the $\nu_\mu N \to \nu_\mu \mu^+\mu^- N$ cross section with $40\%$ uncertainty (obtained after running for $\sim6$
~years in neutrino mode or, equivalently, 3~years in neutrino mode and 3~years in antineutrino mode) at the DUNE \dword{nd} could cover the blue hashed regions (95\% \dword{cl}). These numbers show that a measurement of the SM di-muon trident production at the 40\% level could be possible. Our baseline analysis does not extend the sensitivity into parameter space that is unconstrained by the CCFR measurement. However, it is likely that the use of a magnetized spectrometer, as it is being considered for the DUNE \dword{nd}, able to identify the charge signal of the trident final state, along with a more sophisticated  event selection (e.g., deep-learning-based), will significantly improve separation between neutrino trident interactions and backgrounds. Therefore, we also present the region (blue dashed line) that could be probed by a 25\% measurement of the neutrino trident cross section at DUNE, which would extend the coverage of new physics parameter space substantially.

We consider a class of models that modify the trident cross section through the presence of an additional neutral gauge boson, $Z'$, that couples to neutrinos and charged leptons. A consistent way of introducing such a $Z'$ is to gauge an anomaly-free global symmetry of the \dword{sm}. Of particular interest is the $Z'$ that is based on gauging the difference of muon-number and tau-number, $L_\mu - L_\tau$~\cite{He:1990pn,He:1991qd}. Such a $Z'$ is relatively weakly constrained and can for example address the longstanding discrepancy between \dword{sm} prediction and measurement of the anomalous magnetic moment of the muon, $(g-2)_\mu$~\cite{Baek:2001kca,Harigaya:2013twa}. The $L_\mu - L_\tau$ $Z'$ has also been used in models to explain $B$ physics anomalies~\cite{Altmannshofer:2014cfa} and as a portal to \dword{dm}~\cite{Baek:2008nz,Altmannshofer:2016jzy}. The $\nu_\mu N \to \nu_\mu \mu^+\mu^- N$ trident process has been identified as an important probe of gauged $L_\mu - L_\tau$ models over a broad range of $Z^\prime$ masses~\cite{Altmannshofer:2014cfa,Altmannshofer:2014pba}.

\begin{figure}[htb] 
\centering
\includegraphics[width=0.9\columnwidth]{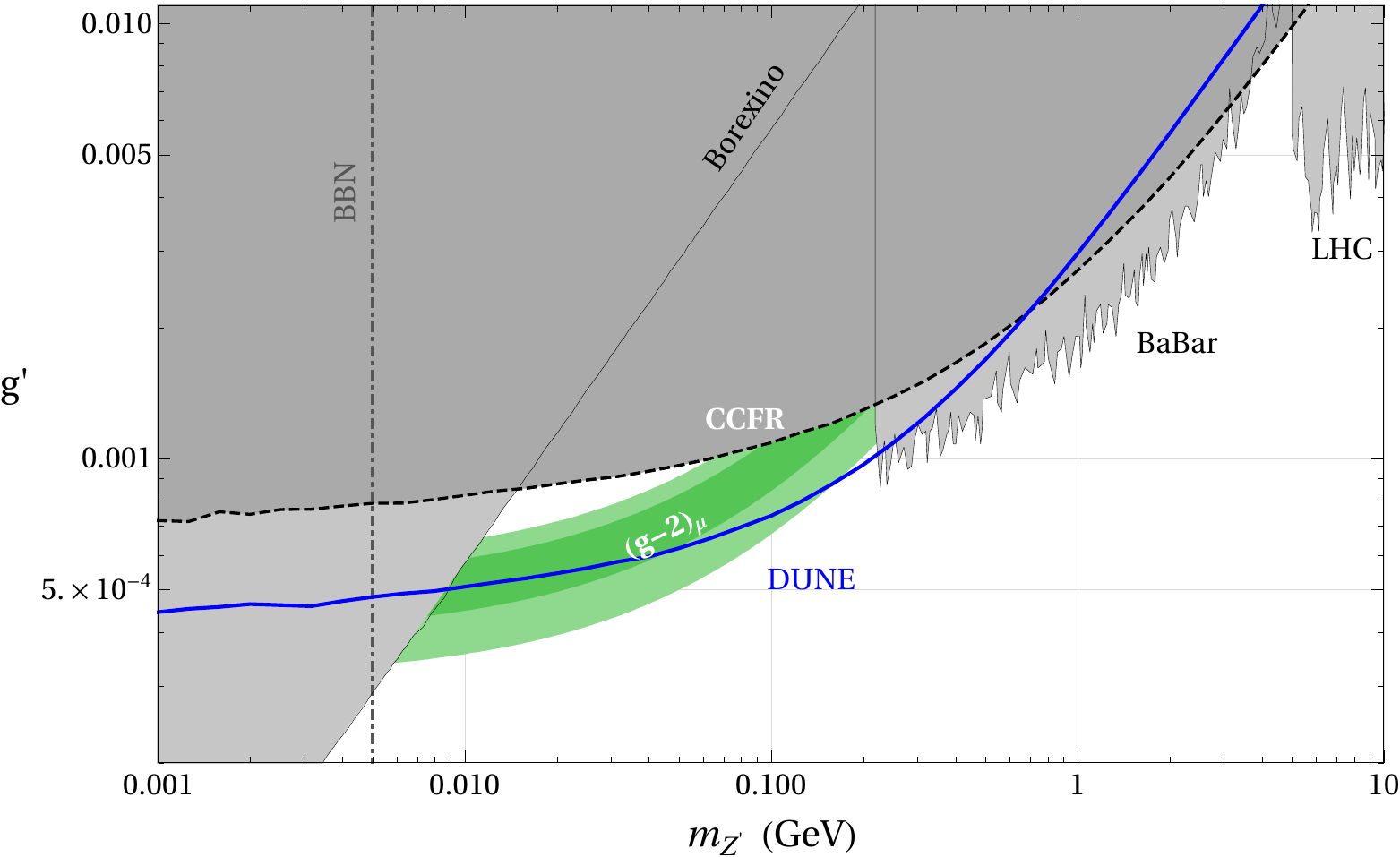}
\caption[Existing constraints and projected sensitivity in the $L_\mu - L_\tau$ parameter space]{Existing constraints and projected DUNE sensitivity in the $L_\mu - L_\tau$ parameter space. Shown in green is the region where the $(g-2)_\mu$ anomaly can be explained at the $2\sigma$ level. The parameter regions already excluded by existing constraints are shaded in gray and correspond to a CMS search for $pp \to \mu^+\mu^- Z' \to \mu^+\mu^-\mu^+\mu^-$~\cite{Sirunyan:2018nnz} (``LHC''), a BaBar search for $e^+e^- \to \mu^+\mu^- Z' \to \mu^+\mu^-\mu^+\mu^-$~\cite{TheBABAR:2016rlg} (``BaBar''), a previous measurement of the trident cross section~\cite{Mishra:1991bv,Altmannshofer:2014pba} (``CCFR''), a measurement of the scattering rate of solar neutrinos on electrons~\cite{Bellini:2011rx,Harnik:2012ni,Agostini:2017ixy} (``Borexino''), and bounds from Big Bang Nucleosynthesis~\cite{Ahlgren:2013wba,Kamada:2015era} (``BBN''). The DUNE sensitivity shown by the solid blue line assumes 6 years of data running in neutrino mode, leading to a measurement of the trident cross section with 40\% precision.}
\label{fig:LmuLtau}
\end{figure}

In Fig.~\ref{fig:LmuLtau} we show the existing CCFR constraint on the model parameter space in the $m_{Z'}$ vs. $g'$ plane, where $g'$ is the $L_\mu - L_\tau$ gauge coupling, and compare it to the region of parameter space where the anomaly in $(g-2)_\mu = 2 a_\mu$ can be explained. The green region shows the $1\sigma$ and $2\sigma$ preferred parameter space corresponding to a shift $\Delta a_\mu = a_\mu^\text{exp}-a_\mu^\text{SM} = (2.71 \pm 0.73) \times 10^{-9}$~\cite{Keshavarzi:2018mgv}.
In addition, constraints from LHC searches for the $Z'$ in the $pp \to \mu^+\mu^- Z' \to \mu^+\mu^-\mu^+\mu^-$ process~\cite{Sirunyan:2018nnz} (see also~\cite{Altmannshofer:2014pba}) and direct searches for the $Z'$ at BaBar using the $e^+e^- \to \mu^+\mu^- Z' \to \mu^+\mu^-\mu^+\mu^-$ process~\cite{TheBABAR:2016rlg} are shown.  
A Borexino bound on non-standard contributions to neutrino-electron scattering~\cite{Harnik:2012ni,Bellini:2011rx,Agostini:2017ixy} has also been used to constrain the $L_\mu - L_\tau$ gauge boson~\cite{Kamada:2015era,Araki:2015mya,Kamada:2018zxi}. Our reproduction of the Borexino constraint is shown in Fig.~\ref{fig:LmuLtau}.
For very light $Z'$ masses of $O$(few MeV) and below, strong constraints from measurements of the effective number of relativistic degrees of freedom during Big Bang Nucleosynthesis (BBN) apply~\cite{Ahlgren:2013wba,Kamada:2015era}.
Taking into account all relevant constraints, parameter space to explain $(g-2)_\mu$ is left below the di-muon threshold $m_{Z'} \lesssim 210$~MeV.  The DUNE sensitivity shown by the solid blue line assumes a measurement of the trident cross section with $40\%$ precision.

\section{Dark Matter Probes}\label{sec:DM}
Dark matter is a crucial ingredient to understand the cosmological history of the universe, and the most up-to-date measurements suggests the existence of \dword{dm} with a density
parameter ($\Omega_{c}$) of 0.264~\cite{Aghanim:2018eyx}. 
In light of this situation, a tremendous amount of experimental effort has gone into  
the search for \dword{dm}-induced signatures, for example, \dword{dm} direct and indirect detections and collider searches. However, no ``smoking-gun'' signals have been discovered thus far while more parameter space in relevant \dword{dm} models is simply ruled out.  
It is noteworthy that most conventional \dword{dm} search strategies are designed to be sensitive to signals from the \dword{wimp}, one of the well-motivated \dword{dm} candidates, whose mass range is from a few GeV to tens of TeV. 
The non-observation of \dword{dm} via non-gravitational interactions actually motivates unconventional or alternative \dword{dm} search schemes. 
One such possibility is  
a search for experimental signatures induced by boosted, hence relativistic, \dword{dm} for which 
a mass range smaller than that of the weak scale is often motivated. 

One of the possible ways to produce and then detect relativistic \dword{dm} particles can be through accelerator experiments, 
for example, neutrino beam experiments~\cite{Alexander:2016aln,Battaglieri:2017aum,LoSecco:1980nf,Acciarri:2015uup,Dutta:2019nbn}. 
Due to highly intensified beam sources, large signal statistics is usually expected 
so that this sort of search strategy can allow for significant
sensitivity to \dword{dm}-induced signals despite the feeble interaction of \dword{dm} with \dword{sm} particles. 
DUNE will perform a search for the relativistic scattering of \dword{ldm}, whose lowest mass particle is denoted as $\chi$ throughout this section, at the \dword{nd}, as it is close enough to the beam source to sample a substantial level of \dword{dm} flux, assuming that \dword{dm} is produced.

Alternatively, it is possible that \dword{bdm} particles are created in the universe under non-minimal dark-sector scenarios~\cite{Agashe:2014yua,Belanger:2011ww}, and can reach terrestrial detectors. 
For example, one can imagine a two-component \dword{dm} scenario in which a lighter component ($\chi$) is usually a subdominant relic with direct coupling to \dword{sm} particles, while the heavier (denoted as $\psi$ throughout this section) is the cosmological \dword{dm} that pair-annihilates directly to a lighter \dword{dm} pair, not to \dword{sm} particles. Other mechanisms such as semi-annihilation in which a \dword{dm} particle pair-annihilates to a (lighter) \dword{dm} particle and a dark sector particle that may decay away are also possible~\cite{DEramo:2010keq,Huang:2013xfa,Berger:2014sqa}.
In typical cases, the \dword{bdm} flux is not large and thus large-volume neutrino detectors are desirable 
to overcome the challenge in statistics (for an  exception, see~\cite{Cherry:2015oca,Giudice:2017zke,Cui:2017ytb,Bringmann:2018cvk}).

Indeed, a (full-fledged) DUNE \dword{fd} with a fiducial mass of \fdfiducialmass and quality detector performance is expected to possess competitive sensitivity to \dword{bdm} signals from various sources in the current universe such as the galactic halo~\cite{Agashe:2014yua,Alhazmi:2016qcs,Kim:2016zjx,Giudice:2017zke,Chatterjee:2018mej,Kim:2018veo,Necib:2016aez}, the sun~\cite{Huang:2013xfa,Berger:2014sqa,Kong:2014mia,Alhazmi:2016qcs,Kim:2018veo}, and dwarf spheroidal galaxies~\cite{Necib:2016aez}.
Furthermore, the \dword{protodune} detectors 
have taken data, and we anticipate preliminary studies with their cosmic data. Interactions of \dword{bdm} with electrons~\cite{Agashe:2014yua} 
and with hadrons (protons)~\cite{Berger:2014sqa}, were investigated for Cherenkov detectors, such as \superk, which recently published a dedicated search for \dword{bdm} in the electron channel~\cite{Kachulis:2017nci}. However, in such detectors the \dword{bdm} signal rate is shown to often be significantly attenuated due to Cherenkov threshold, in particular for hadronic channels.  \lar detectors, such as DUNE's, have the potential to greatly improve the sensitivity for \dword{bdm} compared to Cherenkov detectors. This is due to improved particle identification techniques, as well as a significantly lower energy threshold for proton detection. Earlier studies have shown an improvement with DUNE for \dword{bdm}-electron interaction~\cite{Necib:2016aez}.

We consider several benchmark \dword{dm} models. These describe only couplings of dark-sector states including \dword{ldm} particles.
We consider two example models: i)~a vector portal-type scenario where a (massive) dark-sector photon $V$ mixes with the \dword{sm} photon and ii)~a leptophobic $Z'$ scenario.
\dword{dm} and other dark-sector particles are assumed to be fermionic for convenience.

\paragraph{Benchmark Model i)}
The relevant interaction Lagrangian is given by~\cite{Kim:2016zjx}
\bea
\label{eq:lagrangian}
\mathcal{L}_{\rm int} & \supset  -\frac{\epsilon}{2}V_{\mu\nu}F^{\mu\nu} & +g_D \bar{\chi}\gamma^\mu \chi V_\mu \\ \nonumber
& & +g'_D \bar{\chi}'\gamma^\mu \chi V_\mu +h.c. , \\ \nonumber
\eea
where $V^{\mu\nu}$ and $F^{\mu\nu}$ are the field strength tensors for the dark-sector photon and the \dword{sm} photon, respectively. 
Here we have introduced the kinetic mixing parameter $\epsilon$, while $g_D$ and $g'_D$ parameterize the interaction strengths for flavor-conserving (second operator) and flavor-changing (third operator) couplings, respectively.  
Here $\chi$ and $\chi'$ denote a dark matter particle and a heavier, \textit{un}stable dark-sector state, respectively (i.e., $M_{\chi'}>M_{\chi}$), and the third term allows (boosted) $\chi$ transition to $\chi'$ after a scattering (i.e., an ``inelastic'' scattering process).

This model introduces six new free parameters that may be varied for our sensitivity analysis: dark photon mass $M_V$, \dword{dm} mass $M_{\chi}$, heavier dark-sector state mass $M_{\chi'}$, kinetic mixing parameter $\epsilon$, dark-sector couplings $g_D$ and $g'_D$. 
We shall perform our analyses with some of the parameters fixed to certain values for illustration.

\paragraph{Benchmark Model ii)}
This model employs a leptophobic $Z^\prime$ mediator for interactions with the nucleons. The interaction Lagrangian for this model is~\cite{Berger:2014sqa}
\bea
\label{eq:zprimelag}
\mathcal{L}_{\rm int} & \supset - & g_{\rm Z^\prime} \sum_f Z^\prime_\mu \bar{q}_f \gamma^\mu \gamma^5 q_f -  g_{\rm Z^\prime} Z^\prime_\mu \bar{\chi} \gamma^\mu \gamma^5 \chi \\ \nonumber
& & - Q_\psi g_{\rm Z^\prime} Z^\prime_\mu \bar{\psi} \gamma^\mu \gamma^5 \psi. \\ \nonumber 
\eea
Here, all couplings are taken to be axial. $f$ denotes the quark flavors in the \dword{sm} sector. The dark matter states are denoted by $\chi$ and $\psi$ with $M_\chi < M_\psi$. The coupling $g_{\rm Z^\prime}$ and the masses of the dark matter states are free parameters. The \dword{dm} flux abundance parameter, $Q_\psi$ is taken to be less than 1 and determines the abundance of dark matter in the universe. The hadronic interaction model study presented here is complementary to and has different phenomenology compared to others such as Benchmark Model i).

\begin{table}[htp]
    \centering
    \caption{A summary of the three different studies in this section. }
    \label{tab:summaryDMsignal}
    \begin{tabular}{c|c| c| c}
    \hline 
         & \ref{sec:darkmatter_ND} & \ref{bdmatfd} & \ref{sec:FDsun} \\
    \hline \hline 
    Model & i) & i) & ii) \\
    $\chi$ source& {Beam} & {Galaxy} & {Sun} \\
    Detector & ND & FD & FD \\
    Detection & \multirow{2}{*}{$\chi e^- \to \chi e^-$} & $\chi e^-(p) \to \chi'e^-(p)$, &  \multirow{2}{*}{$\chi N \to \chi X$} \\
    channel & & $\chi'\to \chi e^+e^-$ &  \\
    \hline 
    \end{tabular}
\end{table}

We summarize key information for the three different studies in this section in Table~\ref{tab:summaryDMsignal}. 
The $e^-$ ($p$) outside (inside) the parentheses in the third column imply the electron (proton) scattering channel. $N$ in the last column denotes a nucleon, while $X$ stands for particle(s) created via the $\chi-N$ scattering process.

\subsection{Search for Low-Mass Dark Matter at the Near Detector} \label{sec:darkmatter_ND}
\begin{figure}[htp]
\centering
 \includegraphics[width=0.9\columnwidth]{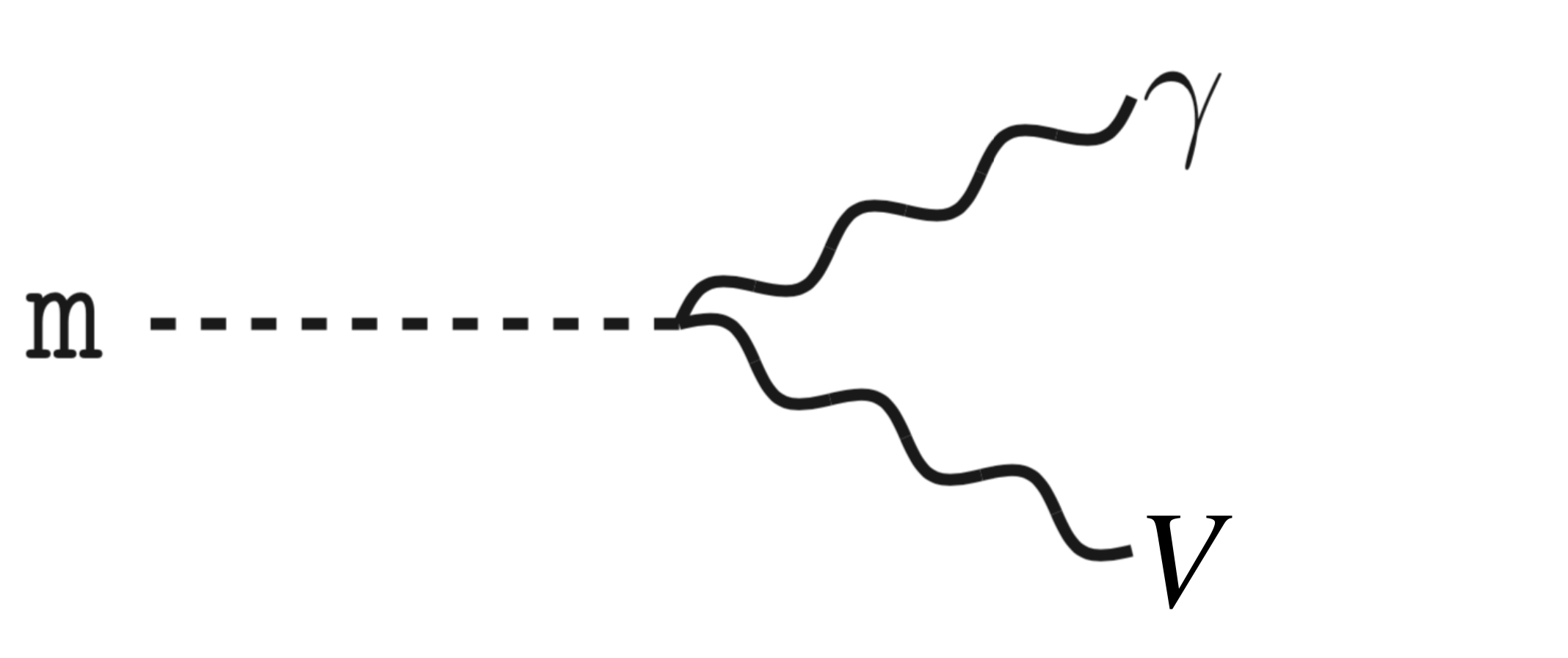}
    \includegraphics[width=0.9\columnwidth]{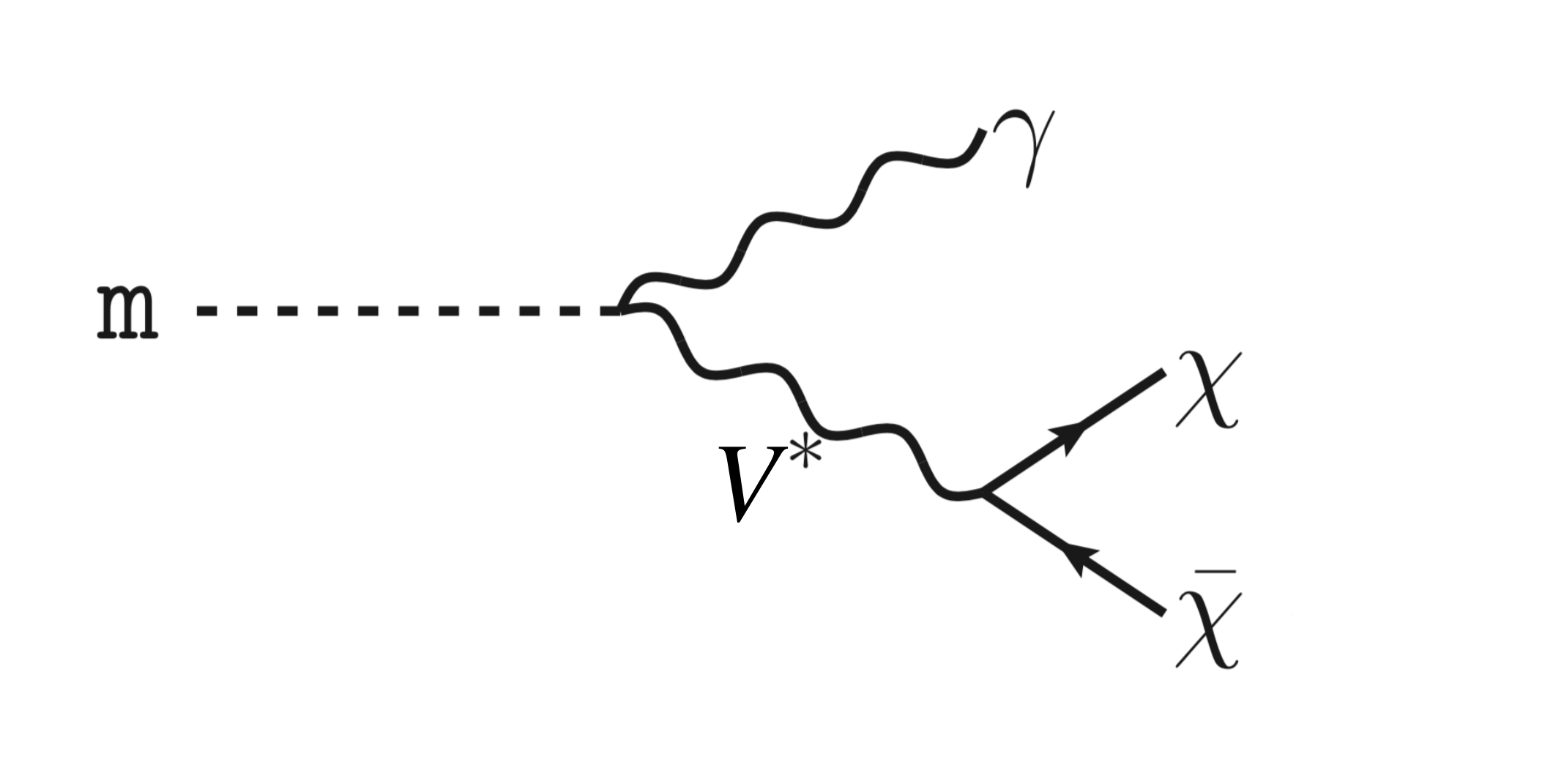}
\includegraphics[width=0.9\columnwidth]{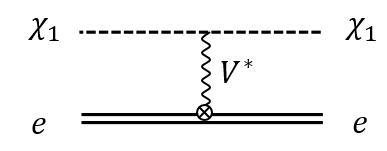}
\caption[DM production via meson decays and DM-e$^-$ elastic scattering]{Production of fermionic \dword{dm} via two-body pseudoscalar meson decay $\mathfrak{m} \to \gamma V$, when $M_{V} < m_\mathfrak{m}$ (top) or via three-body decay $\mathfrak{m} \to \gamma \chi \overline{\chi}$ (center) and \dword{dm}-electron elastic scattering (bottom).}
\label{fig:dm_prod}
\end{figure}

Here, we focus on Benchmark Model i) from Eq.~(\ref{eq:lagrangian}), specifically where only one \dword{dm} particle $\chi$ is relevant. 
We also define the dark fine structure constant $\alpha_D \equiv g_D^2/(4\pi)$. We assume that $\chi$ is a fermionic thermal relic -- in this case, the \dword{dm}/dark photon masses and couplings will provide a target for which the relic abundance matches the observed abundance in the universe. Here, the largest flux of dark photons $V$ and \dword{dm} to reach the DUNE \dword{nd} will come from the decays of light pseudoscalar mesons (specifically $\pi^0$ and $\eta$ mesons) that are produced in the DUNE target, as well as proton bremsstrahlung processes $p + p \to p + p + V$.
For the entirety of this analysis, we will fix $\alpha_D = 0.5$ and assume that the DM mass $M_{\chi}$ is lighter than half the mass of a pseudoscalar meson $\mathfrak{m}$ that is produced in the DUNE target. In this scenario, $\chi$  is produced via two decays, those of on-shell $V$ and those of off-shell $V$. This production is depicted in Fig.~\ref{fig:dm_prod}. 

The flux of \dword{dm} produced via meson decays -- via on-shell $V$ -- may be estimated by\footnote{See Ref.~\cite{DeRomeri:2019kic} for a complete derivation of these expressions, including those for meson decays via off-shell $V$.}
\begin{eqnarray}
    N_\chi & = & 2 N_\mathrm{POT} c_\mathfrak{m} \{ \mathrm{Br}(\mathfrak{m}\to \gamma\gamma) \\ \nonumber 
    & & \times  2 \varepsilon^2 \left(1 - \frac{M_{V}^2}{m_\mathrm{m}^2}\right)^3 \\ \nonumber
    & & \times \mathrm{Br}(V \to \chi\bar{\chi}) \} g(M_\chi, M_{V}), \\ \nonumber
\end{eqnarray}
where $N_\mathrm{POT}$ is the number of protons on target delivered by the beam, $c_\mathfrak{m}$ is the average number of meson $\mathfrak{m}$ produced per POT, the term in braces is the relative branching fraction of $\mathfrak{m} \to \gamma V$ relative to $\gamma\gamma$, and $g(x, y)$ characterizes the geometrical acceptance fraction of \dword{dm} reaching the DUNE \dword{nd}. $g(x, y)$ is determined given model parameters using Monte Carlo techniques. For the range of dark photon and \dword{dm} masses in which DUNE will set a competitive limit, the \dword{dm} flux due to meson decays will dominate over the flux due to proton bremsstrahlung. Considering \dword{dm} masses in the $\sim$1-300 MeV range, this will require production via the $\pi^0$ and $\eta$ mesons. Our simulations using {\sc Pythia} determine that $c_{\pi^0} \approx 4.5$ and $c_\eta \approx 0.5$.

If the \dword{dm} reaches the near detector, it may scatter elastically off nucleons or electrons in the detector, via a $t$-channel dark photon. Due to its smaller backgrounds, we focus on scattering off electrons, depicted in the bottom panel of Fig.~\ref{fig:dm_prod}. The differential cross section of this scattering, as a function of the recoil energy of the electron $E_e$, is
\begin{eqnarray}
\frac{d\sigma_{{\chi}e}}{dE_{e}} &  
= & 4\pi \epsilon^{2}\alpha_D\alpha_{EM} \\ \nonumber
& & \times \frac{2m_{e}E_{\chi}^{2} - (2m_{e}E_{\chi} + M_{\chi}^{2})(E_e-m_{e})}{(E_e^{2}-M_{\chi}^{2})(M_{V}^{2}+2m_{e}E_{e}-2m_{e}^{2})^{2}}, \\ \nonumber
\end{eqnarray}
where $E_{\chi}$ is the incoming \dword{dm} $\chi$ energy. The signal is an event with only one recoil electron in the final state. We can exploit the difference between the scattering angle and the energy of the electron to distinguish between signal and the background from neutrino-electron scattering (discussed in the following) events.

The background to the process shown in the bottom panel of Fig.~\ref{fig:dm_prod} consists of any processes involving an electron recoil. As the \dword{nd} is located near the surface, background events, in general, can be induced by cosmic rays as well as by neutrinos generated from the beam. Since the majority of cosmic-induced events, however, will be vetoed by triggers and timing information, the dominant background will be from neutrinos coming in the DUNE beam.

The two neutrino-related backgrounds are $\nu_\mu -e^-$ scattering, which looks nearly identical to the signal, and $\nu_e$ CCQE scattering, which does not. The latter has a much larger rate ($\sim$ 10 times higher) than the former, however, we expect that using the kinematical variable $E_e \theta_e^2$ of the final state, where $\theta_e$ is the direction of the outgoing electron relative to the beam direction, will enable us to exploit the differences in the scattering angle of the electron from the DM interactions to reduce a substantial fraction of the $\nu_e$ CCQE background~\cite{Marshall:2019vdy}.

While spectral information regarding $E_e$ could allow a search to distinguish between $\chi e$ and $\nu_\mu e$ scattering, we expect that uncertainties in the $\nu_\mu$ flux (both in terms of overall normalization and shape as a function of neutrino energy) will make such an analysis very complicated. For this reason, we include a normalization uncertainty of $10\%$ on the expected background rate and perform a counting analysis. Studies are ongoing to determine how such an analysis may be improved.

For this analysis we have assumed $3.5$ years of data collection each in neutrino and antineutrino modes, analyzing events that occur within the fiducial volume of the DUNE near detector. We compare results assuming either all data is collected with the ND on-axis, or data collection is divided equally among all off-axis positions, $0.7$ year at each position  $i$, between $0$ and $24$ m transverse to the beam direction (in steps of 6 meters).
We assume three sources of uncertainty: statistical, correlated systematic, and an uncorrelated systematic in each bin. 
For a correlated systematic uncertainty, we include a nuisance parameter $A$ that modifies the number of neutrino-related background events in all bins -- an overall normalization uncertainty across all off-axis locations. 

We further include an additional term in our test statistic for $A$, a  Gaussian probability with width $\sigma_A = 10\%$. 
We also include an uncorrelated uncertainty in each bin, which we assume to be much narrower than $\sigma_A$. 
We assume this uncertainty to be parameterized by a Gaussian with width $\sigma_{f_i} = 1\%$. 
After marginalizing over the corresponding uncorrelated nuisance parameters, the test statistic reads

\begin{align}
\label{eq:chisqfull}
-2\Delta \mathcal{L} = & \sum_i \frac{r_i^m\left( \left(\frac{\varepsilon}{\varepsilon_0}\right)^4 N_i^\chi + (A-1)N_i^\nu\right)^2}{A\left(N_i^\nu + (\sigma_{f_i} N_i^\nu)^2 \right)} \\ \nonumber
& + \frac{\left(A-1\right)^2}{\sigma_A^2}. \\ \nonumber
\end{align}

In Eq.~(\ref{eq:chisqfull}), $N_i^\chi$ is the number of \dword{dm} scattering events, calculated assuming $\varepsilon$ is equal to some reference value $\varepsilon_0 \ll 1$. $N_i^\nu$ is the number of $\nu_\mu e^-$ scattering events expected in detector position $i$, and $r_i^m$ is the number of years of data collection in detector position $i$ during beam mode $m$ (neutrino or antineutrino mode). If data are only collected on-axis, then this test statistic will be dominated by the systematic uncertainty associated with $\sigma_A$. If on- and off-axis measurements are combined, then the resulting sensitivity will improve significantly.

We present results in terms of the \dword{dm} or dark photon mass and the parameter $Y$, where

\begin{equation}
Y \equiv \varepsilon^2 \alpha_D \left(\frac{M_\chi}{M_V}\right)^4.    
\end{equation}

\begin{figure*}[t]
 \centering
 \includegraphics[width=0.9\textwidth]{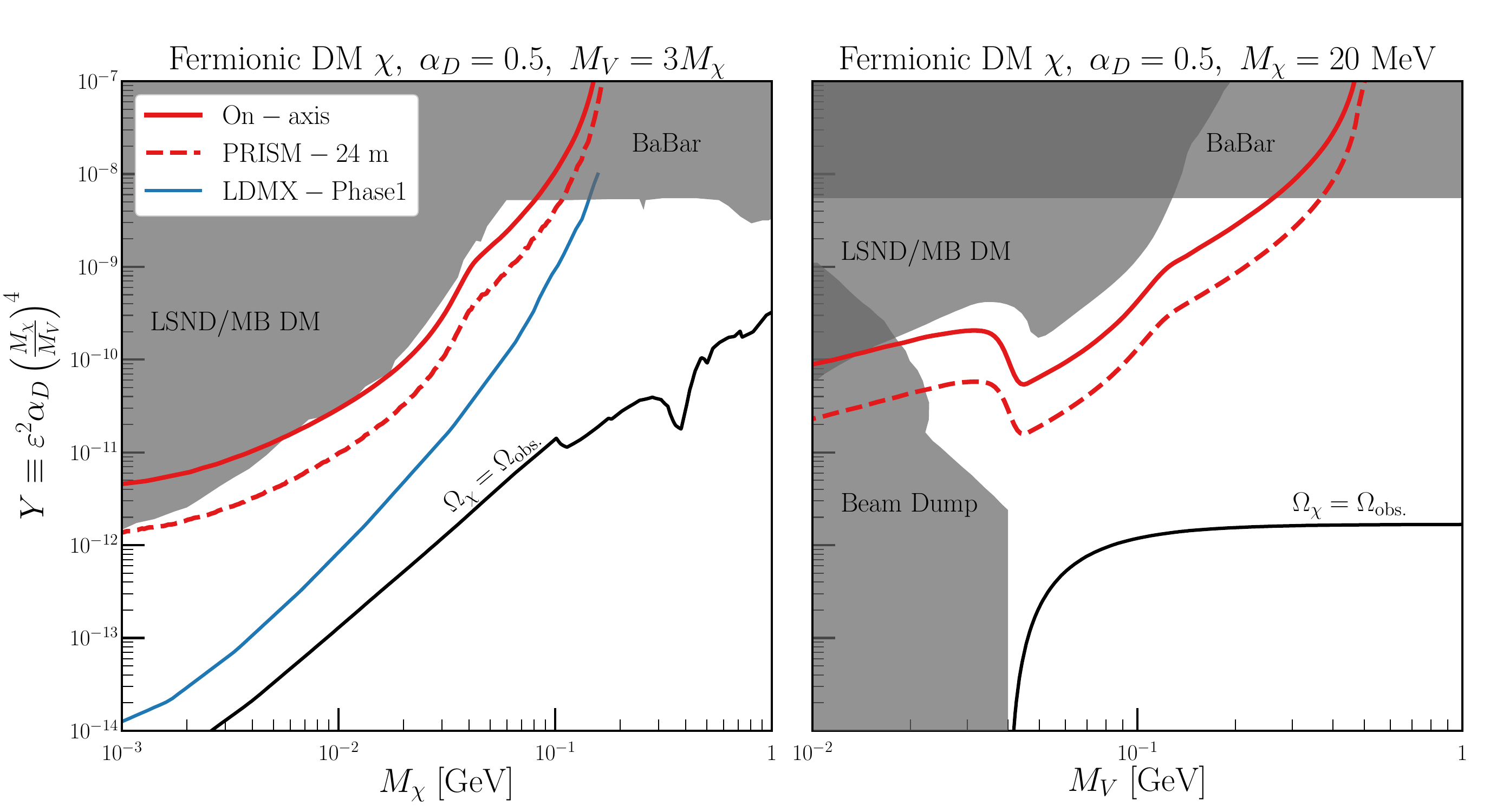}
 \caption[90$\%$ \dword{cl} limit for Y as a function of $m_{\chi}$ at the ND]{\label{fig:chisq} Expected DUNE On-axis (solid red) and PRISM (dashed red) sensitivity using $\chi e^- \to \chi e^-$ scattering. We assume $\alpha_D = 0.5$ in both panels, and $M_V = 3M_\chi$ ($M_\chi = 20$ MeV) in the left (right) panel, respectively. Existing constraints are shown in grey, and the relic density target is shown as a black line. We also show for comparison the sensitivity curve expected for LDMX-Phase I (solid blue)~\cite{Akesson:2018vlm}.
 }
 \end{figure*}

Assuming $M_V \gg M_\chi$, this parameter determines the relic abundance of \dword{dm} in the universe today, and sets a theoretical goal in terms of sensitivity reach. We present the 90\% CL sensitivity reach of the DUNE \dword{nd} in Fig.~\ref{fig:chisq}. 
We assume $\alpha_D = 0.5$ in our simulations and we display the results fixing $M_V = 3M_\chi$ (left panel) and $M_\chi = 20$ MeV (right panel).
We also compare the sensitivity reach of this analysis with other existing experiments, shown as grey shaded regions. We further show for comparison the sensitivity curve expected for a proposed dedicated experiment to search for \dword{ldm}, LDMX-Phase I~\cite{Akesson:2018vlm} (solid blue).

 From our estimates, we see that DUNE can significantly improve the constraints from LSND~\cite{deNiverville:2018dbu} and the MiniBooNE-DM search~\cite{Aguilar-Arevalo:2018wea}, as well as BaBar~\cite{Lees:2017lec} if $M_V \lesssim 200$ MeV. We also show limits in the right panel from beam-dump experiments (where the dark photon is assumed to decay visibly if $M_V < 2 M_\chi$)~\cite{Davier:1989wz,Batley:2015lha,Bjorken:1988as,Riordan:1987aw,Bjorken:2009mm,Bross:1989mp}, as well as the lower limits obtained from matching the thermal relic abundance of $\chi$ with the observed one (black).
 
 The features in the sensitivity curve in the right panel can be understood by looking at the DM production mechanism.
For a fixed $\chi$ mass, as $M_V$ grows, the DM production goes from off-shell to on-shell and back to off-shell. The first transition explains the strong feature near $M_V=2M_\chi = 40$~MeV, while the second is the source for the slight kink around $M_V=m_{\pi^0}$ (which appears also in the left panel).

\subsection{Inelastic Boosted Dark Matter Search at the DUNE FD}
\label{bdmatfd}

We consider an annihilating two-component \dword{dm} scenario~\cite{Belanger:2011ww} in this study. 
The heavier \dword{dm} (denoted $\Psi$) plays a role of cosmological \dword{dm} and pair-annihilates to a pair of lighter \dword{dm} particles (denoted $\chi$) in the universe today. 
The expected flux near the earth is given by~\cite{Agashe:2014yua,Giudice:2017zke,Kim:2018veo}

\begin{align}
\mathcal{F}_1= & 1.6 \times 10^{-6} {\rm cm}^{-2}{\rm s}^{-1}\times \left( \frac{\langle \sigma v\rangle_{\Psi\rightarrow \chi}}{5\times 10^{-26}{\rm cm}^3{\rm s}^{-1}}\right) \nonumber \\
 \times & \left( \frac{10\, {\rm GeV}}{M_{\Psi}}\right)^2, 
\label{eq:flux}
\end{align}
where $m_{\Psi}$ is the mass of $\Psi$ and $\langle \sigma v\rangle_{\Psi\rightarrow \chi}$ stands for the velocity-averaged annihilation cross section of $\Psi\bar{\Psi} \to \chi\bar{\chi}$ in the current universe.
To evaluate the reference value shown as the first prefactor, we take $M_{\Psi} = 10$ GeV and $\langle \sigma v\rangle_{\Psi\rightarrow \chi}=5\times 10^{-26}~{\rm cm}^3{\rm s}^{-1}$, the latter of which is consistent with the current observation of \dword{dm} relic density assuming $\Psi$ and its anti-particle $\bar{\Psi}$ are distinguishable. 
To integrate all relevant contributions over the entire galaxy, we assume the Navarro-Frenk-White (NFW) \dword{dm} halo profile~\cite{Navarro:1995iw,Navarro:1996gj}.
In this section we assume the \dword{bdm} flux with a $M_{\Psi}$ dependence given by Eq.~(\ref{eq:flux}) for the phenomenological analysis. 

\begin{figure}[htp]
\centering
\includegraphics[width=0.9\columnwidth]{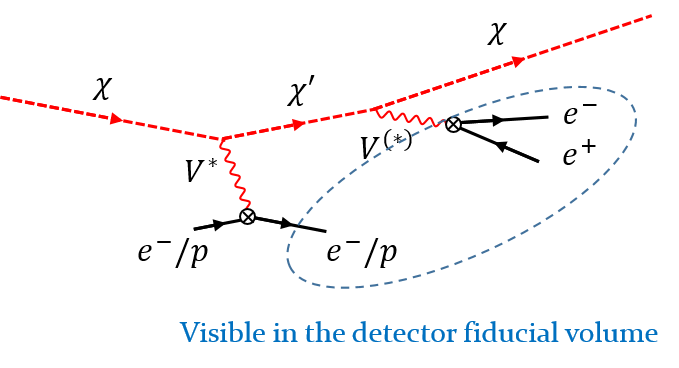}
\caption{The inelastic BDM signal under consideration.}
\label{fig:sig}
\end{figure}

The \dword{bdm} that is created, e.g., at the galactic center, reaches the DUNE \dword{fd} 
detectors and scatters off either electrons or protons energetically. 
In this study, we focus on electron scattering signatures for illustration, under Benchmark Model i) defined in Eq.~\eqref{eq:lagrangian}. 
The overall process is summarized as follows:

\begin{align}
\chi + e^-~({\rm or}~p) & \to \\ \nonumber
e^-~({\rm or}~p) + \chi' & (\to \chi + V^{(*)} \to \chi + e^+ +e^-), \\ \nonumber
\end{align}
where $\chi'$ is a dark-sector unstable particle that is heavier than $\chi$ as described earlier.
A diagrammatic description is shown in Fig.~\ref{fig:sig} where 
particles visible by the detector are circled in blue. 
In the final state of the $e$-scattering case, there exist three visible particles that usually leave sizable ($e$-like) tracks in the 
detectors.  
On the other hand, for the $p$-scattering case we can replace $e^-$ in the left-hand side and the first $e^-$ in the right-hand side of the above process by $p$.
In the basic model, Eq.~\eqref{eq:lagrangian}, and given the source of \dword{bdm} at the galactic center,  the resulting signature accompanies a quasi-elastic proton recoil~\cite{Kim:2020ipj} together with a pair of $e^+e^-$ tracks.

As we have identified a possible inelastic BDM  ($i$\dword{bdm}) signature, we are now in a position to discuss potential \dword{sm} background events. For the DUNE \dwords{detmodule} located $\sim 1480$ m deep underground, the cosmic-induced backgrounds are not an issue except the background induced by atmospheric neutrinos. 
The most plausible scenario for background production is that an atmospheric neutrino event involves the creation of multiple pions that subsequently decay to electrons, positrons, photons, and neutrinos. 
Relevant channels are the resonance production and/or \dword{dis} by the \dword{cc} $\nu_e$ or $\bar \nu_e$ scattering with a nucleon in the \lar target.
Summing up all the resonance production and \dword{dis} events that are not only induced by $\nu_e$ or $\bar \nu_e$ 
but relevant to production of a few pions, we find that the total number of multi-pion production events is at most $\sim 20$ (\si{\ktyr})$^{-1}$~\cite{DeRoeck:2020ntj}, based on the neutrino flux calculated in Ref.~\cite{Honda:2015fha} and the cross section in Ref.~\cite{Formaggio:2013kya}.
In addition, the charged pions often leave long enough tracks inside the detector so that the probability of misidentifying the $e^\pm$ from the decays of $\pi^\pm$ with the \textit{i}\dword{bdm} signal events would be very small.
Some quasi-elastic scattering events by atmospheric neutrinos may involve a detectable proton recoil together with a single $e$-like track, which might behave like backgrounds in the proton scattering channel. 
However, this class of events can be rejected by requiring two separated $e$-like tracks.
Hence, we conclude that it is fairly reasonable to assume that almost no background events exist.
See also Ref.~\cite{DeRoeck:2020ntj} for a more systematic background consideration for the $i$BDM signals.

We finally present the expected experimental sensitivities of 
DUNE, in the searches for $i$\dword{bdm}. 
We closely follow the strategies illustrated in Refs.~\cite{Giudice:2017zke,Kim:2020ipj} to represent phenomenological interpretations. 
In displaying the results, we separate the signal categories into 
%--
\begin{itemize}
\item Scenario 1: $M_V > 2 M_{\chi}$, experimental limits for $V \to$ invisible applied.
\item Scenario 2: $M_V \le 2 M_{\chi}$, experimental limits for $V \to e^+ e^-$ applied.
\end{itemize}

We develop an event simulation code using the \texttt{ROOT} package with the matrix elements for the $\chi$ scattering and the $\chi'$ decays implemented. 
Once an event is generated, we require that all the final state particles should pass the (kinetic) energy threshold (30~MeV for electrons and protons) and their angular separation from the other particles should be greater than the angular resolution ($1^\circ$ for electrons and $5^\circ$ for protons)~\cite{DeRoeck:2020ntj}. 

\begin{figure*}[htp]
\centering
\includegraphics[width=0.45\textwidth]{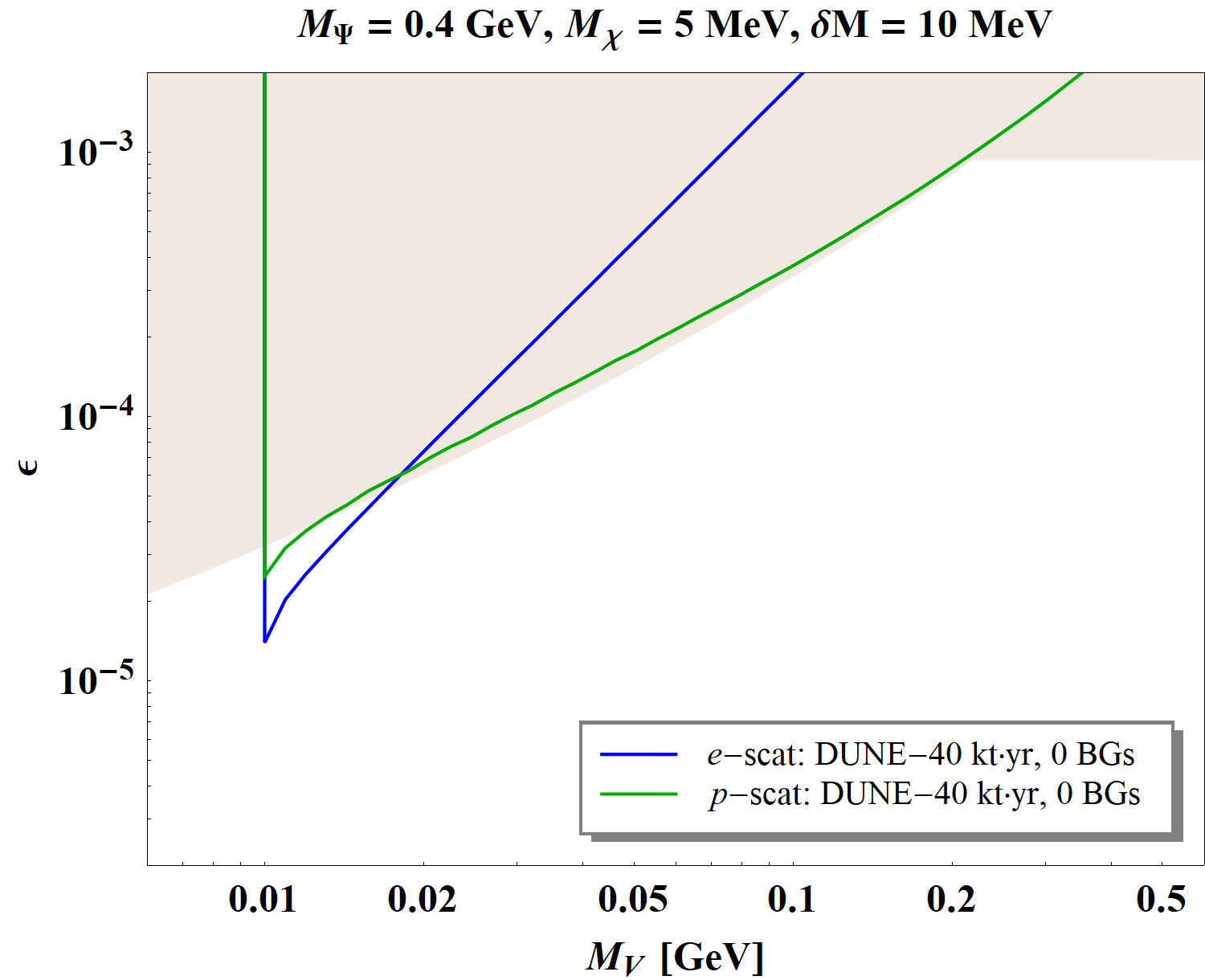} 
\includegraphics[width=0.45\textwidth]{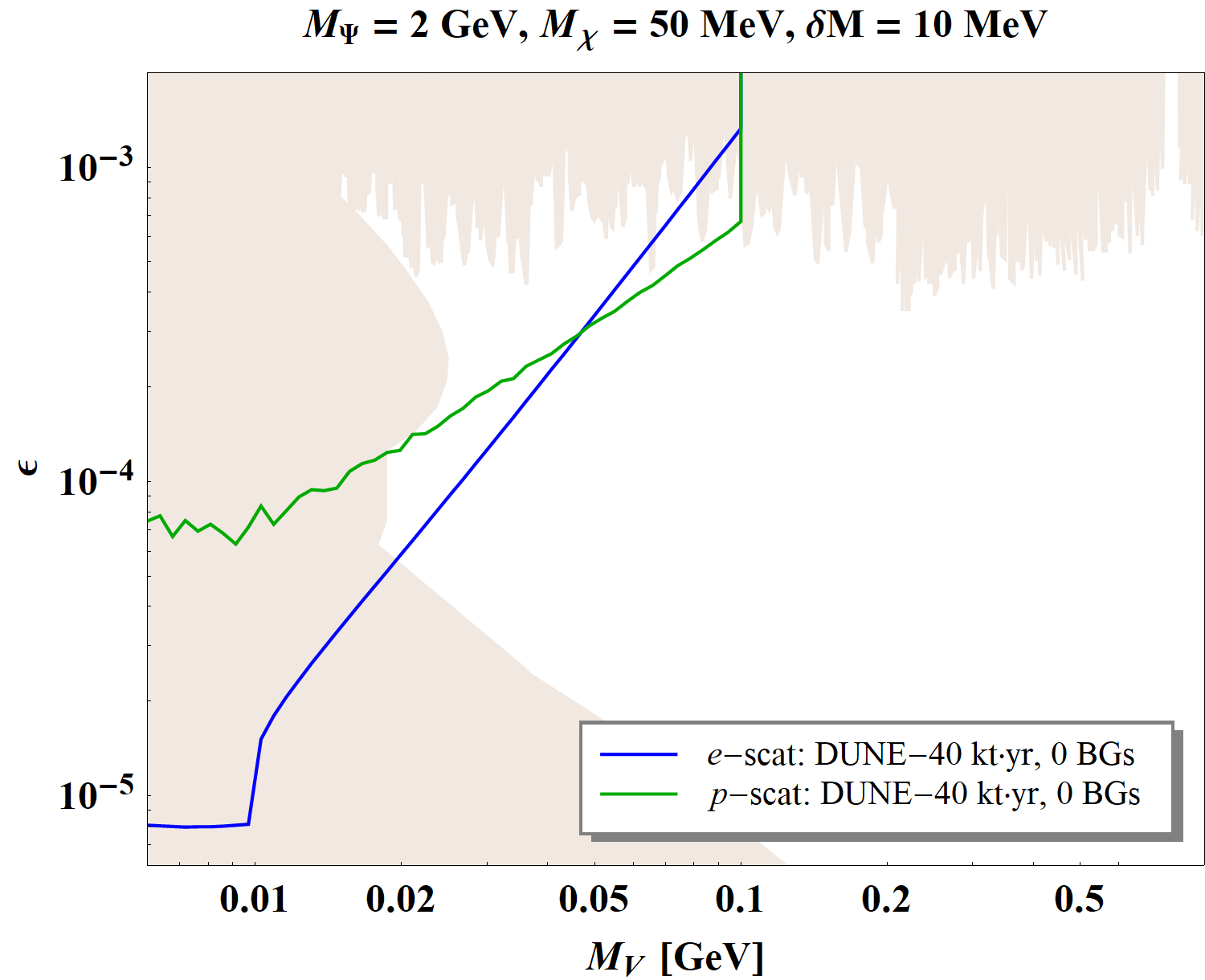} \\
\vspace{0.3cm}
\includegraphics[width=0.45\textwidth]{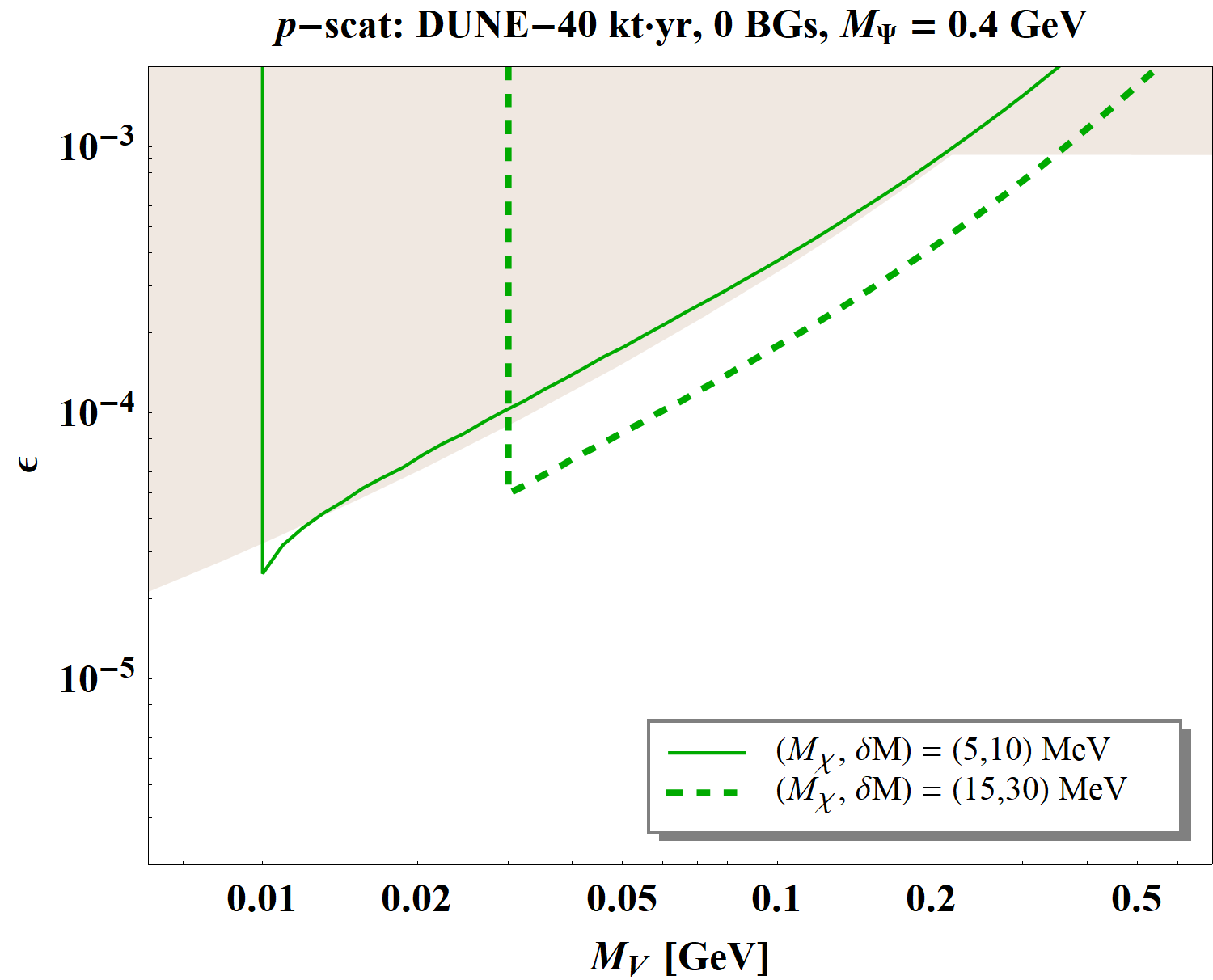}
\includegraphics[width=0.45\textwidth]{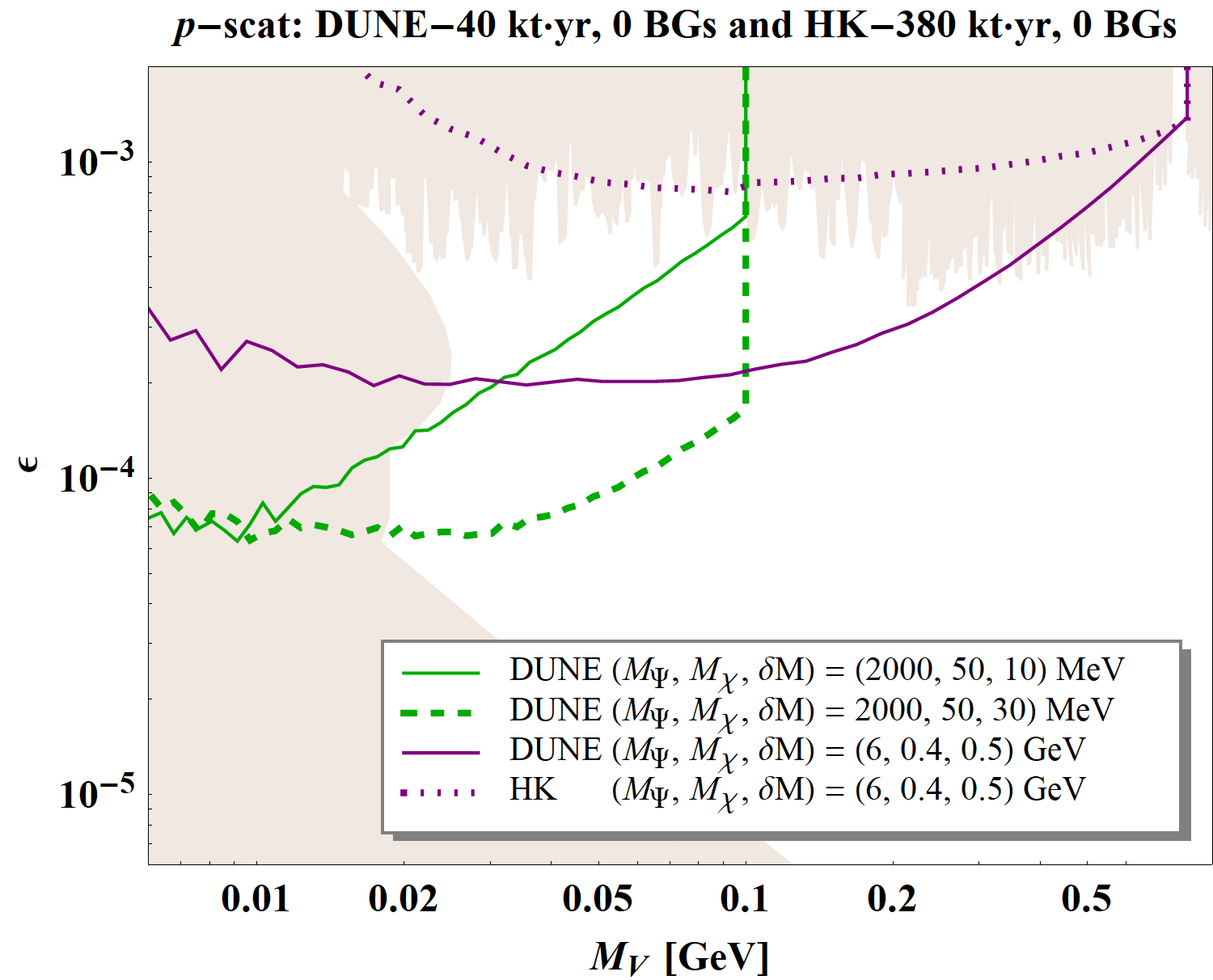}
\caption[Experimental sensitivities for $m_{\chi_n}$ values in terms of $m_V - \epsilon$]{
The experimental sensitivities in terms of reference model parameters $M_V - \epsilon$ 
for $M_{\Psi} = 0.4$ GeV, $M_{\chi} = 5$ MeV, and $\delta M = M_{\chi'} - M_{\chi} = 10$ MeV (top-left panel) and $M_{\Psi} = 2$ GeV, $M_{\chi'} = 50$ MeV, and $\delta M = 10$ MeV (top-right panel).
The left panels are for Scenario 1 and the right ones are for Scenario 2.
The bottom panels compare different reference points in the $p$-scattering channel.
See the text for the details.
\label{fig:darkphotonparameter} }
\end{figure*}

We first show the results for Scenario 1 in the left panels of Fig.~\ref{fig:darkphotonparameter}, taking a parameter set, $M_{\Psi} = 0.4$~GeV, $M_{\chi} = 5$~MeV, $\delta M \equiv M_{\chi'}-M_\chi = 10$~MeV with $g'_D=1$.
The brown-shaded region shows the latest limits set by various experiments such as the fixed-target experiment NA64~\cite{NA64:2019imj} at the CERN SPS and the B-factory experiment BaBar~\cite{Banerjee:2017hhz}.
Note that some of the limits are from ongoing experiments such as NA64 which will collect more data in the next years and improve their sensitivity reaches. 
The blue solid and the green solid lines describe the experimental sensitivity\footnote{This is defined as the boundary of parameter space that can be probed by the dedicated search in a given experiment at 90\% \dword{cl}, practically obtained from Eq.~(\ref{eq:MIsensitivity}).} of DUNE \dword{fd} to the $e$-scattering and $p$-scattering signals, respectively, under a zero background assumption.
The associated exposure is 40~\ktyr, i.e., a total fiducial volume of \fdfiducialmass times one year of running time.

For Scenario 2 (the right panels of Fig.~\ref{fig:darkphotonparameter}), we choose a different reference parameter set: $M_{\Psi} = 2$~GeV, $M_{\chi} = 50$~MeV, $\delta M = 10$~MeV with $g'_D=1$. 
The current limits (brown shaded regions), from various fixed target experiments, B-factory experiments, and astrophysical observations, are taken from Refs.~\cite{Beacham:2019nyx,Banerjee:2019hmi}.

In both scenarios, the proton scattering channel enables us to explore different regions of parameter space as it allows heavier $\chi'$ to be accessible which would be kinematically forbidden to access in the electron scattering channel.
Inspired by this potential of the proton scattering channel, we study other reference parameters and compare them with the original ones in the top panels of Fig.~\ref{fig:darkphotonparameter}, and show the results in the bottom panels. 
We see that different parameter choices in the proton scattering channel allow us to cover a wider or different range of parameter space.

We next discuss model-independent experimental sensitivities. 
The experimental sensitivities are determined by the number of signal events excluded at 90\% \dword{cl} in the absence of an observed signal.
The expected number of signal events, $N_{\rm sig}$, is given by

\begin{align}
N_{\rm sig} = \sigma_\epsilon \mathcal F A(\ell_{\rm lab}) t_{\rm exp} N_T\,,
\label{eq:NS}
\end{align}
where $N_T$ is the number of target particles $T$, $\sigma_\epsilon$ is the cross section of the primary scattering $\chi T \to \chi' T$, $\mathcal F$ is the flux of $\chi$, $t_{\rm exp}$ is the exposure time, and $A(\ell_{\rm lab})$ is the acceptance that is defined as 1 if the event occurs within the fiducial volume and 0 otherwise.
Here we determine the acceptance for an $i$\dword{bdm} signal by the distance between the primary and secondary vertices in the laboratory frame, $\ell_{\rm lab}$, so $A(\ell_{\rm lab}) = 1$ when both the primary and secondary events occur inside the fiducial volume. (Given this definition, obviously, $A(\ell_{\rm lab}) = 1$ for elastic \dword{bdm}.)
Our notation $\sigma_\epsilon$ includes additional realistic effects from cuts, threshold energy, and the detector response, hence it can be understood as the fiducial cross section.

\begin{figure}[htp]
\centering
\includegraphics[width=0.9\columnwidth]{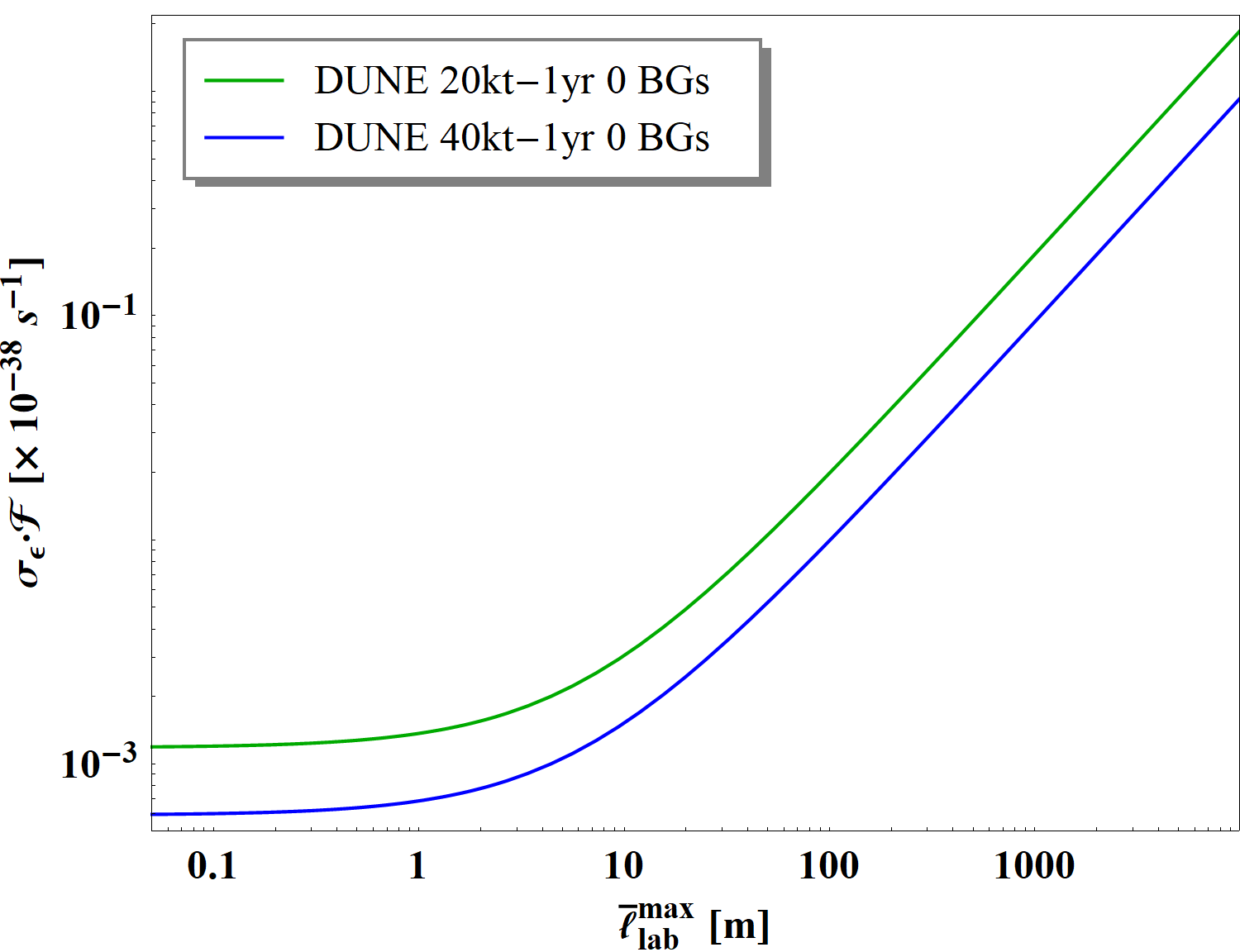}
\includegraphics[width=0.9\columnwidth]{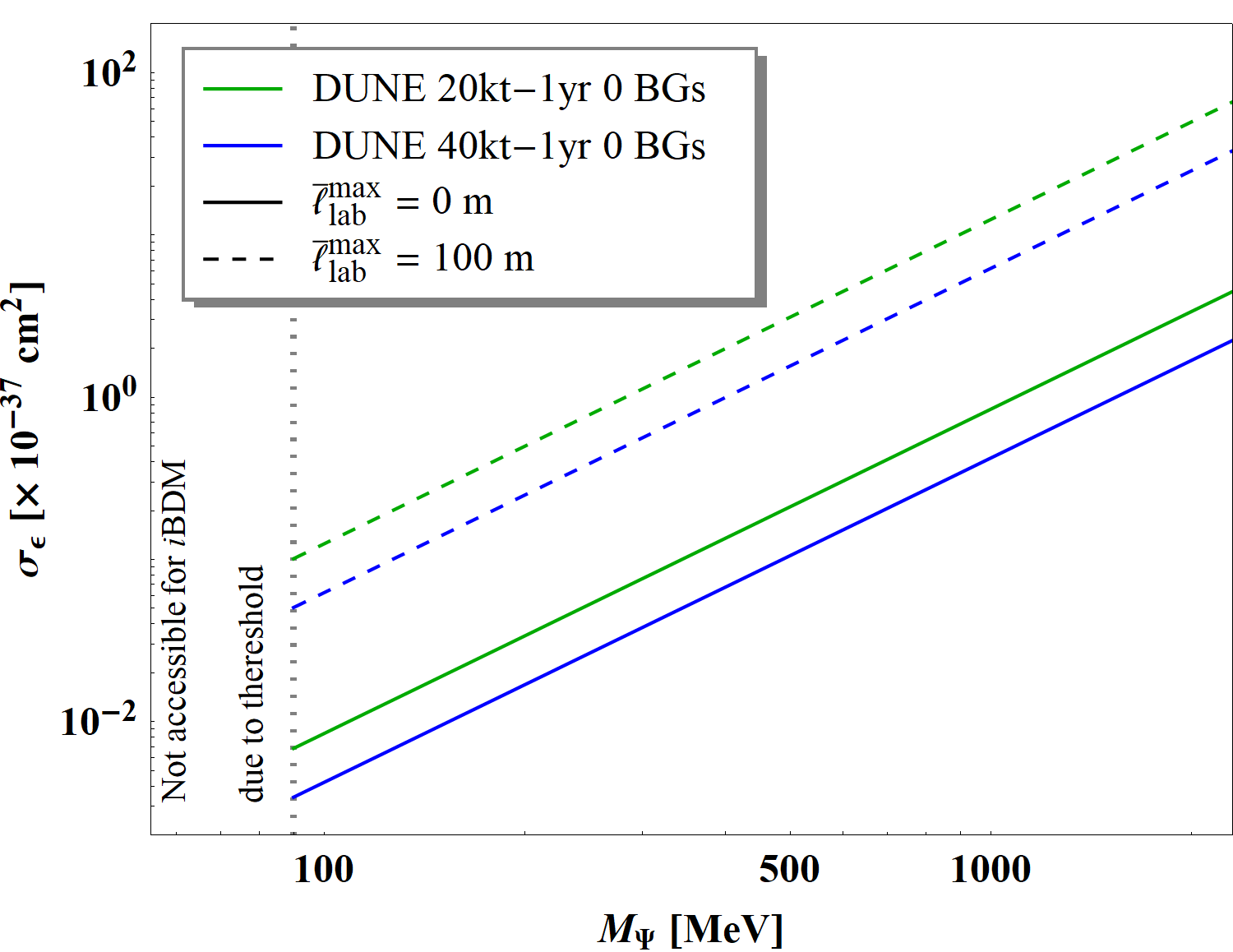}
\caption[Model-independent experimental sensitivities of $i$BDM search]{
Top: model-independent experimental sensitivities of $i$\dword{bdm} search in $\bar{\ell}_{\rm lab}^{\rm max} - \sigma_\epsilon \cdot \mathcal F$ plane. 
The reference experiments are
DUNE \SI{20}{\kt} (green), and DUNE \SI{40}{\kt} (blue) with zero-background assumption for 1-year time exposure. 
Bottom: Experimental sensitivities of $i$\dword{bdm} search in $M_{\Psi} - \sigma_\epsilon$ plane. The sensitivities for $\bar{\ell}_{\rm lab}^{\rm max} = 0$ m and 100 m are shown as solid and dashed lines for each reference experiment in the top panel.
\label{fig:modelindependent} }
\end{figure}

The 90\% \dword{cl} exclusion limit, $N_s^{90}$, can be obtained with a modified frequentist construction~\cite{cls1,cls2}. We follow the methods in Refs.~\cite{Dermisek:2013cxa,Dermisek:2014qca,Dermisek:2016via} in which the Poisson likelihood is assumed. 
An experiment becomes sensitive to the signal model independently if $N_{\rm sig} \ge N_s^{90}$.
Plugging Eq.~\eqref{eq:NS} here, we find the experimental sensitivity expressed by 

\begin{align}
\sigma_\epsilon \mathcal F \ge \frac{N_s^{90}}{A(\ell_{\rm lab}) t_{\rm exp} N_T}\,. 
\label{eq:MIsensitivity}
\end{align}
Since $\ell_{\rm lab}$ differs event-by-event, we take the maximally possible value of laboratory-frame mean decay length, i.e., $\bar{\ell}_{\rm lab}^{\rm max} \equiv \gamma_{\chi'}^{\max} \bar{\ell}_{\rm rest}$ where $\gamma_{\chi'}^{\max}$ is the maximum boost factor of $\chi'$ and $\bar{\ell}_{\rm rest}$ is the rest-frame mean decay length. 
We emphasize that this is a rather conservative approach, because the acceptance $A$ is inversely proportional to $\ell_{\rm lab}$.
We then show the experimental sensitivity of any kind of experiment for a given background expectation, exposure time, and number of targets, in the plane of $\bar{\ell}_{\rm lab}^{\rm max} - \sigma_\epsilon \cdot \mathcal F$. 
The top panel of Fig.~\ref{fig:modelindependent} demonstrates the expected model-independent sensitivities at the DUNE experiment.
The green (blue) line is for the DUNE \dword{fd} with a background-free assumption and 20 (40)~\ktyr exposure.

The bottom panel of Fig.~\ref{fig:modelindependent} reports model-dependent sensitivities for $\bar{\ell}_{\rm lab}^{\rm max} = 0$ m and 100 m corresponding to the experiments in the top panel.
Note that this  
method of presentation is reminiscent of the widely known scheme for showing the experimental reaches in various \dword{dm} direct detection experiments, i.e., $M_{\rm DM} - \sigma_{\rm DM - target}$ where $M_{\rm DM}$ is the mass of \dword{dm} and $\sigma_{\rm DM - target}$ is the cross section between the \dword{dm} and target. 
For the case of non-relativistic \dword{dm} scattering in the direct-detection experiments, $M_{\rm DM}$ determines the kinetic energy scale of the incoming \dword{dm}, just like $M_{\Psi}$ sets out the incoming energy of boosted $\chi$ in the $i$\dword{bdm} search. 

\subsection{Elastic Boosted Dark Matter from the Sun \label{sec:FDsun}}

In this section, we focus on Benchmark Model ii) described by Eq.~\eqref{eq:zprimelag}. This study uses DUNE's full \dword{fd} event generation and detector simulation. 
We focus on \dword{bdm} flux sourced by \dword{dm} annihilation in the core of the sun. \dword{dm} particles can be captured through their scattering with the nuclei within the sun, mostly hydrogen and helium. This makes the core of the sun a region with concentrated \dword{dm} distribution. The \dword{bdm} flux is
\begin{eqnarray} \label{eq:fluxbdm}
\Phi= f \frac{A}{4\pi D^2},
\end{eqnarray}
where $A$ is the annihilation rate, and $D = 1\,\rm{\dword{au}}$ is the distance from the sun. $f$ is a model-dependent parameter, where $f = 2$ for two-component \dword{dm} as considered here.

\begin{figure*}[htp]
  \centering
  \includegraphics[width=0.9\textwidth]{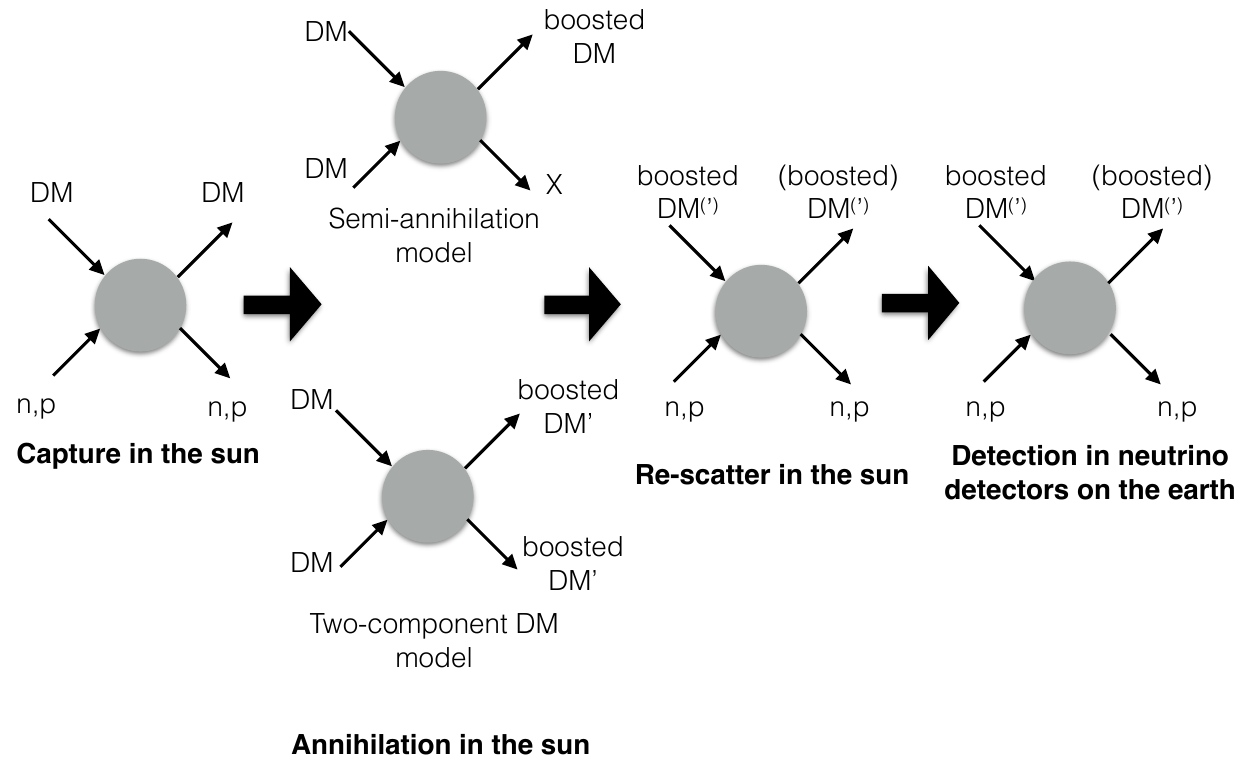}
  \caption[Processes leading to boosted DM signal from the sun]{The chain of processes leading to boosted DM signal from the sun. The semi-annihilation and two-component DM models refer to the two examples of the non-minimal dark-sector scenarios introduced in the beginning of Section~\ref{sec:DM}. DM$'$ denotes the lighter DM in the two-component DM model. $X$ is a lighter dark sector particle that may decay away.}
    \label{fig:processes}
\end{figure*}

For the parameter space of interest,  
assuming that the 
\dword{dm} annihilation cross section is not too small, the \dword{dm} distribution in the sun has reached an equilibrium between capture and annihilation. This helps to eliminate the annihilation cross section dependence in our study. The chain of processes involved in giving rise to the boosted DM signal from the sun is illustrated in Fig.~\ref{fig:processes}.

Two additional comments are in order. First, the \dword{dm} particles cannot be too light, i.e.,  lighter than 4\,GeV~\cite{Griest:1986yu,Gould:1987ju}, otherwise we will lose most of the captured \dword{dm} through evaporation rather than annihilation; this would dramatically reduce the \dword{bdm} flux. Additionally, one needs to check that \dword{bdm} particles cannot lose energy and potentially be recaptured by scattering with the solar material when they escape from the core region after production. Rescattering is found to be rare for the benchmark models considered in this study and we consider the \dword{bdm} flux to be monochromatic at its production energy.

The event rate to be observed at DUNE is 
\begin{equation}
R = \Phi \times \sigma_{\rm{SM} - \chi} \times \varepsilon \times N_T,
\end{equation}
 where $\Phi$ is the flux given by Eq.~\eqref{eq:fluxbdm}, $\sigma_{\rm{SM} - \chi}$ is the scattering cross section of the \dword{bdm} off of \dword{sm} particles, $\varepsilon$ is the efficiency of the detection of such a process, and $N_T$ is the number of target particles in DUNE. The computation of the flux of \dword{bdm} from the sun can be found in~\cite{Berger:2014sqa}. 
 
\begin{figure*}[t]
  \centering
  \includegraphics[width=0.256\textwidth]{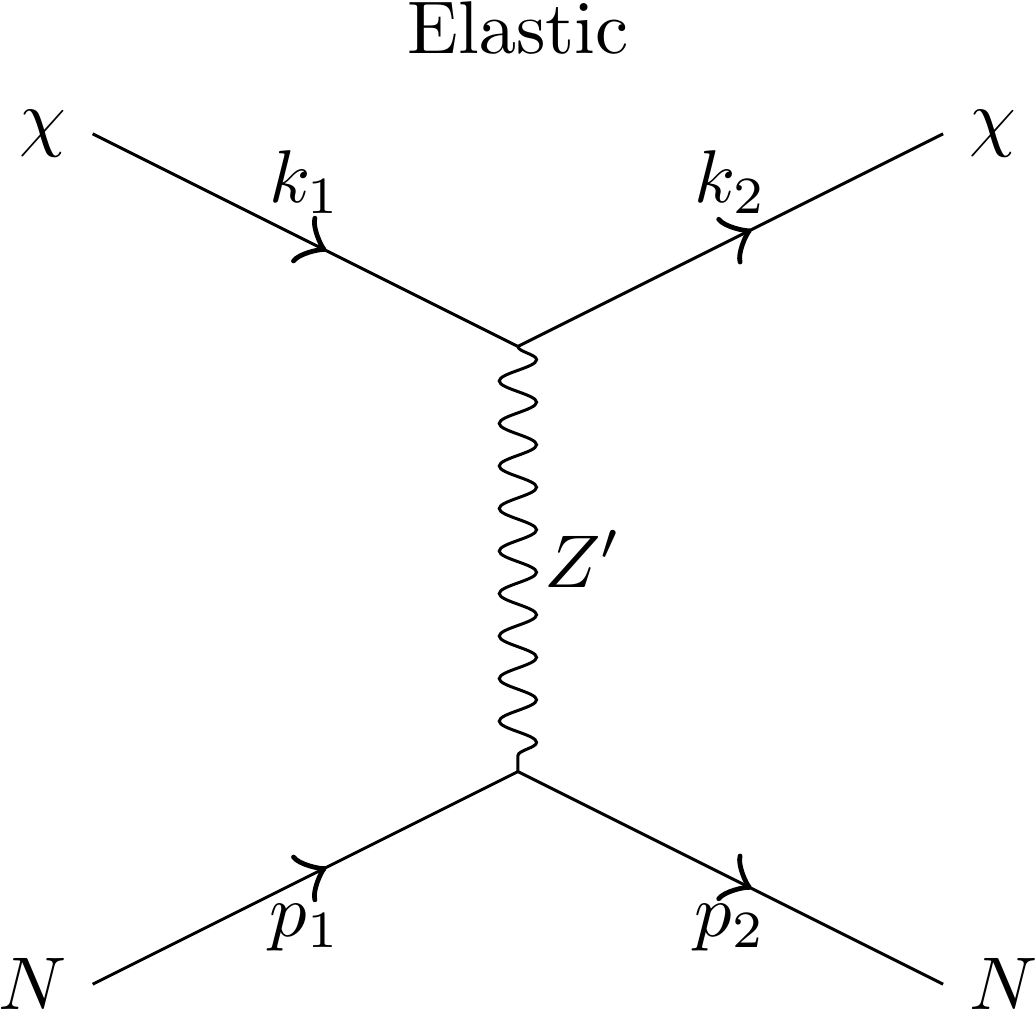}
  \includegraphics[width=0.33\textwidth]{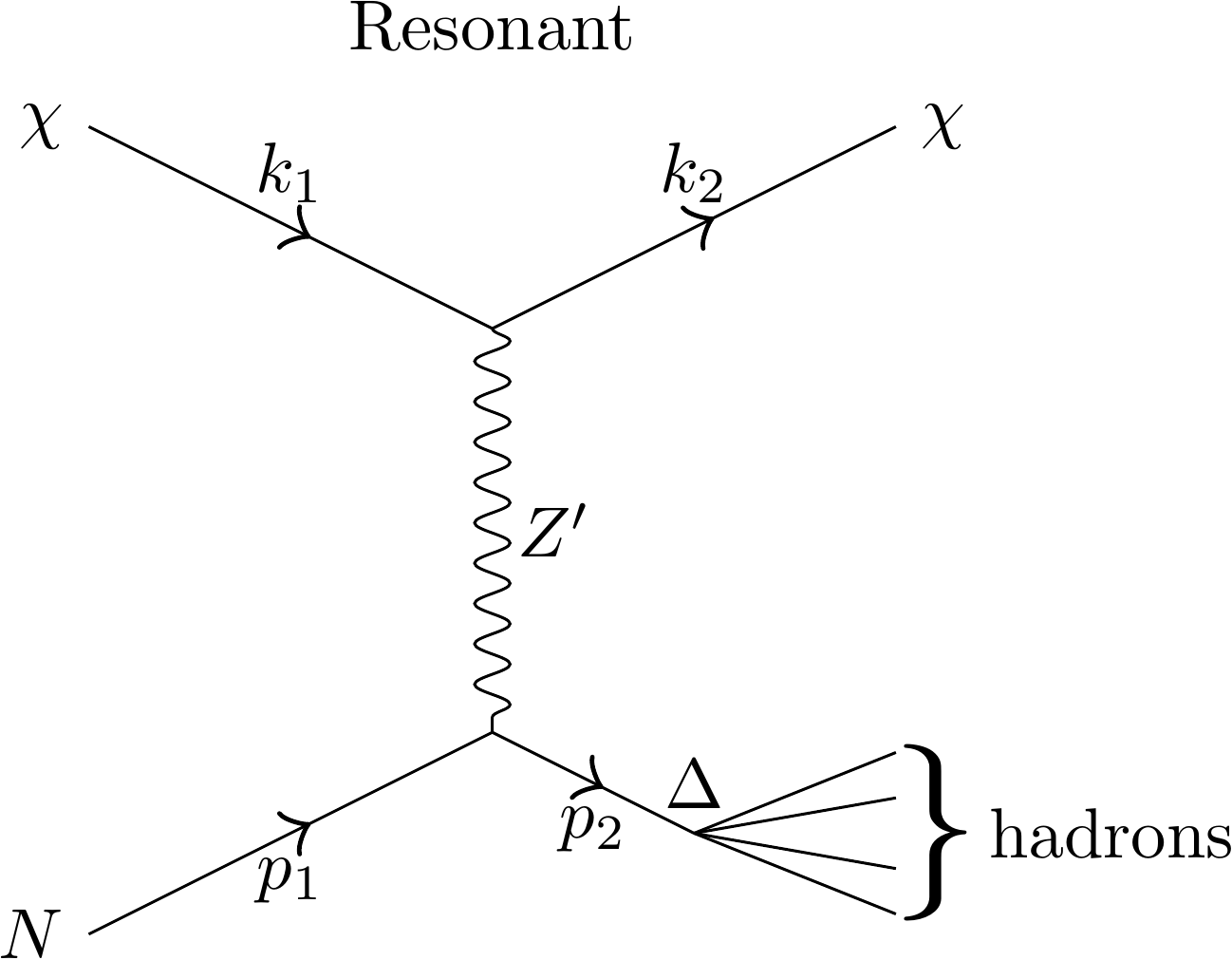}
  \includegraphics[width=0.29\textwidth]{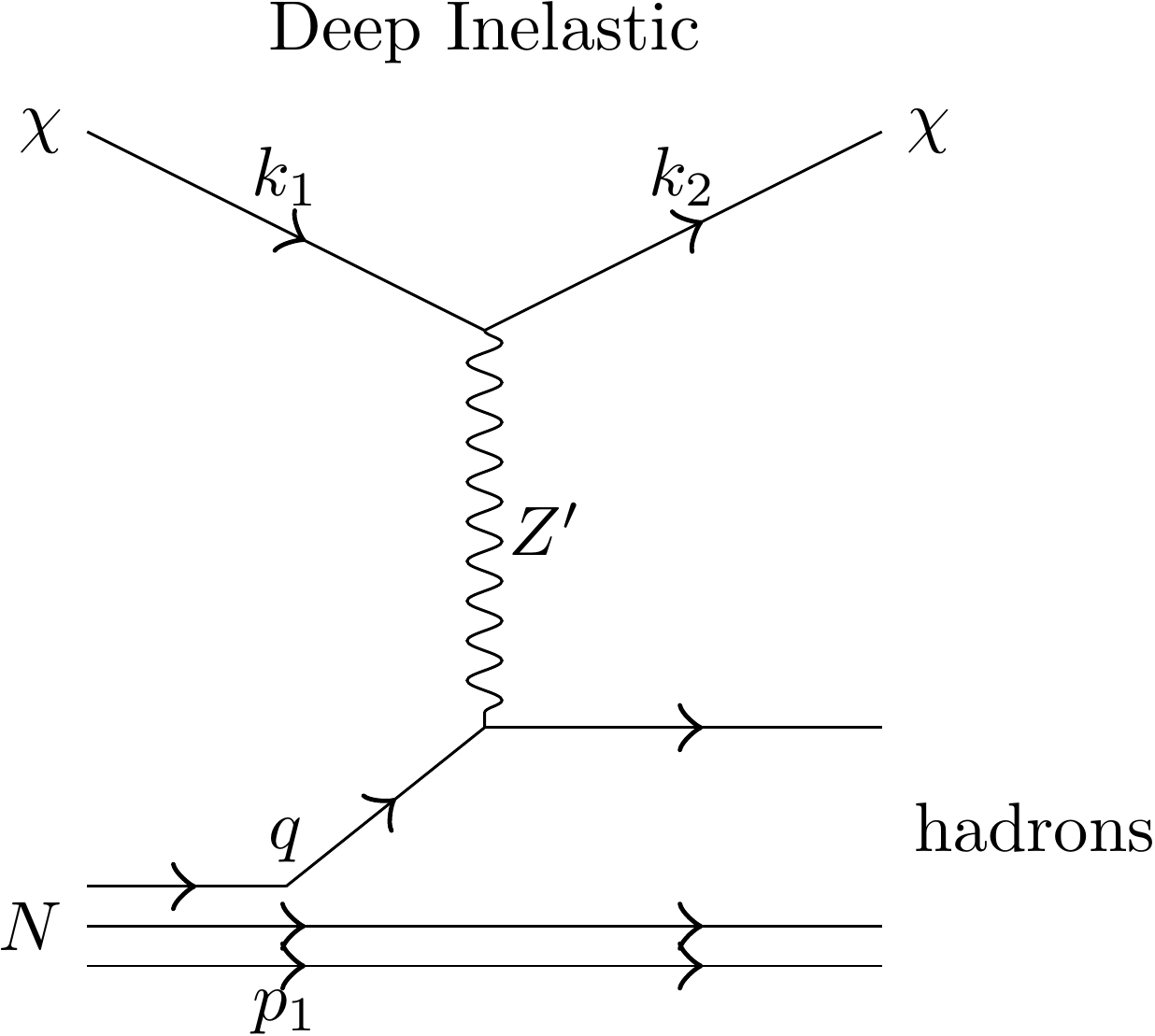}
  \caption{Diagram illustrating each of the three processes contributing to dark matter scattering in argon: elastic (left), baryon resonance (middle), and deep inelastic (right).}
  \label{fig:BDM-argon}
\end{figure*}

 The processes of typical BDM scattering in argon are illustrated in Fig.~\ref{fig:BDM-argon}.
We generate the signal events and calculate interaction cross sections in the detector using a newly developed \dword{bdm} module~\cite{Andreopoulos:2009rq,Andreopoulos:2015wxa,Berger:2018} that includes elastic and deep inelastic scattering, as well as a range of nuclear effects. This conservative event generation neglects the dominant contributions from baryon resonances in the final state hadronic invariant mass range of 1.2 to 1.8 GeV, which should not have a major effect on our main results. The interactions are taken to be mediated by an axial, flavor-universal $Z^\prime$ coupling to both the \dword{bdm} and with the quarks. The axial charge is taken to be 1. 
The events are generated for the \nominalmodsize DUNE detector module~\cite{dunetpc_code}, though we only study the dominant scattering off of the \argon40 atoms therein. The method for determining the efficiency $\varepsilon$ is described below. The number of target argon atoms is $N = 1.5  \times 10^{32}$ assuming a target mass of \nominalmodsize{}.

The main background in this process comes from the \dword{nc} 
interactions of atmospheric neutrinos and argon,
as they share the features that the timing of events is unknown in advance
(unlike events of neutrinos produced by the accelerator),
and that the interactions with argon produce hadronic activity in the detector.
We use \dword{genie}
to generate the \dword{nc} atmospheric
neutrino events.  This simulation predicts 845 events in a \nominalmodsize{} module for one year of
exposure.

\begin{figure}[htb]
\centering
\includegraphics[width=0.9\columnwidth]{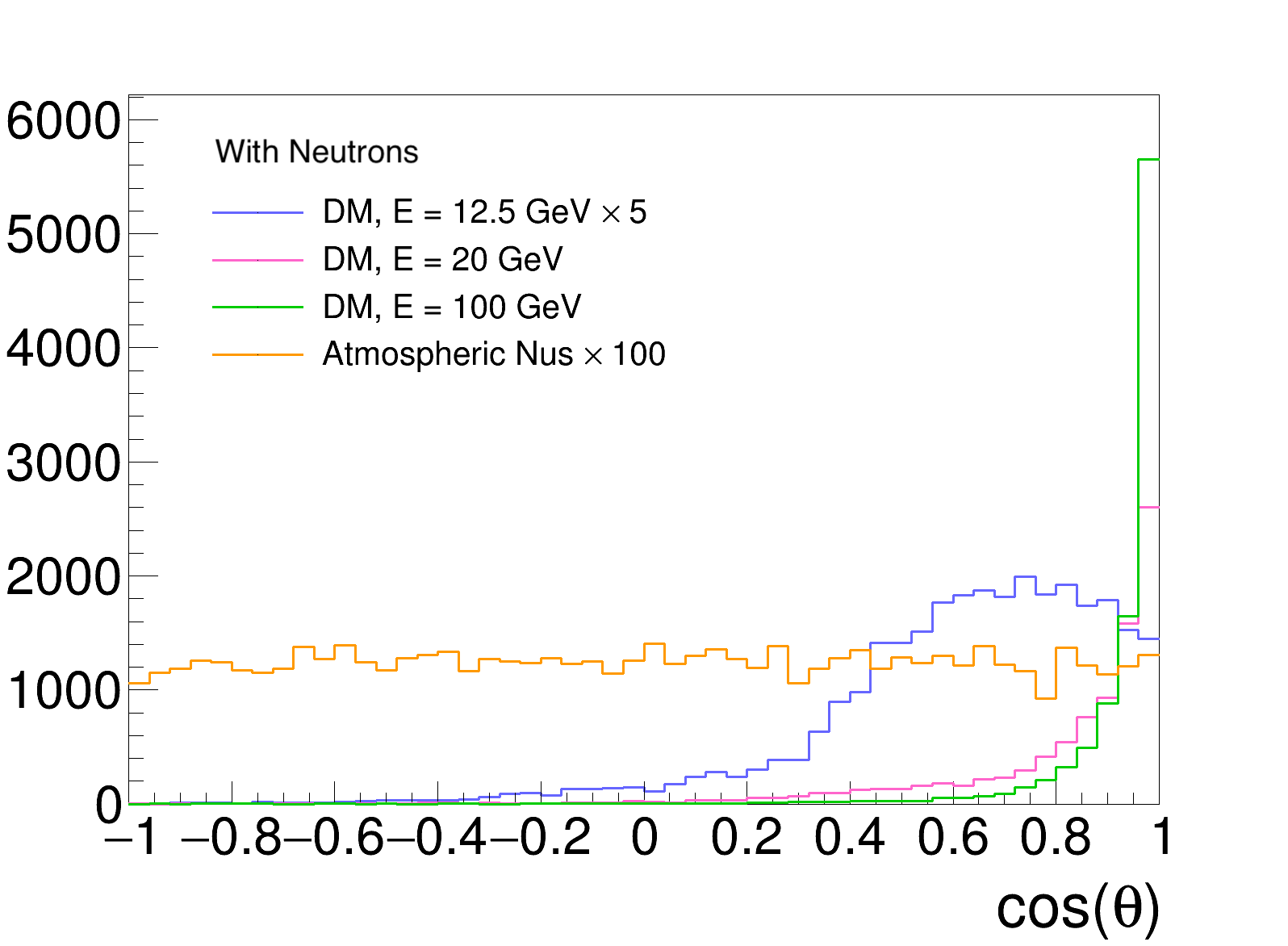}
\includegraphics[width=0.9\columnwidth]{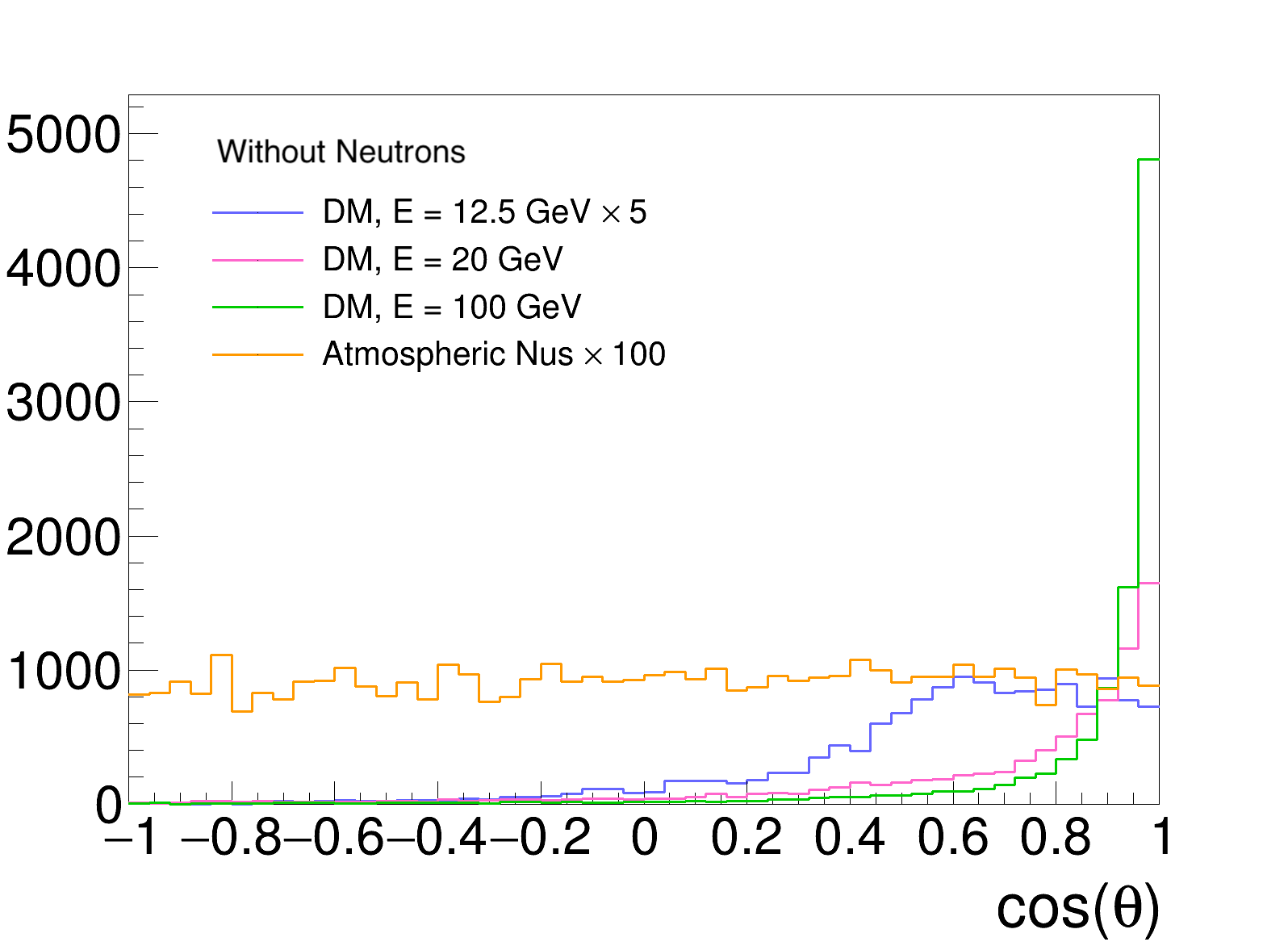}
\caption[Angular distribution of the BDM signal events for a BDM mass of 10\,GeV]{Angular distribution of the \dword{bdm} signal events for a \dword{bdm} mass of 10\,GeV
and different boosted factors, $\gamma$, and of the atmospheric neutrino NC
background events.
$\theta$ represents the angle of the sum over all the stable final state
particles as detailed in the text.
The amount of background represents one-year data collection, magnified by a factor 100,
while the amount of signal reflects the detection efficiency of 10,000 \dword{mc} events. 
%as described in this note.
The top plot shows the scenario where neutrons can be reconstructed,
while the bottom plot represents the scenario without neutrons.}
\label{fig:m10_SmearedReconstructableAngle}
\end{figure}

The finite detector resolution is taken into
account by smearing the direction of the stable final state particles, 
including protons, neutrons, charged pions, muons, electrons, and photons,
with the expected angular resolution,
and by ignoring the ones with kinetic energy below detector threshold,
using the parameters reported in the DUNE \dword{cdr}~\cite{Acciarri:2015uup}.
We form as the observable the total momentum from all the stable final state particles,
and obtain its angle with respect to the direction of the sun.
The sun position is simulated with the SolTrack package~\cite{SolTrack}
including the geographical coordinates of the DUNE \dword{fd}.
%~\cite{DUNE_DocDB136}.
We consider both the scenarios in which we can reconstruct neutrons,
according to the parameters described in the DUNE \dword{cdr}, and in which neutrons will not be reconstructed at all.
Figure~\ref{fig:m10_SmearedReconstructableAngle} shows the angular distributions of
the \dword{bdm} signals with mass of 10\,GeV and different boost factors,
and of the background events.

To increase the signal fraction in our samples, we select events with $\cos\theta > 0.6$,
and obtain the selection efficiency $\varepsilon$ for different \dword{bdm} models.
We predict that $104.0 \pm 0.7$ and $79.4 \pm 0.6$ background events per year, in the scenarios with and without neutrons respectively, survive the selection in a DUNE \nominalmodsize module.

\begin{figure}[htb]
\centering
\includegraphics[width=0.9\columnwidth]{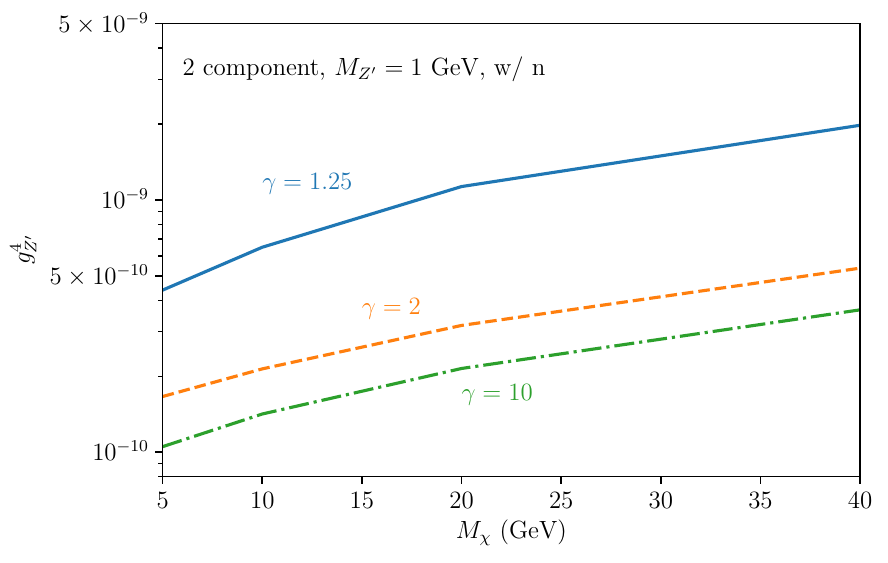}
\includegraphics[width=0.9\columnwidth]{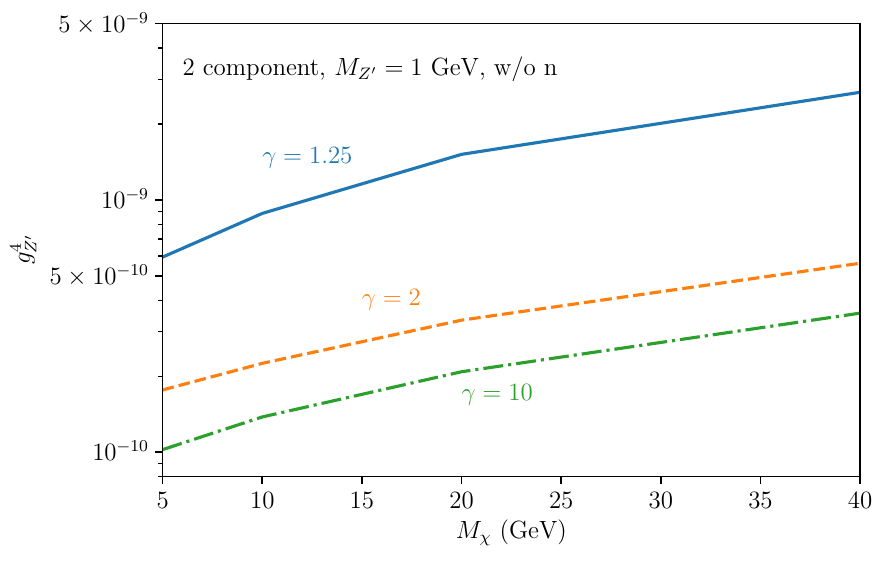}
\caption[Expected $5\sigma$ discovery reach with one year of DUNE livetime]{Expected $5\sigma$ discovery reach with one year of DUNE livetime for one \nominalmodsize module including neutrons in reconstruction (top) and excluding neutrons (bottom).\label{fig:significance}}
\end{figure}
\begin{figure}[htb]
\centering
\includegraphics[width=0.9\columnwidth]{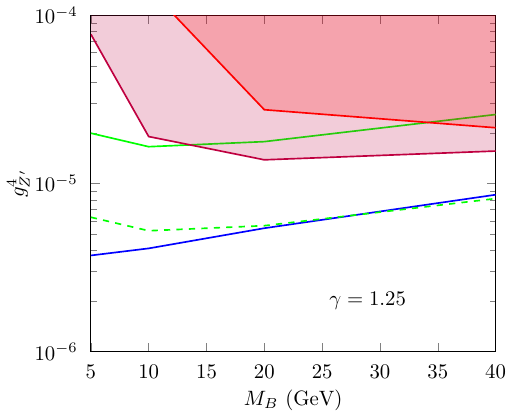}\hspace{0.05\textwidth}
\includegraphics[width=0.9\columnwidth]{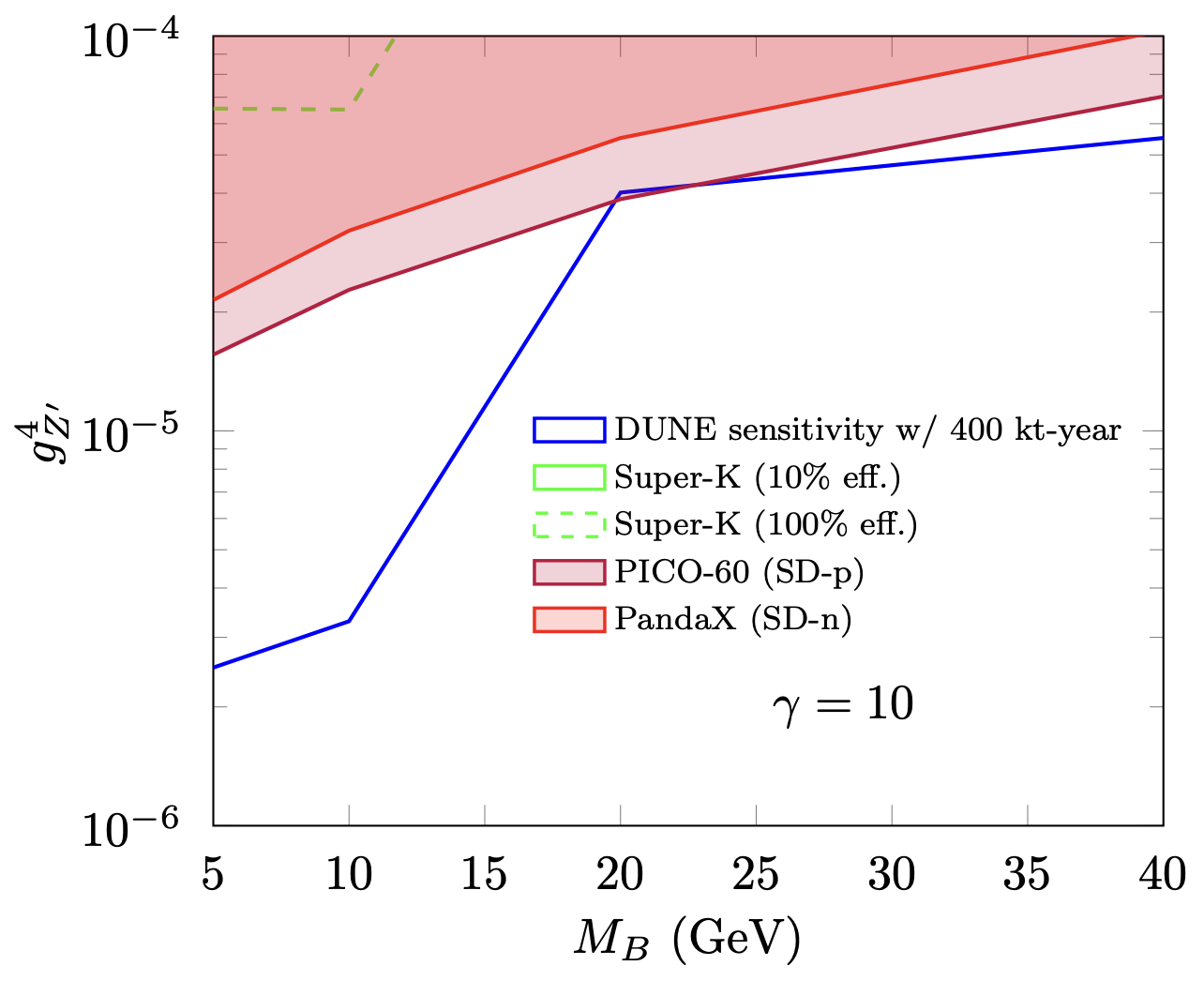}
\caption[Comparison of DUNE (10 year) sensitivity to \superk sensitivity]{Comparison of sensitivity of DUNE for 10 years of data collection and \SI{40}{\kt} of detector mass with Super Kamiokande, assuming 10\% and 100\% of the selection efficiency on the atmospheric neutrino analysis in Ref.~\cite{Fechner:2009aa}, and with the reinterpretations of the current results from PICO-60~\cite{Amole:2019fdf} and PandaX~\cite{Xia:2018qgs}.  The samples with two boosted factors, $\gamma = 1.25$ (top) and $\gamma = 10$ (bottom), are also presented. \label{fig:bdm_sensitivity_comparison}}
\end{figure}

The resulting expected sensitivity is presented in Fig.~\ref{fig:significance} in terms of the \dword{dm} mass and the $Z^\prime$ gauge coupling for potential \dword{dm} boosts of $\gamma = 1.25,2,10$ and for a fixed mediator mass of $M_{Z^\prime} = 1~{\rm GeV}$.  We assume a DUNE livetime of one year for one \nominalmodsize module.  The models presented here are currently unconstrained by direct detection searches if the thermal relic abundance of the \dword{dm} is chosen to fit current observations.
Figure~\ref{fig:bdm_sensitivity_comparison} compares the sensitivity of 10 years of data collected in DUNE (\SI{40}{\kt}) to re-analyses of the results from other experiments, including Super Kamiokande~\cite{Fechner:2009aa} and \dword{dm} direct detection, PICO-60~\cite{Amole:2019fdf} and PandaX~\cite{Xia:2018qgs}. An extension to this study can be found in Ref.~\cite{Berger:2019ttc}.

\subsection{Summary of Dark Matter Detection Prospects}

We have conducted simulation studies of the dark matter models described in Eqs.~\eqref{eq:lagrangian} and \eqref{eq:zprimelag} in terms of their detection prospects at the DUNE \dword{nd} and \dword{fd}. 
Thanks to its relatively low threshold and strong particle identification capabilities, DUNE presents an opportunity to significantly advance the search for \dword{ldm} and \dword{bdm} beyond what has been possible with water Cherenkov detectors.

In the case of the \dword{nd}, we assumed that the relativistic \dword{dm} is being produced directly at the target and leaves an experimental signature through an elastic electron scattering. Using two constrained parameters of the light \dword{dm} model and a range of two free parameters, a sensitivity map was produced. Within the context of the vector portal \dword{dm} model and the chosen parameter constraints along with the electron scattering as the signal event, this result sets stringent limits on \dword{dm} parameters that are comparable or even better than recent experimental bounds in the sub-GeV mass range.

By contrast, in the case of the \dword{fd} modules, we assumed that the signal events are due to \dword{dm} coming from the galactic halo and the sun with a significant boost factor. 
For the inelastic scattering case, the \dword{dm} scatters off either an electron or proton in the detector material into a heavier unstable dark-sector state.
The heavier state, by construction, decays back to \dword{dm} and an electron-positron pair via a dark-photon exchange. 
Therefore, in the final state, a signal event comes with an electron or proton recoil plus an electron-positron pair. 
This distinctive signal feature enabled us to perform (almost) background-free analyses. 

As \dword{protodune} detectors are prototypes of DUNE \dword{fd} modules, the same study was conducted~\cite{Chatterjee:2018mej} and corresponding results were compared with the ones of the DUNE \dword{fd} modules.  
We first investigated the experimental sensitivity in a dark-photon parameter space, dark-photon mass $M_V$ versus kinetic mixing parameter $\epsilon$. 
The results are shown separately for Scenarios 1 and 2 in Fig.~\ref{fig:darkphotonparameter}. 
They suggest that DUNE \dword{fd} modules would probe a broad range of unexplored regions; they would allow for reaching $\sim 1-2$ orders of magnitude smaller $\epsilon$ values than the current limits along MeV to sub-GeV-range dark photons. 
We also examined model-independent reaches at DUNE \dword{fd} modules, providing limits for models that assume the existence of $i$\dword{bdm} (or $i$\dword{bdm}-like) signals (i.e., a target recoil and a fermion pair). 

For the elastic scattering case, we considered the case in which \dword{bdm} comes from the sun. 
With one year of data, the $5\sigma$ sensitivity is expected to reach a coupling of $g_{Z^\prime}^4 = 9.57 \times 10^{-10}$ for a boost of 1.25 and $g_{Z^\prime}^4 = 1.49 \times 10^{-10}$ for a boost of 10 at a \dword{dm} mass of \SI{10}{GeV} without including neutrons in the reconstruction.

\section{Baryon Number Violating Processes}
\label{sect:bnv}

Unifying three of the fundamental forces in the universe, the strong, 
electromagnetic, and weak interactions, is a shared goal for the current 
world-wide program in particle physics. \Dwords{gut}, extending the \dword{sm} to include a unified gauge symmetry 
at very high energies  (more than $10^{15}$~GeV), predict a number of observable 
effects at low energies, such as nucleon  decay \cite{Pati:1973rp,Georgi:1974sy,Langacker:1980js,deBoer:1994dg,Nath:2006ut}. 
Since the early 1980s, supersymmetric \dword{gut} models were preferred for a number of reasons, including gauge-coupling unification, natural embedding in superstring theories, and their ability to solve the fine-tuning problem of the \dword{sm}.  Supersymmetric \dword{gut} models~\cite{Dimopoulos:1981dw,Dimopoulos:1981zb,Sakai:1981pk,Nath:1985ub,Shafi:1999vm,Lucas:1996bc,Pati:2003qia,Babu:1997js,Alciati:2005ur} generically predict that the dominant proton decay mode is \ptoknubar, in contrast to non-supersymmetric \dword{gut} models, which typically predict
the dominant decay mode to be \ptoepizero.  Although the LHC did not find any evidence for \dword{susy} at the electroweak scale, as was expected if \dword{susy} were to solve the gauge hierarchy problem in the \dword{sm}, the appeal of a \dword{gut} still remains. In particular, gauge-coupling unification can still be achieved in non-supersymmetric \dword{gut} models by the introduction of one or more intermediate scales (see, for example, \cite{Altarelli:2013aqa}).
Several experiments have sought signatures of nucleon decay, with the best limits for most decay modes set by the \superk experiment~\cite{Abe:2014mwa,Miura:2016krn,TheSuper-Kamiokande:2017tit}, 
which features the largest sensitive mass and exposure to date. 

The excellent imaging, as well as calorimetric and particle identification capabilities, of the \dword{lartpc} technology implemented for the \dword{dune} \dword{fd} will exploit a number of complementary signatures for a broad range of baryon-number violating processes.  Should nucleon decay rates lie just beyond current limits, observation of even one or two candidate events with negligible background could constitute compelling evidence. In the \dword{dune} era, two other large detectors, \hyperk~\cite{Abe:2018uyc} and JUNO~\cite{Djurcic:2015vqa} will be conducting nucleon decay searches. Should a signal be observed in any single experiment, confirmation from experiments using different detector technologies and nuclear targets, and therefore subject to different backgrounds, would be very powerful.

Neutron-antineutron (\nnbar) oscillation is a baryon number violating process that
has never been observed but is predicted by a number of \dword{bsm} theories~\cite{Phillips:2014fgb}. In this context, baryon number conservation is an accidental
symmetry rather than a fundamental one, which means baryon number violation
does not stand against the fundamental gauge symmetries. Discovering baryon
number violation would have implications on the source of matter-antimatter
symmetry in our universe given Sakharov's conditions for such asymmetry to arise~\cite{Sakharov:1967dj}.
In particular, the neutron-antineutron oscillation (\nnbar) process violates
baryon number by two units and, therefore, could also have further implications for
the smallness of neutrino masses~\cite{Phillips:2014fgb}. 
Since the \nnbar transition operator is a six-quark operator, of dimension \num{9}, with a coefficient function of dimension (mass)$^{-5}$, while the proton decay operator is a four-fermion operator, of dimension \num{6}, with a coefficient function of dimension (mass)$^{-2}$, one might naively assume that \nnbar oscillations would always be suppressed relative
to proton decay as a manifestation of baryon number violation.  However, this is not necessarily the case; indeed, there are models~\cite{Nussinov:2001rb,Arnold:2012sd,Girmohanta:2019fsx,Girmohanta:2020qfd} in which proton decay is very strongly suppressed down to an unobservably small level, while \nnbar oscillations occur at a level comparable to present limits. This shows the
value of a search for \nnbar transitions at DUNE.
 Searches for this process using
both free neutrons and nucleus-bound neutron states have been carried out 
since the 1980s. The current best \num{90}\% \dword{cl} limits on the (free) neutron oscillation lifetime are \SI{8.6e7}{\s} from free \nnbar searches and \SI{2.7e8}{\s} from nucleus-bound \nnbar searches~\cite{BaldoCeolin:1994jz,Abe:2011ky}.  As with nucleon decay, searches for \nnbar oscillations performed by \dword{dune} and those performed by \superk, \hyperk, and the European Spallation Source~\cite{Phillips:2014fgb} are highly complementary.  Should a signal be observed in any one experiment, confirmation from another experiment with a different detector technology and backgrounds would be very powerful.

\subsection{Event Simulation and Reconstruction}

To estimate the sensitivity to baryon number violation in DUNE, simulation of both signal and background events is performed using \dword{genie} version 2.12.10. For nucleon decay, a total of \num{68} single-nucleon exclusive decay channels listed in the 2016 update of the \dword{pdg}~\cite{Tanabashi:2018oca} %Patrignani:2016xqp} 
are available in \dword{genie}. The list includes two-, three-, and five-body decays. 
If a bound nucleon decays, the remaining nucleus can be in an excited state and will typically de-excite by emitting nuclear fission fragments, nucleons, and photons. At present, de-excitation photon emission is simulated only for oxygen. 
The simulation of neutron-antineutron oscillation was developed~\cite{Hewes:2017xtr} and implemented in \dword{genie}. Implementing this process in \dword{genie} used \dword{genie}'s existing modeling of Fermi momentum and binding energy for both the oscillating neutron and the nucleon with which the resulting antineutron annihilates.   Once a neutron has oscillated to an antineutron in a nucleus, the antineutron has a $18/39$ chance of annihilating with a proton in argon, and a $21/39$ chance of annihilating with a neutron. The energies and momenta of the annihilation products are assigned randomly but consistently with four-momentum conservation. The products of the annihilation process follow the branching fractions (shown in Table~\ref{tab:nnbar-br}) measured in low-energy antiproton annihilation on hydrogen~\cite{Hewes:2017xtr}.

The default model in \dword{genie} for the propagation of particles inside the nucleus is $hA2015$, an empirical, data-driven model that does not simulate the cascade of hadronic interactions step by step, but instead uses one effective interaction to represent the effect of \dword{fsi}. Hadron-nucleus scattering data is used to tune the predictions.

The dominant background for these searches is from atmospheric neutrino interactions.  Backgrounds from neutrino interactions are simulated with \dword{genie}, using the Bartol model of atmospheric neutrino flux~\cite{Barr:2004br}. To estimate the event rate, we integrate the product of the neutrino flux and interaction cross section. Table \ref{tab:rate} shows the event rate for different neutrino species for an exposure of
10~\ktyr,
where oscillation effects are not included.
To suppress atmospheric neutrino background to the level of one event per \Mtyr, which would yield \num{0.4} events after ten years of operation with a \SI{40}{\kt} fiducial volume, the necessary background rejection is $1 - (1/288600) = 1 - 3\times10^{-6} = 0.999997$, where background rejection is defined as the fraction of background that is not selected.

\begin{table}
\centering
\caption{Expected rate of atmospheric neutrino interactions in \argon40 for a \SI{10}{\ktyr} exposure (not including oscillations).}
\label{tab:rate}
\begin{tabular}{ccc|c}\hline
     &~CC~&~NC~&~Total \\ \hline\hline
\numu & \num{1038} & \num{398} &\num{1436} \\
\anumu &\num{280} & \num{169} & \num{449} \\
\nue & \num{597} &  \num{206} &\num{803} \\
\anue & \num{126} & \num{72} & \num{198} \\
\hline
Total & \num{2041} & \num{845} & \num{2886} \\ \hline
\end{tabular}
\end{table}

These analyses assume that the detector is successfully triggered on all signal events, and that the \dword{pds} correctly determines the event start time ($t_0$).
Two distinct methods of reconstruction and event selection have been applied in these analyses. One employs \threed track and vertex reconstruction provided by \dword{pma}~\cite{Abi:2020evt}, a standard DUNE reconstruction algorithm. \dword{pma} was designed to address transformation from a set of independently reconstructed \twod projections of objects into a \threed representation. This algorithm uses clusters of hits from \twod pattern recognition as its input. The other reconstruction method involves image classification of \twod images of reconstructed hits using a \dword{cnn}. The two methods are combined in the form of a multivariate analysis, which uses the image classification score with other physical observables extracted from traditional reconstruction.

\subsection{Nucleon Decay}
\label{sect:pdk}

Because of the already stringent limits set by \superk on \ptoepizero and the unique ability to track and identify kaons in a \dword{lartpc}, the initial nucleon decay studies in \dword{dune} have focused on nucleon decay modes featuring kaons, in particular \ptoknubar. The experimental signature of this channel is a single $K^{+}$ originating inside the fiducial volume of the detector. The kaon typically stops and decays at rest with a lifetime of 12 ns. The most common decay mode, $K^{+} \rightarrow \mu^{+}\nu_{\mu}$, results in a monoenergetic muon with momentum of 236 MeV/c. In the next most probable decay, $K^{+} \rightarrow \pi^{+}\pi^{0}$, the two pions are produced back to back. In a water \cherenkov detector, the kaon is typically below \cherenkov threshold, and only the kaon decay products are observed. In DUNE's \dword{lartpc}, the kaon can be detected and identified by its distinctive $dE/dx$ signature, as well as by its decay~\cite{Meddage:2019pbr}. 

For a proton decay at rest, the outgoing kaon is monoenergetic with kinetic energy of $\SI{105}{\MeV}$ and momentum of \SI{339}{\MeV}/c. In bound proton decay, the momentum of the kaon is smeared by the Fermi motion of the protons inside the nucleus. \Dword{fsi} between the outgoing kaon and the residual  nucleus may reduce the kaon momentum, and may also modify the final state, by ejecting nucleons for example. Protons ejected from the nucleus can obscure the $dE/dx$ measurement of the kaon if the tracks overlap. The $K^{+}$ may also charge exchange, resulting in a $K^{0}$ in the final state. The $K^{+}$ cannot be absorbed due to strangeness conservation and the lack of $S=1$ baryons. The residual nucleus may also be in an excited state, producing de-excitation photons.

The main backgrounds in nucleon decay searches are interactions of atmospheric neutrinos. For \ptoknubar, the background is neutrino interactions that mimic a single $K^{+}$ and its decay products. Because the kaon is not detected in a water \cherenkov detector, neutrino interactions that produce a single $K^{+}$ and no other particles above \cherenkov threshold are an irreducible background. This includes charged-current reactions like the Cabibbo-suppressed $\nu_{\mu} n \rightarrow \mu^{-} K^{+} n$, where the final-state muon and kaon are below threshold, as well as neutral-current processes like $\nu p \rightarrow \nu K^{+} \Lambda$ followed by $\Lambda \rightarrow p \pi^{-}$ where the $\Lambda$ decay products are below threshold. Strangeness is always conserved in neutral-currents, so kaons produced in NC interactions are always accompanied by a hyperon or another kaon. 
Water \cherenkov detectors and liquid scintillator detectors like JUNO can also detect neutron captures, which provide an additional handle on backgrounds, many of which have final-state neutrons. However, neutrons can also be present in \ptoknubar signal due to \dword{fsi}, and the rate of nucleon ejection in kaon-nucleus interactions is not well understood. Nuclear de-excitation photons are also typically produced, but these are similar in both proton decay and atmospheric neutrino events.  In the \superk analysis of \ptoknubar the time difference between the de-excitation photons from the oxygen nucleus and the muon from kaon decay was found to be an effective way to reduce backgrounds~\cite{Abe:2014mwa}. In JUNO, the three-fold time coincidence between the kaon, the muon from the kaon decay, and the electron from the muon decay is expected to be an important discriminant between signal and background~\cite{Djurcic:2015vqa}.

The possibility of using the time difference between the kaon scintillation signal and the scintillation signal from the muon from the kaon decay has been investigated in DUNE.  Studies indicate that measuring  time differences on the scale of the kaon lifetime (12~ns) is difficult in DUNE, independent of photon detector acceptance and timing resolution, due to both the scintillation process in argon - consisting of fast (ns-scale) and slow ($\mu$s-scale) components - and Rayleigh scattering over long distances.

In a \dword{lartpc}, a charged particle traveling just a few cm can be detected, and the other particles produced in association with a kaon by atmospheric neutrinos are generally observed. However, with \dword{fsi} the signal process can also include final-state protons, so requiring no other final-state particles will reject some signal events. Furthermore, $\nu_{\mu}$ charged-current quasi-elastic scattering (CCQE), $\nu_{\mu} n \rightarrow \mu^{-} p$, can mimic the $K^{+} \rightarrow \mu^{+} \nu_{\mu}$ decay when the proton is mis-reconstructed as a kaon. 

The kaon reconstruction is especially challenging for very short tracks, which may traverse only a few wires. The $dE/dx$ signature in signal events can be obscured by additional final-state protons that overlap with the start of the kaon track. Without timing resolution sufficient to resolve the 12 ns kaon lifetime, the $dE/dx$ profile is the only distinguishing feature. The background from atmospheric neutrino events without true final-state kaons, which is important given the presence of \dword{fsi}, was neglected in previous estimates of \ptoknubar sensitivity in \dword{lartpc}~\cite{bueno-pdk}. 

Other backgrounds, such as those initiated by cosmic-ray muons, can be controlled by requiring no activity close to the edges of the \dwords{tpc} and by stringent single kaon identification within the energy range of interest~\cite{Adams:2013qkq,Klinger:2015kva}.  

\begin{figure}
\centering
\includegraphics[width=0.9\columnwidth]{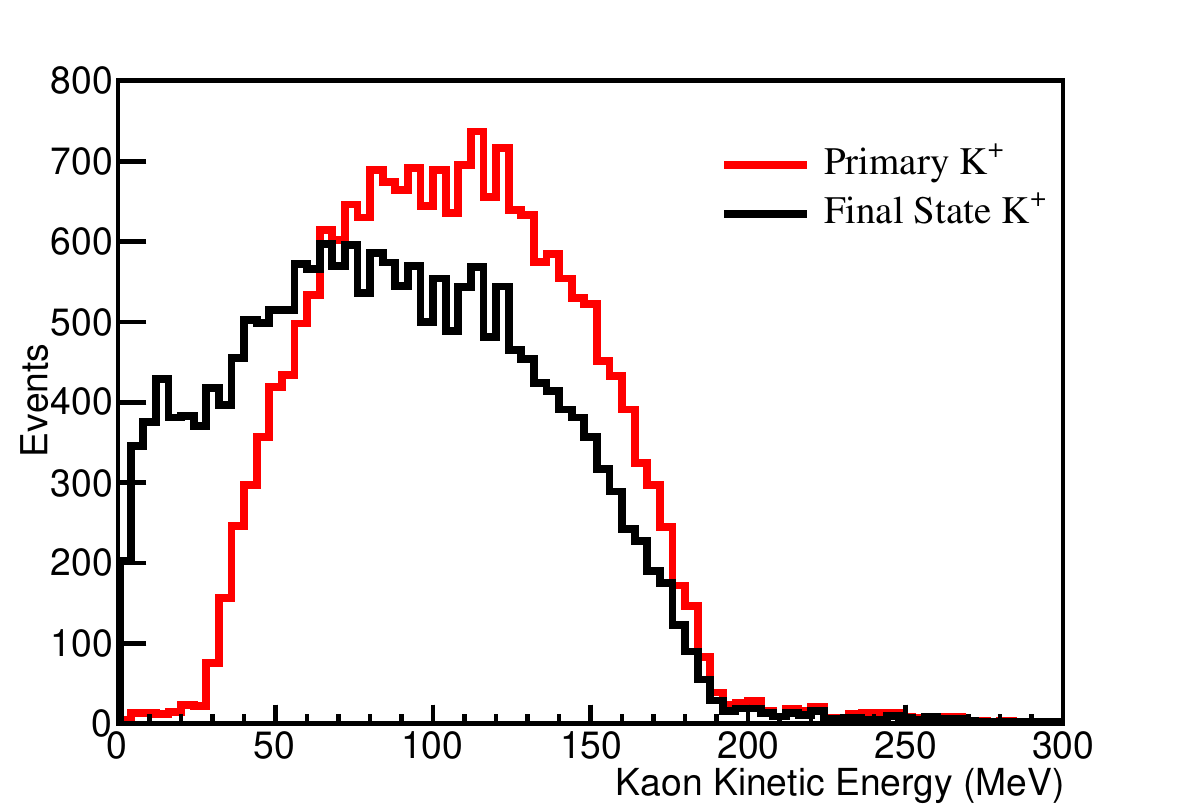}
\caption{Kinetic energy of kaons in simulated proton decay events, \ptoknubar, in \dword{dune}.  The kinetic energy distribution is shown before and after final state interactions in the argon nucleus.}
\label{fig:K-wFSI-hA2015}
\end{figure}

\dword{fsi} significantly modify the observable distributions in the detector. 
For charged kaons, the $hA2015$ model includes only elastic scattering and nucleon knock-out, tuned to $K^{+}-$C data~\cite{Bugg:1968zz,Friedman:1997eq}. Charge exchange is not included, nor are strong processes that produce kaons inside the nucleus, such as $\pi^{+}n \rightarrow K^{+} \Lambda$. 
Figure~\ref{fig:K-wFSI-hA2015} shows the kinetic energy of a kaon from \ptoknubar before and after \dword{fsi} as simulated with $hA2015$. Kaon interactions always reduce the kaon energy, and the kaon spectrum becomes softer on average with \dword{fsi}. Of the kaons, \num{31.5}\%  undergo elastic scattering resulting in events with very low kinetic energy;  \num{25}\% of kaons have a kinetic energy of $\le\SI{50}{\MeV}$. When the kaon undergoes elastic scattering, a nucleon can be knocked out of the nucleus. Of decays via this channel, \num{26.7}\%  have one neutron coming from \dword{fsi}, \num{15.3}\% have at least one proton, and \num{10.3}\% have two protons coming from \dword{fsi}. These secondary nucleons are detrimental to reconstructing and selecting $K^{+}$.

Other \dword{fsi} models include the full cascade, and predict slightly different final states, but existing data lack power to favor one model over another. MINERvA has measured the differential cross section for charged-current $K^{+}$ production by neutrinos on plastic scintillator (CH) as a function of kaon energy, which is sensitive to \dword{fsi}, and shows a weak preference for the \dword{genie} $hA2015$ \dword{fsi} model over a prediction with no \dword{fsi}~\cite{Marshall:2016rrn}. Compared to the kaon energy spectrum measured by MINERVA, \dword{fsi} have a much larger impact on \ptoknubar in argon, and the differences between models are less significant than the overall effect. 

The kaon \dword{fsi} in \superk's simulation of \ptoknubar in oxygen seem to have a smaller effect on the outgoing kaon momentum distribution~\cite{Abe:2014mwa} than is seen here with the \dword{genie} simulation on argon.  Some differences are expected due to the different nuclei, but differences in the \dword{fsi} models are under investigation.

Kaon \dword{fsi} have implications on the ability to identify \ptoknubar events in \dword{dune}. Track reconstruction efficiency for a charged particle $x^{\pm}$ is defined as 

\begin{equation}
\epsilon_{x^{\pm}} = \frac{\mbox{$x^{\pm}$ particles with a reconstructed track}}{\mbox{events with $x^{\pm}$ particle }}.
\end{equation}
The denominator includes events in which an $x^{\pm}$ particle was created and has deposited energy within any of the \dwords{tpc}.  The numerator includes events in which an $x^{\pm}$ particle was created and has deposited energy within any of the \dwords{tpc}, and a reconstructed track can be associated to the $x^{\pm}$ particle based on the number of hits generated by that particle along the track. This efficiency can be calculated as a function of true kinetic energy and true track length.

\begin{figure}
\centering
\includegraphics[width=0.9\columnwidth]{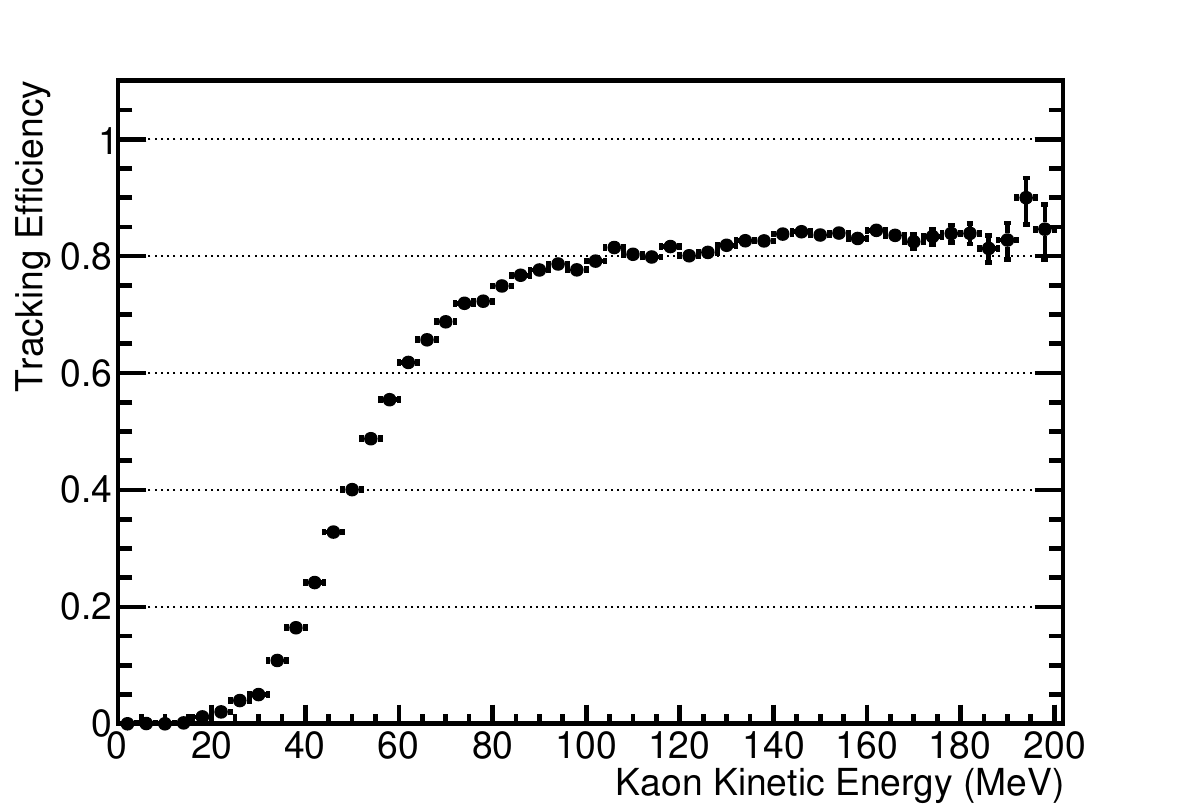}
\includegraphics[width=0.9\columnwidth]{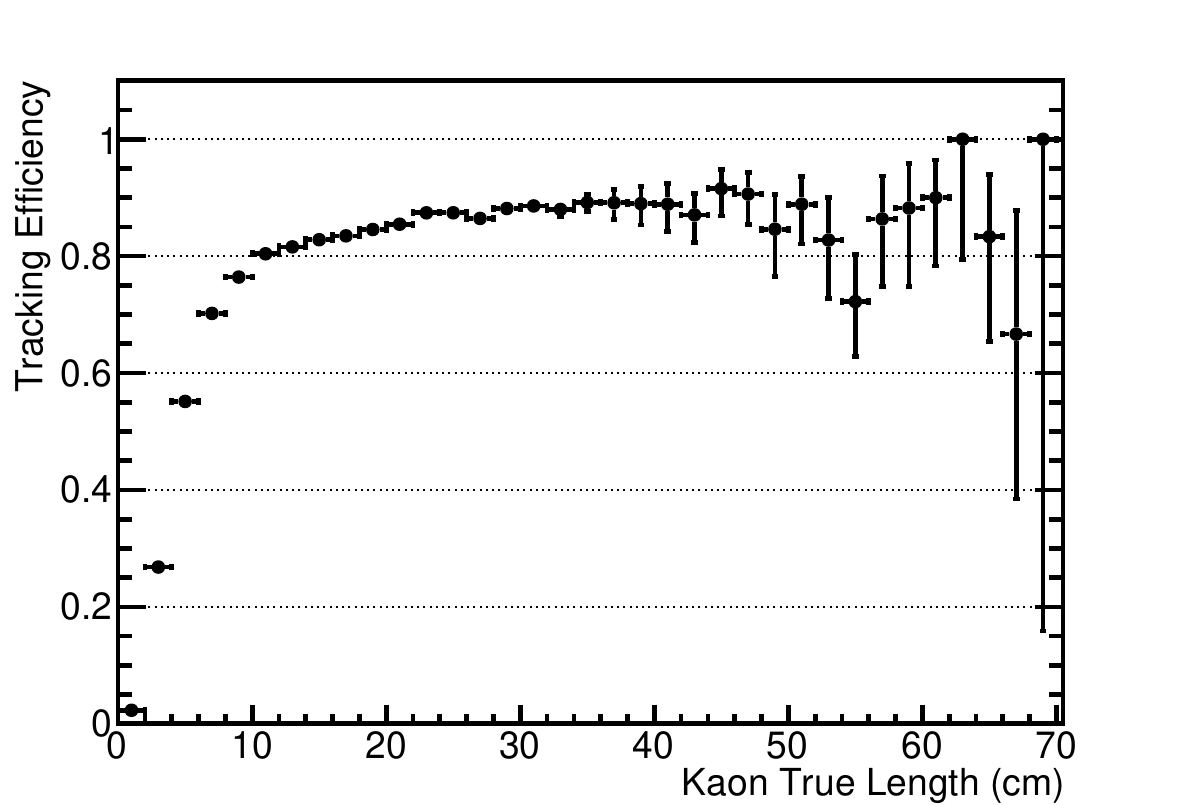}
\caption{Tracking efficiency for kaons in simulated proton decay events, \ptoknubar, as a function of kaon kinetic energy (top) and true path length (bottom).}
\label{fig:k-trk-eff}
\end{figure}

Figure~\ref{fig:k-trk-eff} shows the tracking efficiency for $K^{+}$  from proton decay via \ptoknubar as a function of true kinetic energy and true path length. The overall tracking efficiency for kaons from proton decay is \num{58.0}\%, meaning that \num{58.0}\% of all the simulated kaons are associated with a reconstructed track in the detector.  From Fig.~\ref{fig:k-trk-eff}, the tracking threshold is approximately $\sim\SI{40}{\MeV}$ of kinetic energy, which translates to $\sim\SI{4.0}{\cm}$ in true path length.  The biggest loss in tracking efficiency is due to kaons with $<\SI{40}{\MeV}$ of kinetic energy due to scattering inside the nucleus. The efficiency levels off to approximately \num{80}\% above \SI{80}{\MeV} of kinetic energy; this inefficiency even at high kinetic energy is due mostly to kaons that decay in flight.
Both kaon scattering in the \dword{lar} and charge exchange are included in the detector simulation but are relatively small effects (\num{4.6}\% of kaons scatter in the \dword{lar} and \num{1.2}\% of kaons experience charge exchange).  The tracking efficiency for muons from the decay of the $K^{+}$ in \ptoknubar is \num{90}\%.

Hits associated with a reconstructed track are used to calculate the energy loss of charged particles, which provides valuable information on particle energy and species.
If the charged particle stops in the \dword{lartpc} active volume, a combination of $dE/dx$ and the reconstructed residual range ($R$, the path length to the end point of the track) is used to define a parameter for \dword{pid}.  The parameter, $PIDA$, 
is defined as~\cite{Acciarri:2013met}  

\begin{equation}
PIDA = \left\langle \left(\frac{dE}{dx}\right)_{i}R^{0.42}_{i}\right\rangle,\label{eqn:PIDA}
\end{equation}
where the median is taken over all track points $i$ for which the residual range $R_i$ is less than \SI{30}{\cm}.

\begin{figure}
\centering
\includegraphics[width=0.9\columnwidth]{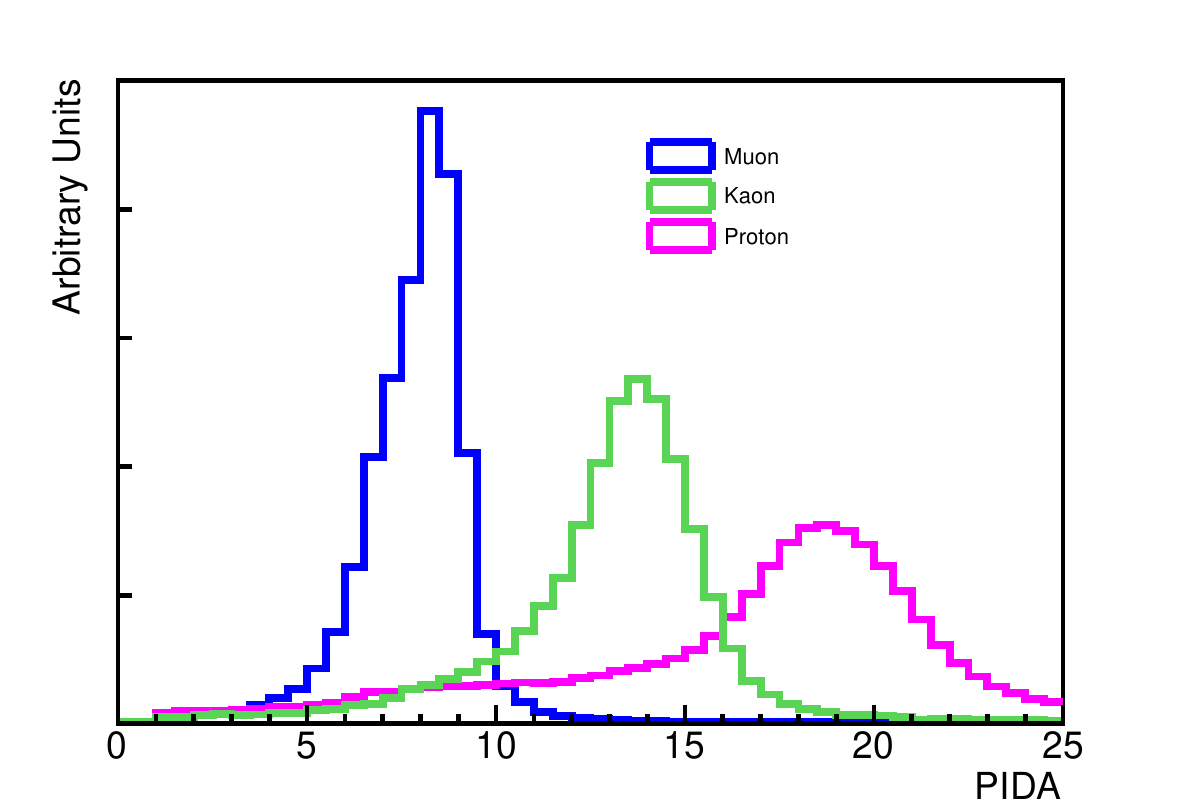}
\caption{Particle identification using $PIDA$ for muons and kaons in simulated proton decay events, \ptoknubar, and protons in simulated atmospheric neutrino background events.  The curves are normalized by area.}
\label{fig:PIDA}
\end{figure}

Figure~\ref{fig:PIDA} shows the $PIDA$ performance for kaons (from proton decay), muons (from kaon decay), and protons produced by atmospheric neutrino interactions. The tail with lower values in each distribution is due to cases where the decay/stopping point was missed by the track reconstruction. The tail with higher values is caused when a second particle overlaps at the decay/stopping point causing higher values of $dE/dx$ and resulting in higher values of $PIDA$. In addition, ionization fluctuations smear out these distributions.

\dword{pid} via $dE/dx$ becomes complicated when the reconstructed track direction is ambiguous, in particular if additional energy is deposited  at the vertex in events where FSI is significant.
The dominant background to \ptoknubar in \dword{dune} is atmospheric neutrino \dword{cc} \dword{qe}
scattering, $\nu_{\mu} n \rightarrow \mu^{-} p$.  When the muon happens to have very close
to the \SI{236}{\MeV$/c$} momentum expected from a $K^{+}$ decay at rest and is not captured, it is indistinguishable from the muon resulting from \ptoknubar followed by $K^{+} \rightarrow \mu^{+}\nu_{\mu}$. When
the proton is also mis-reconstructed as a kaon, this background mimics the signal process.  

The most important difference between signal and this background source is the direction of the hadron track. For an atmospheric neutrino, the proton and muon originate from the same neutrino interaction point, and the characteristic Bragg rise occurs at the end of the proton track farthest from the muon-proton vertex. In signal, the kaon-muon vertex is where the $K^{+}$ stops and decays at rest, so its ionization energy deposit is highest near the kaon-muon vertex.  To take advantage of this difference, a log-likelihood ratio discriminator is used to distinguish signal from background.  Templates are formed by taking the reconstructed and calibrated energy deposit as a function of the number of wires from both the start and end of the $K^{+}$ candidate hadron track. 
Two log-likelihood ratios are computed separately for each track. The first begins at the hadron-muon shared vertex and moves along the hadron track (the ``backward'' direction). The second begins at the other end of the track, farthest from the hadron-muon shared vertex, moves along the hadron track the other way (the ``forward'' direction). For signal events, this effectively looks for the absence of a Bragg rise at the $K^{+}$ start, and the presence of one at the end, and vice versa for background.  At each point, the probability density for signal and background, $P^{sig}$ and $P^{bkg}$, are determined from the templates. Forward and backward log-likelihood ratios are computed as

\begin{align}
 \mathcal{L}_{fwd(bkwd)} = \sum_{i} \log\frac{P^{sig}_i}{P^{bkg}_i}, 
\end{align}
where the summation is over the wires of the track, in either the forward or backward direction.  Using either the forward or backward log-likelihood ratio alone gives some discrimination between signal and background, but using the sum gives better discrimination. While the probability densities are computed based on the same samples, defining one end of the track instead of the other as the vertex provides more information. The discriminator is the sum of the forward and backward log-likelihood ratios:

\begin{align}
    \mathcal{L} = \mathcal{L}_{fwd} + \mathcal{L}_{bkwd}.\label{eqn:L}
\end{align}
Applying this discriminator to tracks with at least ten wires gives a signal efficiency of roughly \num{0.4} with a background rejection of \num{0.99}.

\begin{figure}
\centering
\includegraphics[width=0.9\columnwidth]{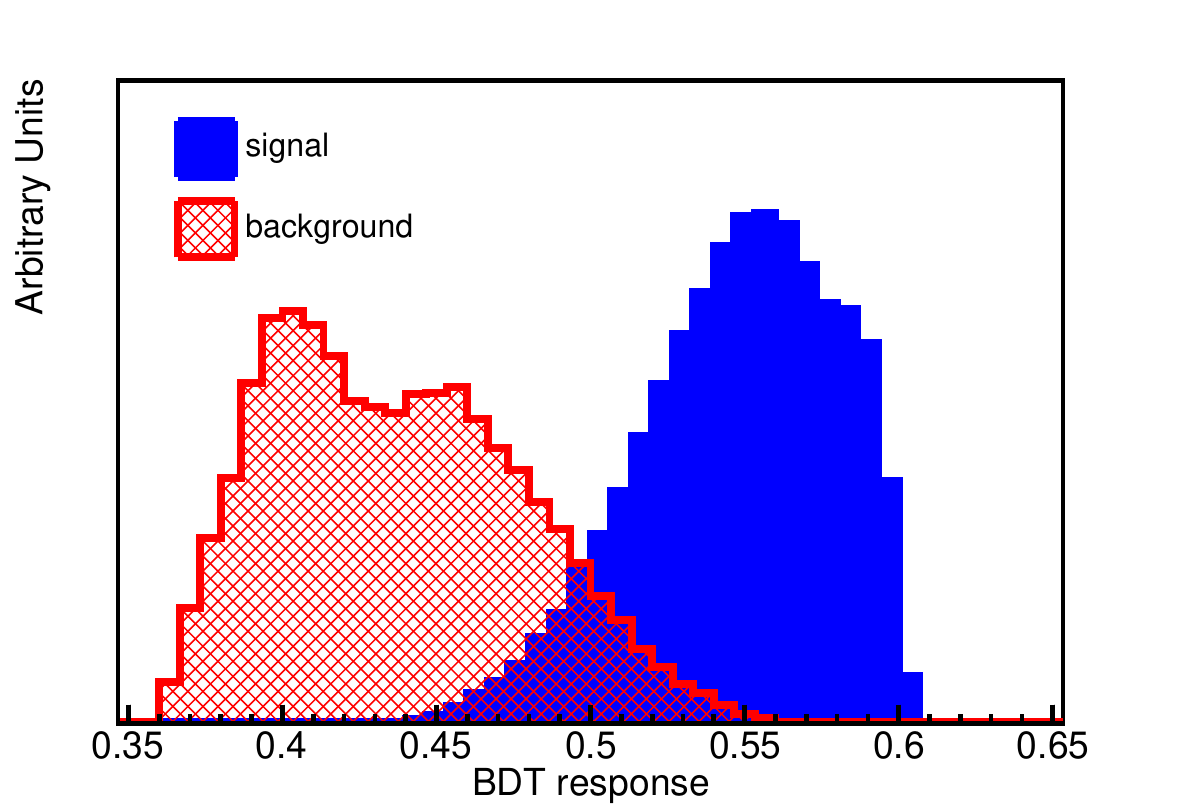}
\caption{Boosted Decision Tree response for \ptoknubar for signal (blue) and background (red).}
\label{fig:BDT_response}
\end{figure}

A \dword{bdt}
classifier is used for event selection in the analysis presented here. The software package Toolkit for Multivariate Data Analysis with ROOT (TMVA4)~\cite{Hocker:2007ht}
is used with AdaBoost as the boosted algorithm.  The \dword{bdt} is trained on a sample of \dword{mc} events (\num{50000} events for signal and background) that is statistically independent from the sample of \dword{mc} events used in the analysis (approximately \num{100000} events for signal and \num{600000} events for background).
Image classification using a \dword{cnn}
is performed using \twod images of \dword{dune} \dword{mc} events. The image classification provides a single score value as a metric of whether any given event is consistent with a proton decay, and this score can be used as a powerful discriminant for event identification.  In the analysis presented here, the \dword{cnn} technique alone does not discriminate between signal and background as well as a \dword{bdt}, so the \dword{cnn} score is used as one of the input variables to the \dword{bdt} in this analysis.
The other variables in the \dword{bdt} include numbers of reconstructed objects (tracks, showers, vertices), variables related to visible energy deposition, \dword{pid} variables [$PIDA$, Eq.~\eqref{eqn:PIDA}, and $\mathcal{L}$, Eq.~\eqref{eqn:L}], reconstructed track length, and reconstructed momentum.
Figure~\ref{fig:BDT_response} shows the distribution of the \dword{bdt} output for signal and background.
Backgrounds from atmospheric neutrinos are weighted by the oscillation probability in the \dword{bdt} input distributions.

\begin{figure}
\centering
\includegraphics[width=0.9\columnwidth]{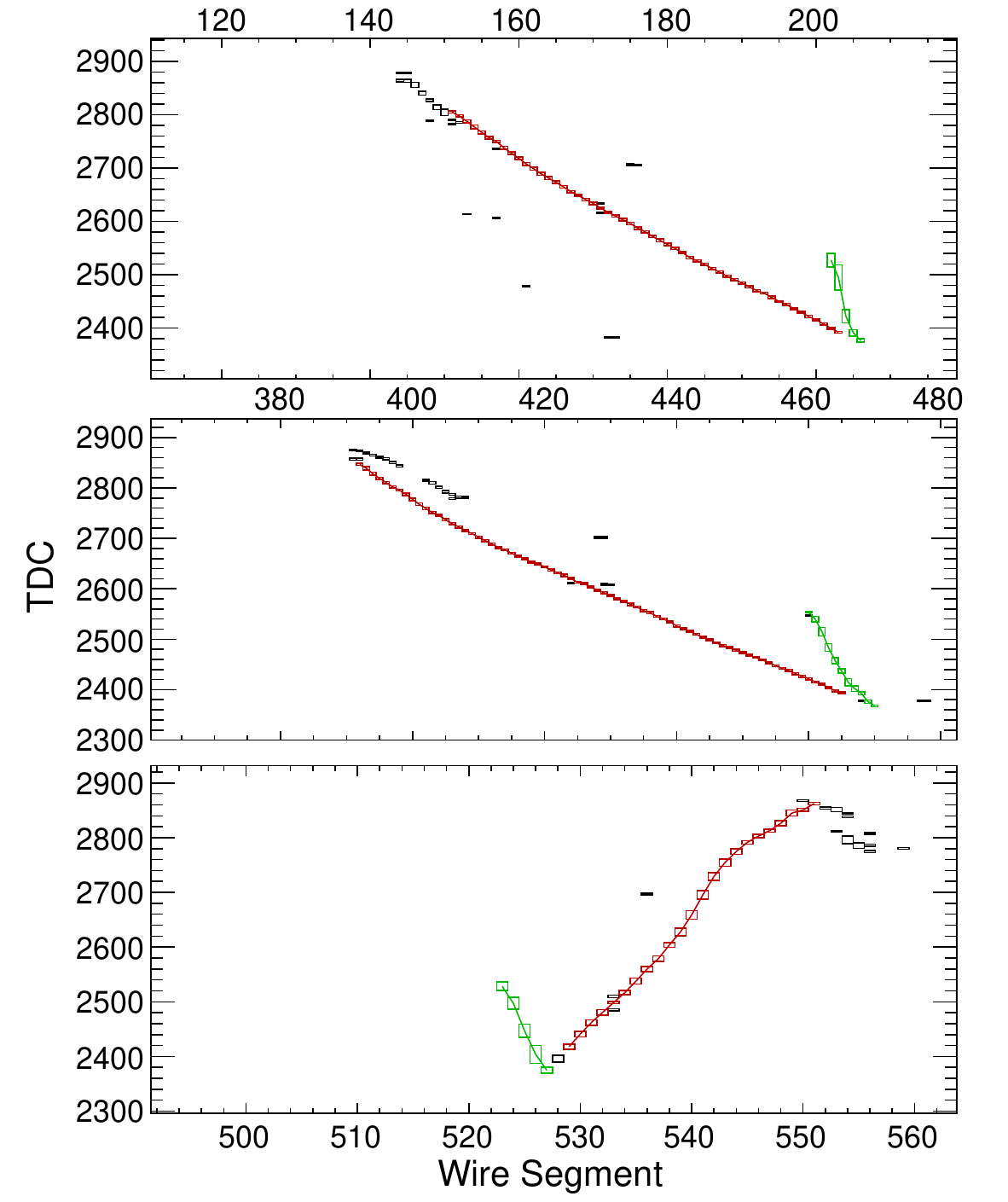}
\caption{Event display for an easily recognizable \ptoknubar signal event.  The vertical axis is TDC value, and the horizontal axis is wire number. The bottom view is induction plane one, the middle is induction plane two, and the top is the collection plane. Hits associated with the reconstructed muon track are shown in red, and hits associated with the reconstructed kaon track are shown in green.  Hits from the decay electron can be seen at the end of the muon track.}
\label{fig:event_signal}
\end{figure}

\begin{figure}
\centering
\includegraphics[width=0.9\columnwidth]{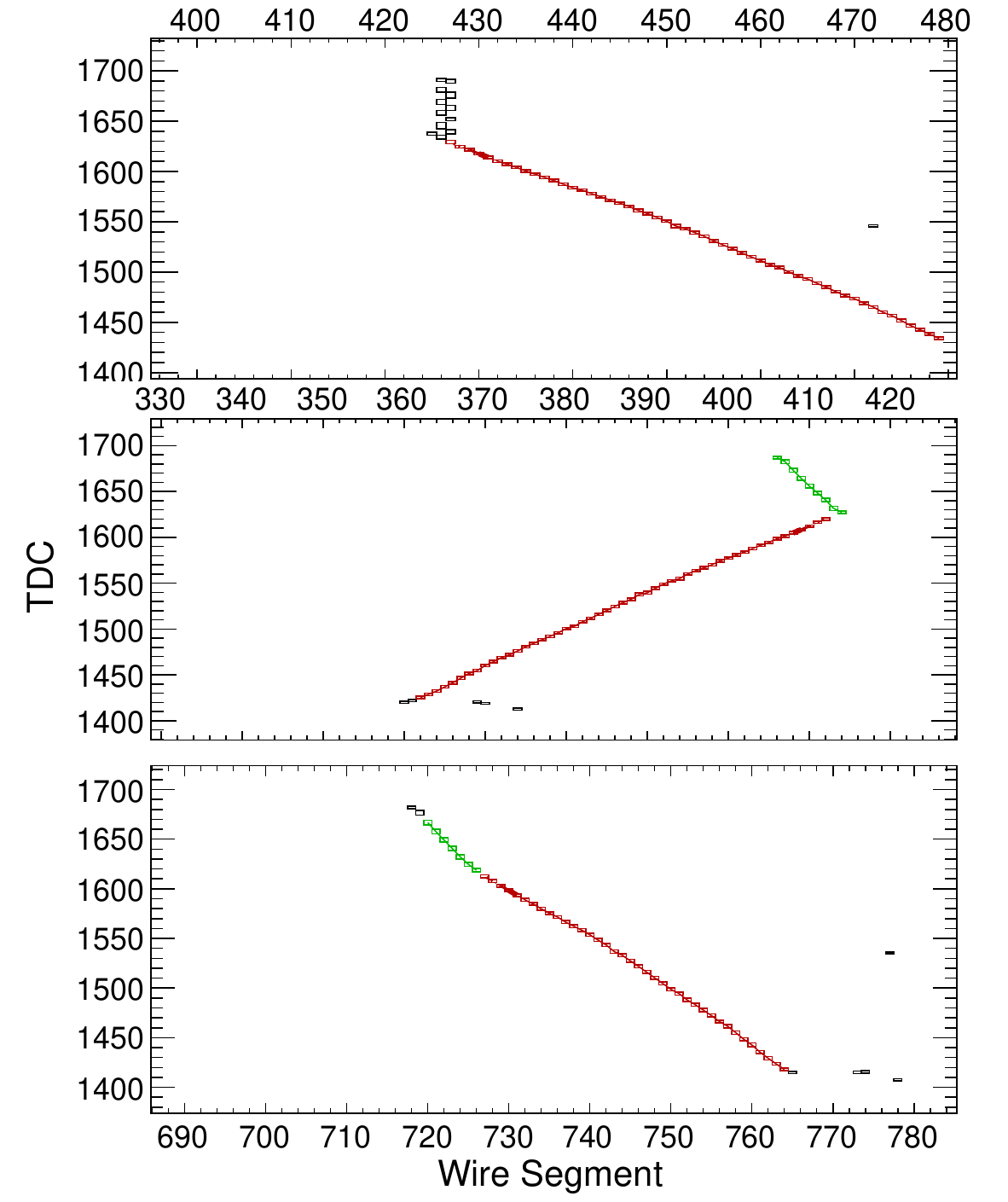}
\caption{Event display for an atmospheric neutrino interaction, $\nu_{\mu} n \rightarrow \mu^{-} p$, which might be selected in the \ptoknubar sample if the proton is misidentified as a kaon. The vertical axis is TDC value, and the horizontal axis is wire number. The bottom view is induction plane one, the middle is induction plane two, and the top is the collection plane. Hits associated with the reconstructed muon track are shown in red, and hits associated with the reconstructed proton track are shown in green.  Hits from the decay electron can be seen at the end of the muon track.}
\label{fig:event_bkgd}
\end{figure}

Figure~\ref{fig:event_signal} shows a \ptoknubar signal event. The event display shows the reconstructed kaon track in green and the reconstructed muon track from the kaon decay in red; hits from the Michel electron coming from the muon decay can be seen at the end of the muon track. Figure~\ref{fig:event_bkgd} shows an event with a similar topology produced by an atmospheric neutrino interaction, $\nu_{\mu} n \rightarrow \mu^{-} p$. This type of event can be selected in the \ptoknubar sample if the proton is misidentified as a kaon. Hits associated with the reconstructed muon track are shown in red, and hits associated with the reconstructed proton track are shown in green.  Hits from the decay electron can be seen at the end of the muon track.

The proton decay signal and atmospheric neutrino background events are processed using the same reconstruction chain and subject to the same selection criteria. There are two preselection cuts to remove obvious background. One cut requires at least two tracks, which aims to select events with a kaon plus a kaon decay product (usually a muon).  The other cut requires that the longest track be less than \SI{100}{\cm}; this removes backgrounds from high energy neutrino interactions.  After these cuts, \num{50}\% of the signal and \num{17.5}\% of the background remain in the sample.  The signal inefficiency at this stage of selection is due mainly to the kaon tracking efficiency.
Optimal lifetime sensitivity is achieved by combining the preselection cuts with a \dword{bdt} cut that gives a signal efficiency of \num{0.15} and a background rejection of 
$0.999997$,
which corresponds to approximately one background event per \Mtyr.

The limiting factor in the sensitivity is the kaon tracking efficiency.  
The reconstruction is not yet optimized, and the kaon tracking efficiency should increase with improvements in the reconstruction algorithms.  
To understand the potential improvement, a visual scan of simulated decays of kaons into muons was performed. For this sample of events, with kaon momentum in the \SIrange{150}{450}{\MeV$/c$} range, scanners achieved greater than \num{90}\% efficiency at recognizing the $K^{+} \rightarrow \mu^{+} \rightarrow e^{+}$ decay chain.  The inefficiency came mostly from short kaon tracks (momentum below \SI{180}{\MeV$/c$}) and kaons that decay in flight. Note that the lowest momentum kaons ($<$\SI{150}{\MeV$/c$}) were not included in the study; the path length for kaons in this range would also be too short to track.  Based on this study, the kaon tracking efficiency could be improved to a maximum value of approximately \num{80}\% with optimized reconstruction algorithms, where the remaining inefficiency comes from low-energy kaons and kaons that charge exchange, scatter, or decay in flight.
Combining this tracking performance improvement with some improvement in the $K/p$ separation performance for short tracks, the overall signal selection efficiency improves from \num{15}\% to approximately \num{30}\%.

The analysis presented above is inclusive of all possible modes of kaon decay; however, the current version of the \dword{bdt} preferentially selects kaon decay to muons, which has a branching fraction of roughly \num{64}\%. The second most prominent kaon decay is $K^{+} \rightarrow \pi^{+}\pi^0$, which has a branching fraction of \num{21}\%.  Preliminary studies that focus on reconstructing a $\pi^{+}\pi^0$ pair with the appropriate kinematics indicate that the signal efficiency for kaons that decay via the $K^{+} \rightarrow \pi^{+}\pi^0$ mode is approximately the same as the signal efficiency for kaons that decay via the $K^{+} \rightarrow \mu^{+}\nu_{\mu}$ mode.  This assumption is included in our sensitivity estimates below.

Because the DUNE efficiency to reconstruct a kaon track is strongly dependent on the kaon kinetic energy as seen in Fig.~\ref{fig:k-trk-eff}, the \dword{fsi} model is an important source of systematic uncertainty.
To account for this uncertainty, kaon-nucleon elastic scattering ($K^{+}p(n)\rightarrow K^{+}p(n)$) is re-weighted by $\pm \num{50}\%$ in the simulation. 
The absolute uncertainty on the efficiency with this re-weighting is \num{2}\%, which is taken as the systematic uncertainty on the signal efficiency.
The dominant uncertainty in the background 
is due to the absolute normalization of the atmospheric neutrino rate. The Bartol group has carried out a detailed study of the systematic uncertainties, where the absolute neutrino fluxes have uncertainties of approximately \num{15}\%~\cite{Barr:2006it}.
The remaining uncertainties are due to the cross section models for neutrino interactions.
The uncertainty on the \dword{cc}0$\pi$ cross section in the energy range relevant for these backgrounds is roughly \num{10}\%~\cite{Mahn:2018mai}.
Based on these two effects, a conservative \num{20}\% systematic uncertainty in the background is estimated.

With a \num{30}\% signal efficiency and an expected background of one event per \si{\Mtyr}, a \num{90}\% \dword{cl} lower limit on the proton lifetime in the \ptoknubar channel of \SI{1.3e34}{years} can be set, assuming no signal is observed over ten years of running with a total of \SI{40}{\kt} of fiducial mass. This calculation assumes constant signal efficiency and background rejection over time and for each of the \dword{fd} modules.  Additional running improves the sensitivity proportionately if the experiment remains background-free.

Another potential mode for a baryon number violation search is the decay of the neutron into a charged lepton plus meson, i.e.,~\ntoek. In this mode, $\Delta B = -\Delta L$, where $B$ is baryon number and $L$ is lepton number.  The current best limit on this mode is \SI{3.2e31}{years} from the FREJUS collaboration~\cite{Berger:1991fa}. The reconstruction software for this analysis is the same as for the \ptoknubar analysis; the analysis again uses a \dword{bdt} that includes an image classification score as an input. To calculate the lifetime sensitivity for this decay mode the same systematic uncertainties and procedures are used. The selection efficiency for this channel including the expected tracking improvements is \num{0.47}
with a background rejection of 
$0.99995$,
which corresponds to \num{15} background events per \si{\Mtyr}. The lifetime sensitivity for a \SI{400}{\ktyr} exposure is \SI{1.1e34}{years}.

\subsection{Neutron-Antineutron Oscillation}
\label{sect:nnbar}

Neutron-antineutron oscillations can be detected via the subsequent antineutron annihilation with a neutron or a proton. Table~\ref{tab:nnbar-br} shows the effective branching ratios for the antineutron annihilation modes applicable to intranuclear searches, modified from~\cite{Abe:2011ky}.  It is known that other, more fundamentally consistent branching fractions exist~\cite{Golubeva:2018mrz,Barrow:2019viz}, but the effects of these on final states is believed to be minimal. The annihilation event will have a distinct, roughly spherical signature of a vertex with several emitted light hadrons (a so-called ``pion star"), with total energy of twice the nucleon mass and roughly zero net momentum.  Reconstructing these hadrons correctly and measuring their energies is key to identifying the signal event. The main background for these \nnbar annihilation events is caused by atmospheric neutrinos. 
As with nucleon decay, nuclear effects and \dword{fsi} make the picture more complicated.
As shown in Table~~\ref{tab:nnbar-br}, every decay mode contains at least one charged pion and one neutral pion.
The pion \dword{fsi} in the $hA2015$ model in \dword{genie} include pion elastic and inelastic scattering, charge exchange and absorption.

\begin{table}
\centering
\caption[\nnbar annihilation modes]{Effective branching ratios for antineutron annihilation in \argon40, as implemented
in \dword{genie}.}
\label{tab:nnbar-br}
\begin{tabular}{l |c}\hline
Channel & Branching ratio \\ \hline\hline
\multicolumn{2}{l}{$\bar{n}+p$:} \\
         $\pi^{+}\pi^{0}$ & 1.2\%  \\ 
         $\pi^{+}2\pi^{0}$ & 9.5\% \\
         $\pi^{+}3\pi^{0}$ & 11.9\% \\ 
         $2\pi^{+}\pi^{-}\pi^{0}$ & 26.2\% \\  $2\pi^{+}\pi^{-}2\pi^{0}$ & 42.8\% \\ $2\pi^{+}\pi^{-}2\omega$ & 0.003\% \\ $3\pi^{+}2\pi^{-}\pi^{0}$ & 8.4\% \\ \hline 
\multicolumn{2}{l}{$\bar{n}+n$:}\\ 
         $\pi^{+}\pi^{-}$ & 2.0\% \\ 
         $2\pi^{0}$ & 1.5\% \\ 
         $\pi^{+}\pi^{-}\pi^{0}$ & 6.5\% \\ 
         $\pi^{+}\pi^{-}2\pi^{0}$ & 11.0\% \\  $\pi^{+}\pi^{-}3\pi^{0}$ & 28.0\% \\ 
         $2\pi^{+}2\pi^{-}$ & 7.1\% \\ 
         $2\pi^{+}2\pi^{-}\pi^{0}$ & 24.0\% \\ 
         $\pi^{+}\pi^{-}\omega$ & 10.0\% \\ 
         $2\pi^{+}2\pi^{-}2\pi^{0}$ & 10.0\% \\ \hline
\end{tabular}
\end{table}

\begin{figure}
\centering
\includegraphics[width=0.9\columnwidth]{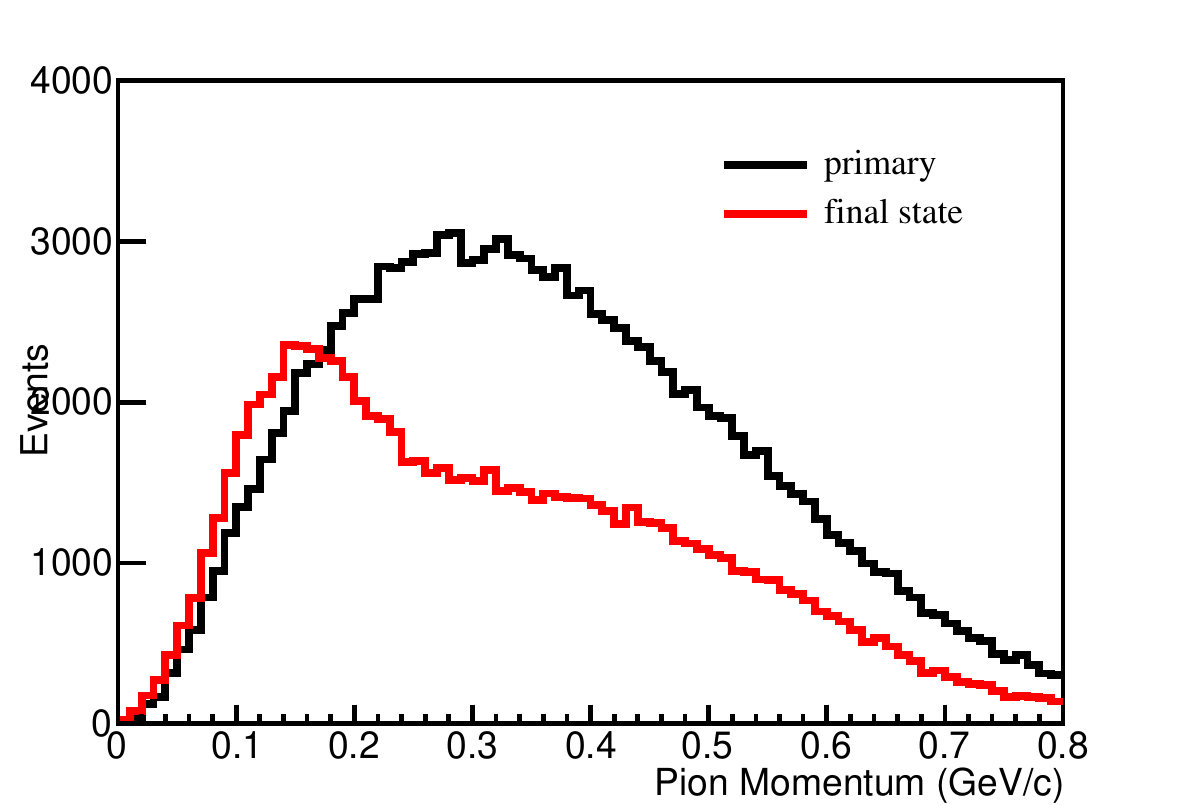}
\includegraphics[width=0.9\columnwidth]{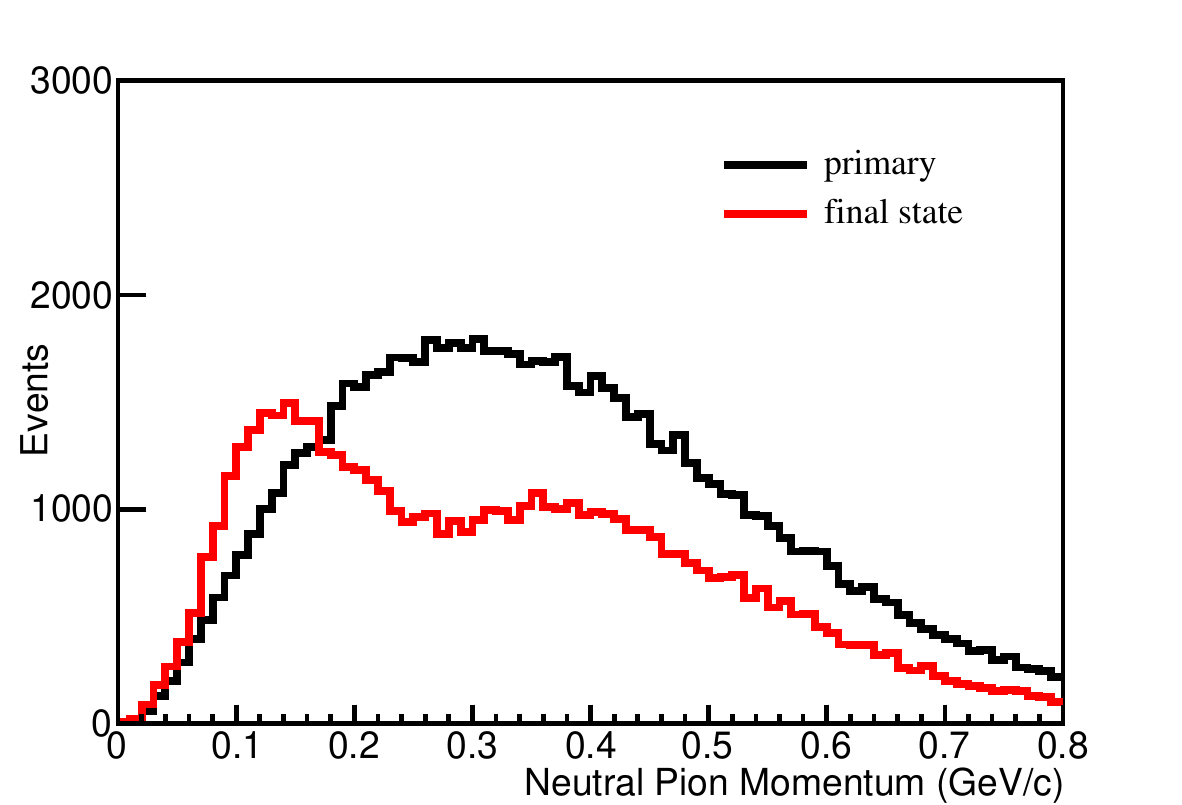}
\caption{Top: momentum of an individual charged pion before and after final state interactions. Bottom: momentum of an individual neutral pion before and after final state interactions.}
\label{fig:pi_FSI_m}
\end{figure}

Figure~\ref{fig:pi_FSI_m} shows the momentum distributions for charged and neutral pions before \dword{fsi} and after \dword{fsi}. These distributions show the \dword{fsi} makes both charged and neutral pions less energetic.  The effect of \dword{fsi} on pion multiplicity is also rather significant; \num{0.9}\% of the events have no charged pions before \dword{fsi}, whereas after \dword{fsi} \num{11.1}\% of the events have no charged pions. In the case of the neutral pion, \num{11.0}\% of the events have no neutral pions before \dword{fsi}, whereas after \dword{fsi}, \num{23.4}\% of the events have no neutral pions. The decrease in pion multiplicity is primarily due to pion absorption in the nucleus. Another effect of \dword{fsi} is nucleon knockout from pion elastic scattering. Of the events, \num{94}\% have at least one proton from \dword{fsi} and \num{95}\% of the events have at least one neutron from \dword{fsi}. Although the kinetic energy for these nucleons peak at a few tens of \si{\MeV}, the kinetic energy can be as large as hundreds of \si{\MeV}.  In summary, the effects of \dword{fsi} in \nnbar become relevant because they modify the kinematics and topology of the event. For instance, even though the decay modes of Table \ref{tab:nnbar-br} do not include nucleons in their decay products, nucleons appear with high probability after \dword{fsi}.

\begin{figure}
\centering
\includegraphics[width=0.9\columnwidth]{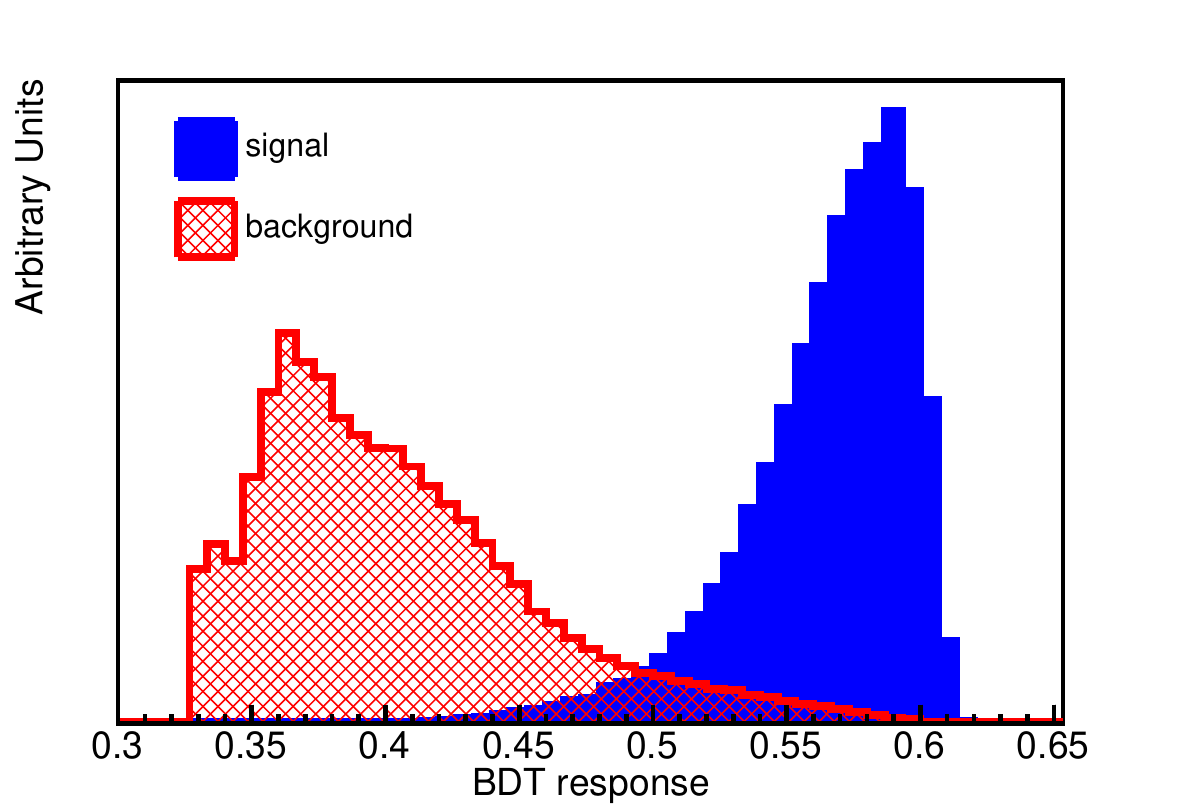}
\caption{Boosted Decision Tree response for \nnbar oscillation for signal (blue) and background (red).}
\label{fig:bdt_nnbar}
\end{figure}

A \dword{bdt} classifier is used. Ten variables are used in the \dword{bdt} as input for event selection, including number of reconstructed tracks and showers, variables related to visible energy deposition, $PIDA$ and $dE/dx$, reconstructed momentum, and CNN score.  Figure~\ref{fig:bdt_nnbar} shows the distribution of the \dword{bdt} output for signal and background.

\begin{figure}
\centering
\includegraphics[width=0.9\columnwidth]{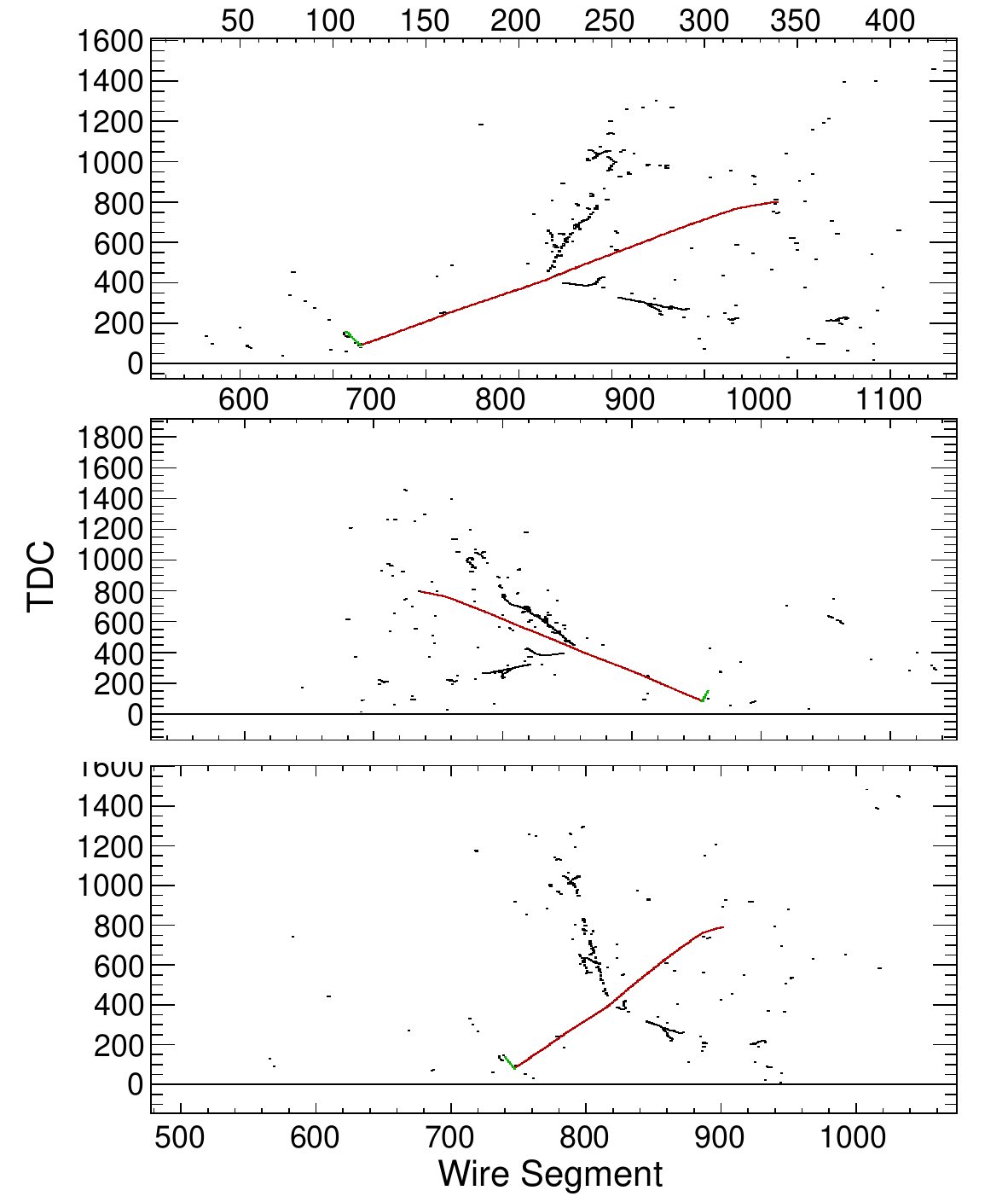}
\caption{Event display for an \nnbar signal event, $n \bar{n} \rightarrow n \pi^0 \pi^0 \pi^{+} \pi^{-}$.  The vertical axis is TDC value, and the horizontal axis is wire number. The bottom view is induction plane one, the middle is induction plane two, and the top is the collection plane. Hits associated with the back-to-back tracks of the charged pions are shown in red.  The remaining hits are from the showers from the neutral pions, neutron scatters, and low-energy de-excitation gammas.
}
\label{fig:nnbar_sig}
\end{figure}

\begin{figure}
\centering
\includegraphics[width=0.9\columnwidth]{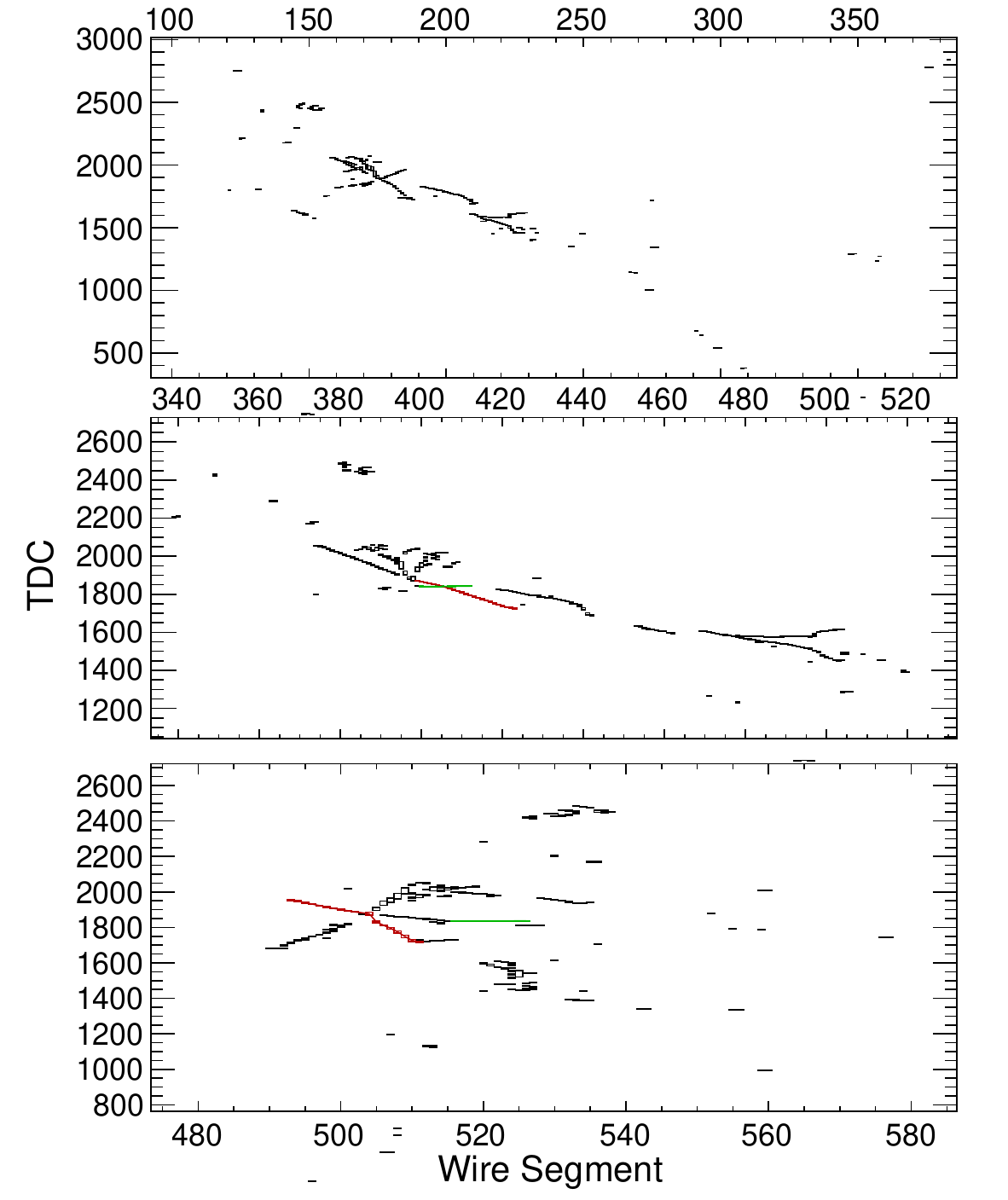}
\caption{Event display for a \dword{nc} \dword{dis} interaction initiated by an atmospheric neutrino. The vertical axis is TDC value, and the horizontal axis is wire number. The bottom view is induction plane one, the middle is induction plane two, and the top is the collection plane. This event mimics the \nnbar signal topology by having multi-particle production and electromagnetic showers.}
\label{fig:nnbar_bkgd}
\end{figure}

Figure~\ref{fig:nnbar_sig} shows an \nnbar signal event, $n \bar{n} \rightarrow n \pi^0 \pi^0 \pi^{+} \pi^{-}$.  Hits associated with the back-to-back tracks of the charged pions are shown in red.  The remaining hits are from the showers from the neutral pions, neutron scatters, and low-energy de-excitation gammas.  The topology of this event is consistent with charged pion and neutral pion production. 
Figure~\ref{fig:nnbar_bkgd} shows an event with a similar topology produced by a \dword{nc} \dword{dis} atmospheric neutrino interaction. This background event mimics the signal topology by having multi-particle production and an electromagnetic shower. 

The sensitivity to the \nnbar oscillation lifetime can be calculated for a given exposure, the efficiency of selecting signal events, and the background rate along with their uncertainties. The lifetime sensitivity is obtained at \num{90}\% \dword{cl} for the bound neutron. Then, the lifetime sensitivity for a free neutron is acquired using the conversion from nucleus bounded neutron to free neutron \nnbar oscillation~\cite{Friedman:2008es}.  
The uncertainties on the signal efficiency and background rejection are conservatively estimated to be \num{25}\%.  A detailed evaluation of the uncertainties is in progress.

The free \nnbar oscillation lifetime, $\tau_{n-\bar{n}}$, and bounded \nnbar oscillation lifetime, $T_{n-\bar{n}}$, are related to each other through the intranuclear suppression factor $R$ as

\begin{equation}
    \tau^{2}_{n-\bar{n}} = \frac{T_{n-\bar{n}}}{R} ~.
    \label{eq:tau}
\end{equation}
The suppression factor $R$ varies for different nuclei. This suppression factor was calculated for $^{16}$O and $^{56}$Fe~\cite{Friedman:2008es}. The $R$ for $^{56}$Fe, \SI{0.666e23}{\per\s}, is used in this analysis for \argon40 nuclei.  More recent work~\cite{Barrow:2019viz} gives a value of $R$ for \argon40 of \SI{0.56e23}{\per\s}, which will be applied in future analyses.

The best bound neutron lifetime limit is achieved using a signal efficiency of \num{8.0}\% at the background rejection probability of \num{99.98}\%, which corresponds to approximately 23~atmospheric neutrino background events for a \SI{400}{\ktyr} exposure. The \num{90}\% \dword{cl} limit of a bound neutron lifetime is \SI{6.45e32}{years} for a \SI{400}{\ktyr} exposure. The corresponding limit for the oscillation time of free neutrons is calculated to be \SI{5.53e8}{\s}. This is approximately an improvement by a factor of two from the current best limit, which comes from \superk~\cite{Abe:2011ky}.  

\section{Other BSM Physics Opportunities}
\label{sec:otheropps}

\subsection{BSM Constraints with Tau Neutrino Appearance} 
With only 19 $\nu_{\tau}$-\dword{cc} and $\bar{\nu}_{\tau}$-\dword{cc} candidates detected with high purity, we have less direct experimental knowledge of tau neutrinos than of any other \dword{sm} particle. Of these, nine $\nu_{\tau}$-\dword{cc} and $\bar{\nu}_{\tau}$-\dword{cc} candidate events with a background of 1.5 events, observed by the DONuT experiment~\cite{Kodama:2000mp,Kodama:2007aa}, were directly produced though $D_S$ meson decays.  The remaining 10 $\nu_{\tau}$-\dword{cc} candidate events with an estimated background of two events, observed by the OPERA experiment~\cite{Guler:2000bd,Agafonova:2018auq}, were produced through the oscillation of a muon neutrino beam. From this sample, a 20\% measurement of $\Delta m^{2}_{32}$ was performed under the assumption that $\sin^22\theta_{23} = 1$.  The \superk and IceCube experiments developed methods to statistically separate samples of $\nu_{\tau}$-\dword{cc} and $\bar{\nu}_{\tau}$-\dword{cc} events in atmospheric neutrinos to exclude the no-tau-neutrino appearance hypothesis at the 4.6$\sigma$ level and 3.2$\sigma$ level respectively~\cite{Abe:2012jj,Li:2017dbe,Aartsen:2019tjl}, but limitations of Cherenkov detectors constrain the ability to select a high-purity sample and perform precision measurements.

\begin{figure}[htb]
 \centering
        \includegraphics[width=0.9\columnwidth]{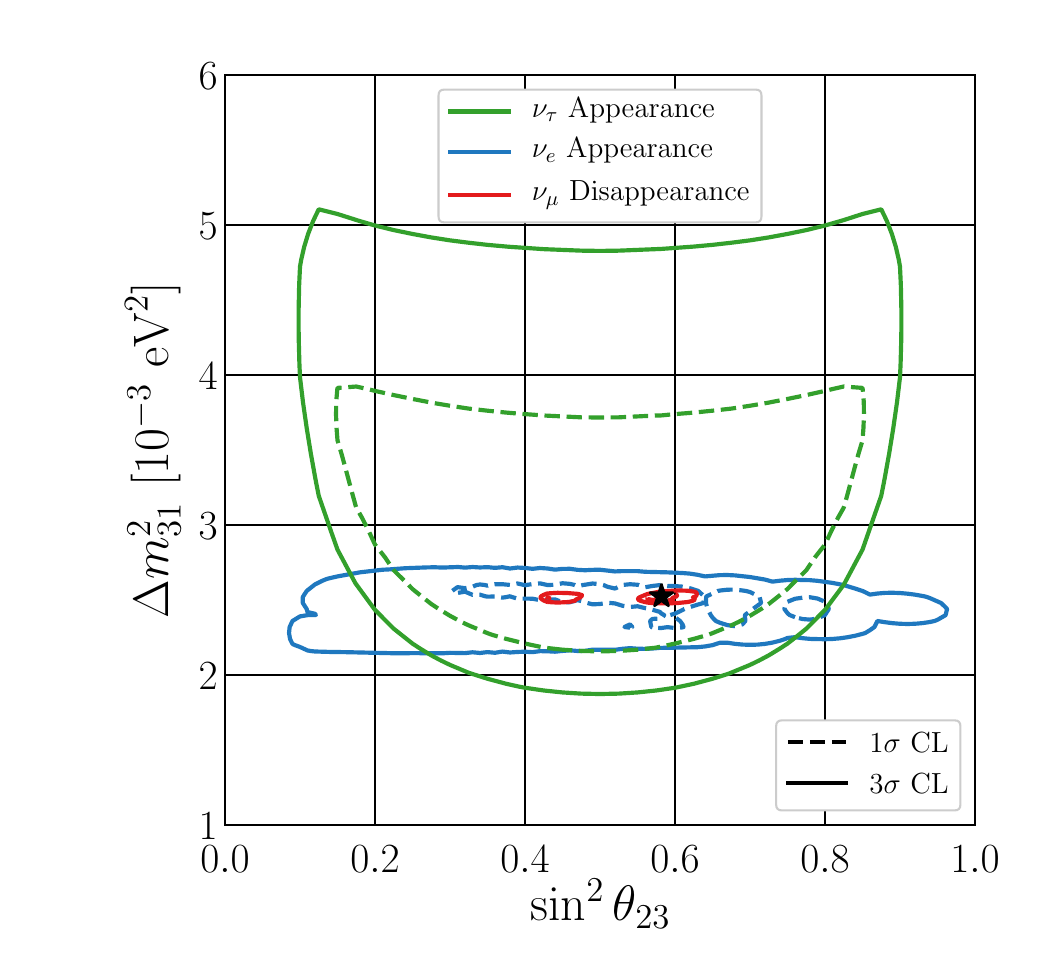}
        \includegraphics[width=0.9\columnwidth]{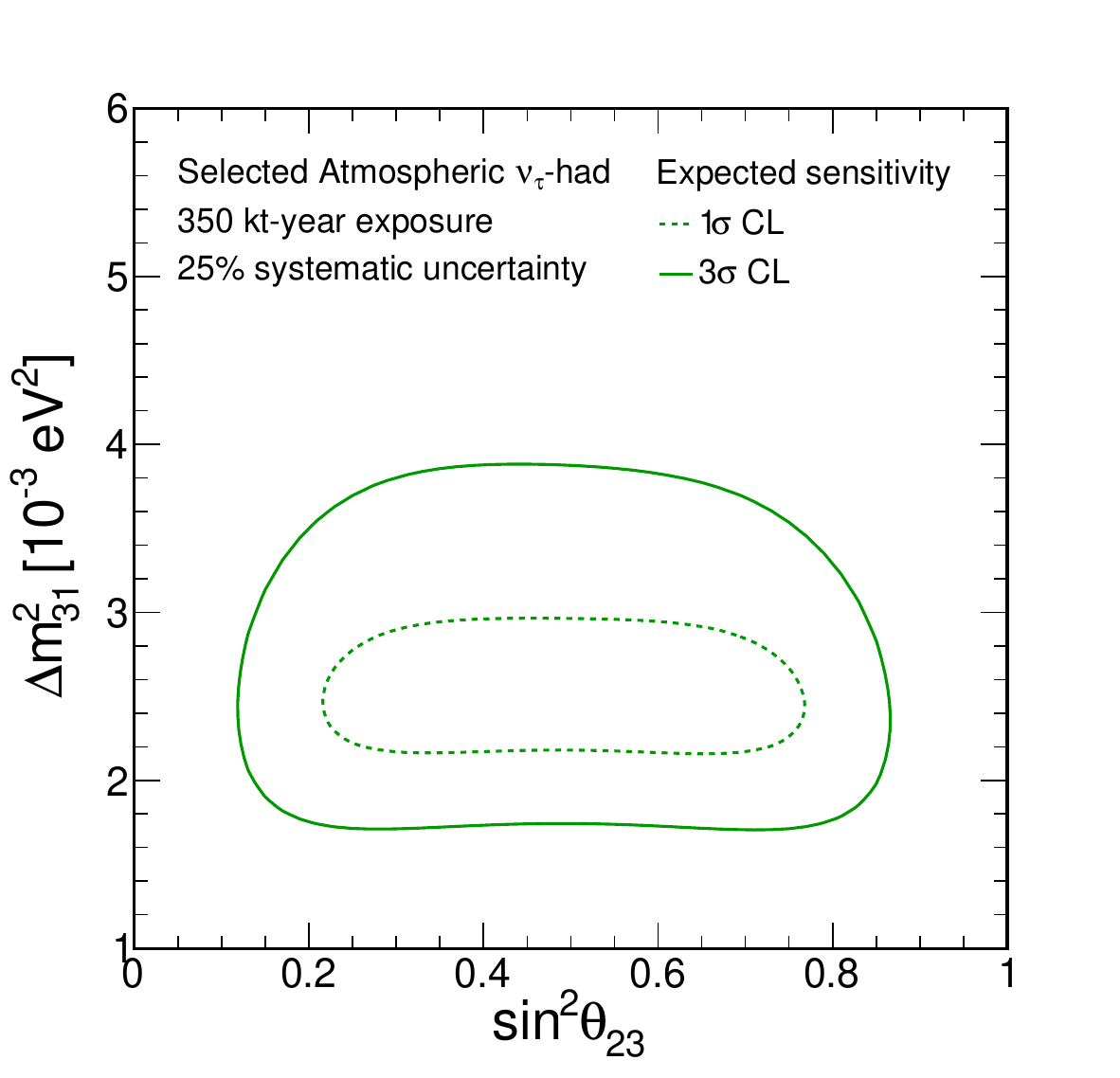}
	\caption[The 1$\sigma$ and 3$\sigma$ expected sensitivity for measuring $\Delta m^2_{31}$ and $\sin^2\theta_{23}$] 
	{The 1$\sigma$ (dashed) and 3$\sigma$ (solid) expected sensitivity for measuring $\Delta m^2_{31}$ and $\sin^2\theta_{23}$ using a variety of samples. Top: The expected sensitivity for seven years of beam data collection, assuming 3.5 years each in neutrino and antineutrino modes, measured independently using $\nu_{e}$ appearance (blue), $\nu_{\mu}$ disappearance (red), and $\nu_{\tau}$ appearance (green). Adapted from Ref.~\cite{deGouvea:2019ozk}. Bottom: The expected sensitivity for the $\nu_{\tau}$ appearance channel using 350~\ktyr of atmospheric exposure.}
	\label{fig:nutauContours}
\end{figure}

The DUNE experiment has the possibility of significantly improving the experimental situation~\cite{machado2020tau}. Tau-neutrino appearance can potentially improve the discovery potential for sterile neutrinos, \dword{nc} \dword{nsi}, and non-unitarity. This channel could also be used as a probe of secret couplings of neutrinos to new light bosons \cite{Bakhti:2018avv}. For model independence, the first goal should be measuring the atmospheric oscillation parameters in the $\nu_{\tau}$ appearance channel and checking the consistency of this measurement with those performed using the $\nu_{\mu}$ disappearance channel.  A truth-level study of $\nu_{\tau}$ selection in atmospheric neutrinos in a large, underground LArTPC detector suggested that $\nu_{\tau}$-\dword{cc} interactions with hadronically decaying $\tau$-leptons, which make up 65\% of total $\tau$-lepton decays~\cite{Tanabashi:2018oca}, can be selected with high purity~\cite{Conrad:1008}.  This analysis suggests that it may be possible to select up to 30\% of $\nu_{\tau}$-\dword{cc} events with hadronically decaying $\tau$-leptons with minimal neutral-current background.  Under these assumptions, we expect to select $\sim$25 $\nu_{\tau}$-\dword{cc} candidates per year using the \dword{cpv} optimized beam. The physics reach of this sample has been studied in Ref.~\cite{deGouvea:2019ozk} and \cite{Ghoshal:2019pab}. As shown in Fig.~\ref{fig:nutauContours} (top), this sample is sufficient to simultaneously constrain $\Delta m^2_{31}$ and $\sin^22\theta_{23}$. Independent measurements of $\Delta m^2_{31}$ and $\sin^22\theta_{23}$ in the $\nu_{e}$ appearance, $\nu_{\mu}$ disappearance, and $\nu_{\tau}$ appearance channels should allow DUNE to constrain $|U_{e3}|^2+|U_{\mu 3}|^2+|U_{\tau 3}|^2$ to 6\%~\cite{deGouvea:2019ozk}, a significant improvement over current constraints~\cite{Parke:2015goa}.

However, all of the events in the beam sample occur at energies higher than the first oscillation maximum due to kinematic constraints.  Only seeing the tail of the oscillation maximum creates a partial degeneracy between the measurement of $\Delta m^2_{31}$ and $\sin^22\theta_{23}$.  Atmospheric neutrinos, due to sampling a much larger $L/E$ range, allow for measuring both above and below the first oscillation maximum with $\nu_{\tau}$ appearance. Although we only expect to select $\sim$70 $\nu_{\tau}$-\dword{cc} and $\bar{\nu}_{\tau}$-\dword{cc} candidates in 350~\ktyr in the atmospheric sample, as shown in Fig.~\ref{fig:nutauContours} (bottom), a direct measurement of the oscillation maximum breaks the degeneracy seen in the beam sample. The complementary shapes of the beam and atmospheric constraints combine to reduce the uncertainty on $\sin^2\theta_{23}$, directly leading to improved unitarity constraints.  Finally, a high-energy beam option optimized for $\nu_{\tau}$ appearance should produce $\sim$150 selected  $\nu_{\tau}$-\dword{cc} candidates in one year~\cite{Acciarri:2015uup}.  These higher energy events are further in the tail of the first oscillation maximum, but they will permit a simultaneous measurement of the $\nu_{\tau}$ cross section. When analyzed within the non-unitarity framework described in Section~\ref{sec:nonUnitarity}, the high-energy beam significantly improves constraints on the parameter $\alpha_{\tau\tau}$ due to increased matter effects~\cite{deGouvea:2019ozk}.

%%%%%%%%%%%%%%%%%%%%%%%%%
\subsection{Large Extra-Dimensions}
DUNE can search for or constrain the size of large extra-dimensions (LED) 
by looking for distortions of the oscillation pattern predicted by the three-flavor paradigm. These distortions arise through mixing between the right-handed neutrino Kaluza-Klein modes, which propagate in the compactified extra dimensions, and the active neutrinos, which exist only in the four-dimensional brane~\cite{Dienes:1998sb,ArkaniHamed:1998vp,Davoudiasl:2002fq}. Such distortions are determined by two parameters in the model, specifically $R$, the radius of the circle where the extra-dimension is compactified, and $m_0$, defined as the lightest active neutrino mass ($m_1$ for normal mass ordering, and $m_3$ for inverted mass ordering). Searching for these distortions in, for instance, the $\nu_\mu$~\dword{cc} disappearance spectrum, should provide significantly enhanced sensitivity over existing results from the MINOS/MINOS+ experiment~\cite{Adamson:2016yvy}.

\begin{figure}[ht]
\centerline{
\includegraphics[width=0.9\linewidth]{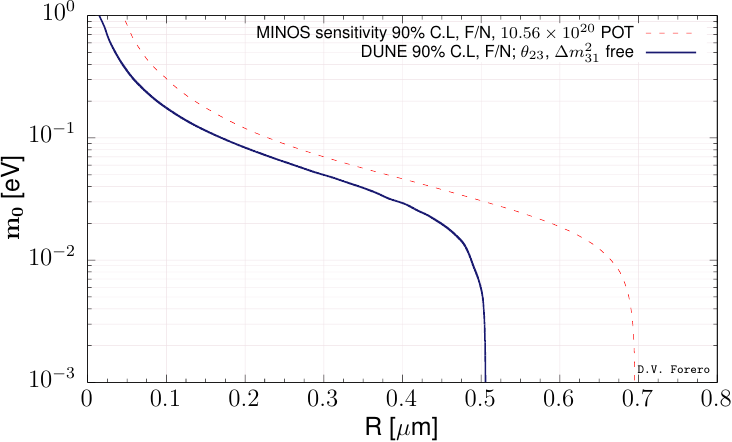}
}
\caption[DUNE sensitivity to the LED model]{Sensitivity to the LED model in Ref.~\cite{Dienes:1998sb,ArkaniHamed:1998vp,Davoudiasl:2002fq} through its impact on the neutrino oscillations expected at 
DUNE. For comparison, the MINOS sensitivity~\cite{Adamson:2016yvy} is also shown.}
\label{fig:ledsensitivity}
\end{figure}

Figure~\ref{fig:ledsensitivity} shows a comparison between the DUNE and MINOS~\cite{Adamson:2016yvy} 
sensitivities to LED at $90\%$ \dword{cl} for 2 d.o.f represented by the solid and dashed lines, respectively. 
In the case of DUNE, an exposure of 300~\ktMWyr was assumed and spectral information from the four oscillation channels, (anti)neutrino 
appearance and disappearance, were included in the analysis. The muon (anti)neutrino 
fluxes, cross sections for the neutrino interactions in argon, detector energy 
resolutions, efficiencies and systematical errors were taken into account by the use of 
\dword{globes} files prepared for the DUNE LBL studies. In the analysis, we assumed DUNE 
simulated data as compatible with the standard three neutrino hypothesis (which corresponds to the limit $R\to 0$) and we have 
tested the LED model. The solar parameters were kept fixed, and also the reactor mixing 
angle, while the atmospheric parameters were allowed to float free. In general, DUNE 
improves over the MINOS sensitivity for all values of $m_0$ and this is more noticeable 
for $m_0\sim 10^{-3}$~eV, where the most conservative sensitivity limit to $R$ is 
obtained. 

%%%%%%%%%%%%%%%%%%%%%%%%%
\subsection{Heavy Neutral Leptons}
\label{sec:hnl}
The high intensity of the LBNF neutrino beam and the production of charm mesons in the beam enables DUNE to search for a wide variety of lightweight long-lived, exotic particles, by looking for topologies of rare event interactions and decays in the fiducial volume of the DUNE \dword{nd}. These particles include weakly interacting heavy neutral leptons (HNLs) as right-handed partners of the active neutrinos, light super-symmetric particles, or vector, scalar, and/or axion portals to a Hidden Sector containing new interactions and new particles. 
Assuming the heavy neutral leptons are the lighter particles of their hidden sector, they will only decay into \dword{sm} particles. The parameter space that can be explored by the DUNE \dword{nd} extends into the cosmologically relevant region, and will be complementary to the LHC heavier mass  searches.

Thanks to small mixing angles, the particles can be stable enough to travel from the production in the proton target 
to the detector and decay inside the active region.
It is worth noting that, differently from a light neutrino beam, an HNL beam is not polarized, due to the large mass of the HNLs.
The correct description of the helicity components in the beam is important for predicting the angular distributions
of HNL decays, as they might depend on the initial helicity state.
More specifically, there is a different phenomenology if the decaying HNL is a Majorana or a Dirac fermion~\cite{Balantekin:2018ukw,Ballett:2019bgd}.
Typical decay channels are two-body decays into a charged lepton and a pseudo-scalar meson, or a vector meson if
the mass allows it, and three-body leptonic decays.

\begin{figure*}[htp]
	\begin{center}
	  	\includegraphics[width=0.9\textwidth]{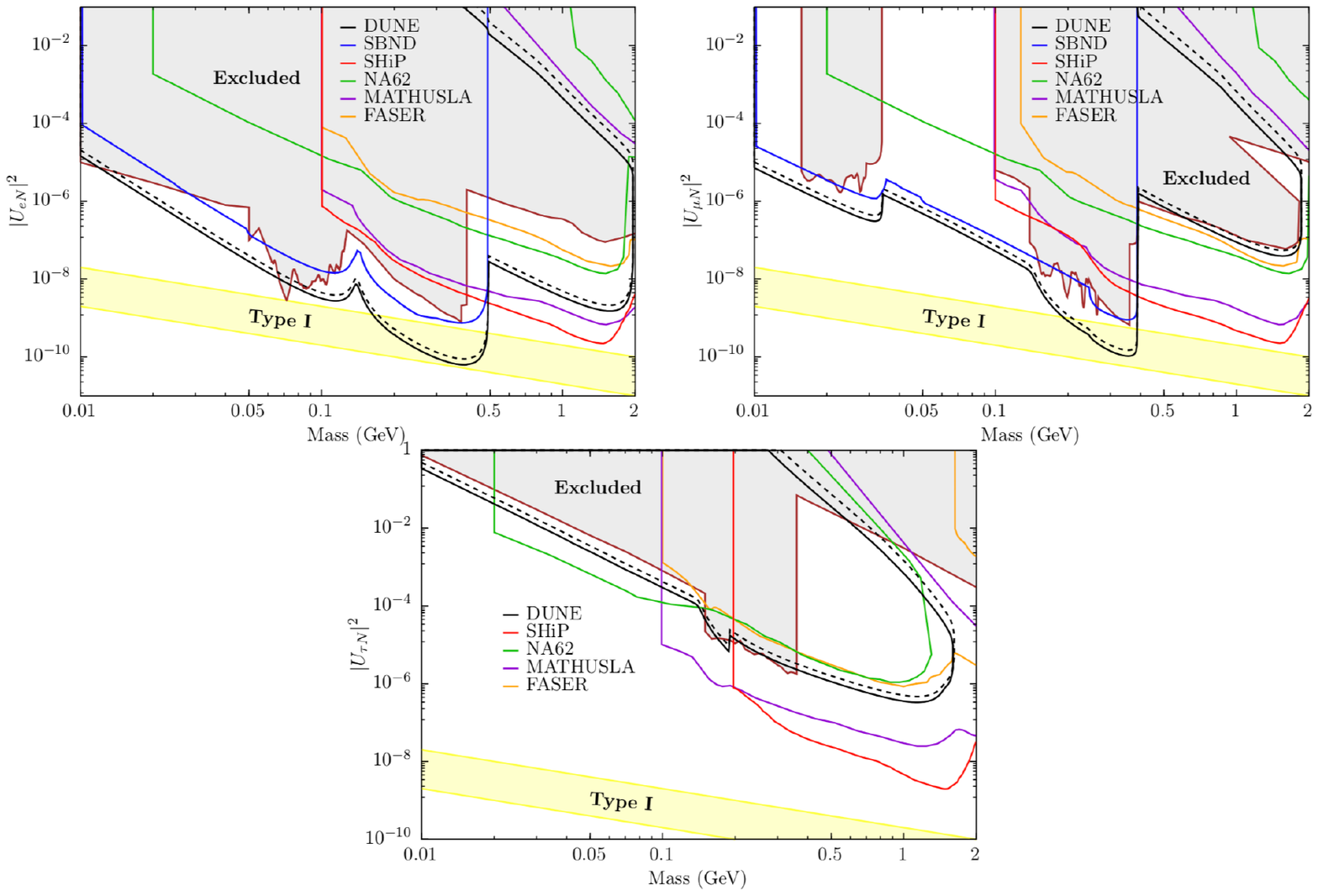}
	\end{center}
\caption[The 90\,\% \dword{cl} sensitivity regions for dominant mixings
		$|U_{\alpha N}|^2$]
		{The 90\,\% \dword{cl} sensitivity regions for dominant mixings %
		$|U_{e N}|^2$ (top~left), $|U_{\mu N}|^2$ (top right), and $|U_{\tau N}|^2$ (bottom) are presented for DUNE ND (black)~\cite{Ballett:2019bgd}.
		The regions are a combination of the sensitivity to HNL decay channels with good detection prospects.These are $N\to\nu e e$, $\nu e \mu$, $\nu \mu \mu$, $\nu \pi^0$, $e \pi$, and $\mu \pi$.The study is performed for Majorana neutrinos (solid) and Dirac neutrinos (dashed), %
		assuming no background. The region excluded by experimental constraints (grey/brown) is obtained by combining the results from PS191~\cite{Bernardi:1985ny,Bernardi:1987ek}, %
		peak searches~\cite{Artamonov:2014urb,Britton:1992pg,Britton:1992xv,Aguilar-Arevalo:2017vlf,Aguilar-Arevalo:2019owf}, %
		CHARM~\cite{Vilain:1994vg}, NuTeV~\cite{Vaitaitis:1999wq}, DELPHI~\cite{Abreu:1996pa}, and T2K~\cite{Abe:2019kgx}. The sensitivity for DUNE ND is compared to the predictions of future experiments, SBN~\cite{Ballett:2016opr} (blue), %
		SHiP~\cite{Alekhin:2015byh} (red), NA62~\cite{Drewes:2018gkc} (green), MATHUSLA~\cite{Curtin:2018mvb} (purple), and the Phase II of FASER~\cite{Kling:2018wct}.
		For reference, a band corresponding to the contribution light neutrino masses between 20~meV and 200~meV in a single generation see-saw type I model is shown (yellow).
		Larger values of the mixing angles are allowed if an extension to see-saw models is invoked,
		for instance, in an inverse or extended see-saw scheme.}
\label{fig:sensa_hnl}
\end{figure*}

A recent study illustrates the potential sensitivity for  HNL searches with the DUNE \dword{nd}~\cite{Ballett:2019bgd}. The sensitivity for HNL particles with masses in the range of 10 MeV to 2 GeV, from decays of mesons produced
in the proton beam dump that produces the pions for the neutrino beam production, was studied. The production
of $D_s$ mesons gives access to high mass part of the HNL production. The dominant HNL decay modes to SM particles
have been included, as well as the basic detector constraints, and dominant background processes have 
been considered. 

The experimental signature for these decays is a decay-in-flight event with no interaction vertex, typical of
neutrino--nucleon scattering, and a rather forward direction with respect to the beam.
The main background to this search comes from SM neutrino--nucleon scattering events in which the hadronic activity
at the vertex is below threshold.
Charged-current quasi-elastic events with pion emission from resonances are background to the semi-leptonic decay channels,
whereas misidentification of long pion tracks as muons can constitute a background to three-body leptonic decays.
Neutral pions are often emitted in neutrino scattering events and can be a challenge for decays into a %
neutral meson or channels with electrons in the final state.

We report in Fig.~\ref{fig:sensa_hnl} the physics reach of the DUNE ND in its current configuration %
without backgrounds for a Majorana and a Dirac HNL.
The sensitivity was estimated assuming a total of 1.32 x $10^{22}$ POT, i.e., for a running scenario with 6 years with a 80 GeV proton beam of 1.2 MW, followed by six years of a beam with 2.4 MW, but using only the neutrino mode configuration, which corresponds to half of the total
runtime.
As a result, a search can be conducted for HNLs with masses up to 2 GeV 
%can be searched for 
in all flavor-mixing channels.

The results show that DUNE will have an improved sensitivity to small values of the
mixing parameters $|U_{\alpha N}|^2$, where $\alpha=e,\,\mu,\,\tau$, compared to the presently available experimental
limits on mixing of HNLs with the three lepton flavors. At 90\% \dword{cl} sensitivity, DUNE can probe mixing parameters as low as 
$10^{-9}-10^{-10}$ in the mass range of 300-500 MeV for  mixing with the electron or muon neutrino flavors. In the region above 500 MeV the sensitivity
is reduced to $10^{-8}$ for $eN$ mixing and $10^{-7}$ for $\mu N$ mixing. The $\tau N$ mixing 
sensitivity is weaker but still covering a new unexplored regime. A large fraction of the covered parameter space for all neutrino flavors falls in the region that is relevant for explaining the baryon asymmetry in the universe.

Studies are ongoing with full detector simulations to validate these 
encouraging results.

\subsection{Dark Matter Annihilation in the Sun}
DUNE's large \dword{fd} LArTPC modules provide an excellent setting to conduct searches for neutrinos arising from \dword{dm} annihilation in the core of the sun. These would typically result in a high-energy neutrino signal almost always accompanied by a low-energy neutrino component, which has its origin in a hadronic cascade that develops in the dense solar medium and produces large numbers of light long-lived mesons, such as $\pi^+$ and $K^+$ that then stop and decay at rest. The decay of each $\pi^+$ and $K^+$ will produce monoenergetic neutrinos with an energy \SI{30}{MeV} or \SI{236}{MeV}, respectively.
The  \SI{236}{MeV} flux can be measured with the DUNE \dword{fd}, thanks to its excellent energy resolution, and importantly, will benefit from directional information. By selecting neutrinos arriving from the direction of the sun, large reduction in backgrounds can be achieved.
This directional resolution for sub-GeV neutrinos will enable DUNE to be competitive with experiments with even larger fiducial masses, but less precise angular information, such as Hyper-K~\cite{ref:DMannihilation}.

\section{Conclusions and Outlook}
\label{sect:conclusion}
DUNE will be a powerful discovery tool for a variety of physics topics under very active exploration today, from the potential discovery of new particles beyond those predicted in the \dword{sm}, to precision neutrino measurements that may uncover deviations from the present three-flavor mixing paradigm and unveil new interactions and symmetries.
The \dword{nd} alone will offer excellent opportunities to search for light \dword{dm} and to measure rare processes such as neutrino trident interactions. Besides enabling its potential to place leading constraints on deviations from the three-flavor oscillation paradigm, such as light sterile neutrinos and non-standard interactions, DUNE's massive high-resolution \dword{fd} will probe the possible existence of baryon number violating processes and \dword{bdm}. The flexibility of the LBNF beamline opens prospects for high-energy beam running, providing access to probing and measuring tau neutrino physics with unprecedented precision.
Through the ample potential for BSM physics, DUNE offers an opportunity for strong collaboration between theorists and experimentalists and will provide significant opportunities for breakthrough discoveries in the coming decades.

\begin{acknowledgements}
This document was prepared by the DUNE collaboration using the
resources of the Fermi National Accelerator Laboratory 
(Fermilab), a U.S. Department of Energy, Office of Science, 
HEP User Facility. Fermilab is managed by Fermi Research Alliance, 
LLC (FRA), acting under Contract No. DE-AC02-07CH11359.
%
% Funding agencies, alphabetical by country, then alphabetical by agency name
%
This work was supported by
CNPq, FAPERJ, FAPEG and FAPESP,              Brazil;
CFI, IPP and NSERC,                          Canada;
CERN;
M\v{S}MT,	                                 Czech Republic;
ERDF, H2020-EU and MSCA,                     European Union;
CNRS/IN2P3 and CEA,                          France;
INFN,                                        Italy;
FCT,                                         Portugal;
NRF,                                         South Korea;
CAM, Fundaci\'{o}n ``La Caixa'' and MICINN,  Spain;
SERI and SNSF,                               Switzerland;
T\"UB\.ITAK,                                 Turkey;
The Royal Society and UKRI/STFC,             United Kingdom;
DOE and NSF,                                 United States of America.
\end{acknowledgements}